\documentclass[twocolumn,superscriptaddress,floatfix,longbibliography,aps,pra,nopreprintnumbers,noeprint,10pt]{revtex4-2}

\usepackage[utf8]{inputenc}
\usepackage[english]{babel}
\usepackage{graphicx}
\usepackage{bm}
\usepackage{bbm}
\usepackage{xcolor}
\usepackage{amsmath}
\usepackage{amssymb}
\usepackage{epstopdf}
\usepackage[normalem]{ulem} %strikethrough \sout
\usepackage{enumerate}
\usepackage[urlcolor=blue,colorlinks=true,citecolor=blue,linkcolor=blue,pdfstartview={FitH},bookmarks=false]{hyperref}
\graphicspath{{Figures/}{./Figures/}{.}}

\begin{document}

\title{Dynamical Quantum Phase Transitions Following Double Quenches:\\Persistence of the Initial State vs Dynamical Phases}

\author{Hadi Cheraghi}
\affiliation{Institute of Physics, Maria Curie-Sk\l{}odowska University, 20-031 Lublin, Poland}
\author{Nicholas Sedlmayr}
\email[e-mail: ]{sedlmayr@umcs.pl}
\affiliation{Institute of Physics, Maria Curie-Sk\l{}odowska University, 20-031 Lublin, Poland}

\date{\today}

\begin{abstract}
Dynamical quantum phase transitions can occur following quenches in quantum systems when the rate function, a dynamical analogue of the free energy, becomes non-analytic at critical times. Here we exhaustively investigate in an exemplary model how the dynamically evolving state responds to a second quench. We demonstrate that for quenches where the initial and final Hamiltonian belong to different phases always result in dynamical quantum phase transitions, irrespective of the intermediate quench and dynamics or the time of the second quench. However, if the initial and final Hamiltonian belong to the same equilibrium phase then the intermediate Hamiltonian must belong to a different phase. In this case, the second quench time in relation to the critical times of the first quench becomes crucial to the existence of dynamical quantum phase transitions.
\end{abstract}

\maketitle

%%%%%%%%%%%%%%%%%%%%%%%%%%%%%%%%%%%%%%%%%%%%%%%%
\section{Introduction}
%%%%%%%%%%%%%%%%%%%%%%%%%%%%%%%%%%%%%%%%%%%%%%%%

In recent years impressive progress in the precise control and manipulation of synthetic quantum  matter has provided a platform to investigate quantum systems in non-equilibrium states. In particular, those which originate from a quantum quench, allowing access to new regimes of behavior not possible in equilibrium~\cite{Streltsov2017,Basov2017,Martinez2016,Halimeh2017}. One example of such behavior is a dynamical quantum phase transition (DQPT), when non-analyticities show up in an analog of the free energy at critical \emph{times}~\cite{Heyl2013,Heyl2018a,Sedlmayr2019a}. The first work focused on the Ising model~\cite{Heyl2013}, but it was soon shown to hold for a wide variety of models and a wealth of theoretical work has followed~\cite{Heyl2015,Sharma2015,Homrighausen2017,Halimeh2018,Shpielberg2018,Zunkovic2018,Srivastav2019,Hagymasi2019,Huang2019,Gurarie2019,Abdi2019,Cao2020,Puebla2020,Soriente2020,Porta2020,Mishra2020,Link2020,Sun2020,Rylands2020,Uhrich2020,Syed2021,Trapin2021,Yu2021,Zhou2021c,Sadrzadeh2021,Halimeh2021,Halimeh2021a,DeNicola2021,Rylands2021,Peotta2021,Modak2021,Cheraghi2021,Cao2021,Bandyopadhyay2021}, including for topological matter~\cite{Vajna2015,Schmitt2015,Jafari2016,Jafari2017a,Sedlmayr2018,Jafari2018,Zache2019,Maslowski2020,Okugawa2021}, for non-Hermitian dynamics~\cite{Zhou2018,Zhou2021,Mondal2022}, and for Floquet systems~\cite{Sharma2014,Yang2019,Zamani2020,Zhou2021a,Jafari2021,Hamazaki2021,Zamani2022,Luan2022,Jafari2022}. In topological band insulators and superconductors~\cite{Hasan2010,Teo2010} DQPTs have been related to dynamical order parameters~\cite{Budich2016,Heyl2017,Bhattacharya2017a} and a dynamical bulk-boundary correspondence~\cite{Sedlmayr2018,Sedlmayr2019a,Maslowski2020}. Though these models tend to be simple one dimensional two band models, multi-band models have also been considered~\cite{Huang2016,Jafari2019,Mendl2019,Maslowski2020}, along with two dimensional systems~\cite{Vajna2015,DeNicola2022,Hashizume2022,Brange2022}. Experimentally DQPTs have been investigated in various ion trap, cold atom, and quantum simulator platforms~\cite{Jurcevic2017,Flaschner2018,Zhang2017b,Guo2019,Smale2019,Nie2020,Tian2020}

Although performing a quench across a quantum critical point is often enough to induce DQPTs~\cite{Heyl2013,Karrasch2013,Heyl2014}, there is in general no one-to-one correspondence between the equilibrium phase diagram and the occurrence of DQPTs~\cite{Vajna2014,Andraschko2014,Vajna2015,Karrasch2017,Jafari2017,Jafari2017a,Cheraghi2018,Jafari2019,Wrzesniewski2022}. Alternative protocols to a sudden quench include using slow quenches~\cite{Puskarov2016,Sharma2016,Bhattacharya2017} or a second quench following the first after some time has elapsed~\cite{Kennes2018,Hou2022}. The typical protocol to induce the dynamics is to quench from a ground state, though mixed state, finite temperatures and open systems have also been investigated~\cite{Mera2017,Sedlmayr2018b,Bhattacharya2017a,Heyl2017,Abeling2016,Lang2018,Lang2018a,Kyaw2020,Starchl2022,Naji2022,Kawabata2022}.

DQPTs are defined via the temporal behaviour of the rate function, an analogue of the free energy defined for the Loschmidt echo rather than the partition function. The Loschmidt echo is the overlap between an initial and a time evolved state. Here, as in most works, we focus on ground states of a Hamiltonian $H_i$ as the initial states and then time evolve with another Hamiltonian $H_m$. During time evolution the Loschmidt echo develops zeros that correspond to non-analyticities in the rate function. These are referred to as dynamical quantum phase transitions by analogy with the non-analytic points in the free energy as a function of a control parameter which define phase transitions. In this case, the seminal work was done the case of the nearest-neighbor transverse-field Ising chain~\cite{Heyl2013} where non-analyticities occur only when the phases of the initial and the driving Hamiltonian are chosen from two different phases. The non-analyticities in the real-time dynamics of the rate function emerge at particular critical times.

On the other hand, borrowing an idea from equilibrium statistical mechanics, one can consider the notion of a local order parameter in the long-time steady state of a quantum many-body system~\cite{Halimeh2017a}. This can be observed experimentally in a trapped ion based quantum simulator~\cite{Zhang2017b}. The main idea here comes back to recognizing the DQPT via a suitable order parameter in a prethermal regime before relaxation to thermal equilibrium has occurred. It is therefore natural to ask whether the DQPTs divide dynamical phases in any meaningful sense. As there is often, though by no means always, a simple relation between the equilibrium phases of the initial and time-evolving Hamiltonian, and the existence of DQPTs, which occurs due to a second quench following the first sometime later.

Multiple quenches have been studied in only several works~\cite{Kennes2018,Hou2022}. In Ref.~\cite{Kennes2018} it has been shown that for quenches of a type A$\to$B$\to$A, where A and B label different equilibrium phases, that DQPTs is by no means inevitable and can be tuned away by choosing the second quench time $T$ appropriately. Here we extend this analysis by considering all possible quench scenarios. For an exemplary model, we demonstrate which double quench scenarios inevitably do or do not lead to DQPTs and the conditions under which the other possibilities may. It is shown that, much as in Ref.~\cite{Kennes2018}, some properties of the initial state remain important for all times, and that some properties belong to the dynamically evolving state. Although it is not really possible to talk about dynamical phases that change across the DQPTs, the critical times where DQPTs occur are important to the possibility of DQPTs following a second quench. By considering the second quench time $T$ as an independent variable we see that for $T$ near critical times particular behavior emerges.

In Ref.~\cite{Hou2022} so called metamorphic DQPTs are studied, where the final state becomes orthogonal to the initial state for all times following the second quench, and hence the rate function is non-analytic for these times~\footnote{We note here however that the discontinuities in the rate function seen in Ref.~\cite{Hou2022} are hard to understand from a unitary time evolution perspective, and we see no evidence for such behavior in our results.}. It is straightforward to prove that for this to occur either the initial state must be an eigenstate of the final Hamiltonian and the second quench must occur precisely at a critical time of the first quench, or the state at the critical time must be an eigenstate of the second quench Hamiltonian and the second quench must still occur precisely at a critical time of the first quench. Strictly speaking, if all three Hamiltonians can be block diagonalised, then this is only necessary for a single sub-block, however in this case the critical time must belong to this sub-block also.  For a typical model however this will in any case require that the second quench Hamiltonian and the Hamiltonian defining the initial state are, up to a scale factor, identical. In this case there are no dynamics in the Loschmidt echo following the second quench. Our model indeed recovers this behaviour, but in practice the second quench time must be absolutely precisely on the critical time for it to occur, which is in practice impossible to achieve. Therefore we will not focus further on metamorphic DQPTs in this work.

This paper is organized as follows. In section \ref{sec:model} we introduce, in both spin and fermionic languages, the exemplary model we will solve, and in section \ref{sec:quenches} we solve the quench dynamics for this model for a single and double quenches. In section \ref{sec:dqpt} we find the conditions for DQPTs following the second quench for all possible scenarios for our model system, and discuss the general conclusion that one can draw. In section \ref{sec:conc} we conclude.

%%%%%%%%%%%%%%%%%%%%%%%%%%%%%%%%%%%%%%%%%%%%%%%%
\section{The model}\label{sec:model}
%%%%%%%%%%%%%%%%%%%%%%%%%%%%%%%%%%%%%%%%%%%%%%%%

Here we focus on an exemplary one-dimensional model, the spin-$\frac{1}{2}$ XY model in the presence of a transverse magnetic field, with ferromagnetic exchange coupling $J>0$, an anisotropy parameter $\delta$ which we will set equal to 1, and an applied magnetic field of magnitude $h$~\cite{Lieb1961}:
\begin{eqnarray}
H&=& -\frac{J}{2} \sum_{n=1}^N \big[(1+\delta)\sigma_n^x \sigma_{n+1}^x+(1-\delta)\sigma_n^y \sigma_{n+1}^y\big] \nonumber\\
&-& h \sum_{n=1}^N \sigma_n^z.      
\end{eqnarray}
Here $\sigma_n^{x,y,z}$ is the Pauli spin operator at site $n$. Following a Jordan-Wigner transformation to spinless fermionic creation and annihilation operators, $c_n^\dag$ and $c_n$, this can be written as
\begin{align}
{\cal H}=&  -J \sum_{n = 1}^N \left[ c_n^\dag c_{n + 1} + \delta c_n^\dag c_{n + 1}^\dag  + \textrm{H.c.} \right] \nonumber\\
& -h\sum_{n = 1}^N
\left(2c_n^\dag {c_n}-1 \right),
\end{align}
which is equivalent to the Kitaev chain~\cite{Kitaev2001}, a one dimensional topological superconductor. Throughout we will focus on the case of periodic boundary conditions, $c_{N+n}=c_n$. The model exhibits a quantum (XY) or topological (Kitaev) phase transition at a critical field $h_c = J$, from a ferromagnetic (FM) phase ($h < J$) to a paramagnetic (PM) spin-polarized phase ($h > J$). Equivalently from a topologically non-trivial to trivial phase for the Kitaev chain.  Because we are interested in bulk properties and the Loschmidt echo, all results that follow are equivalent whether thought of in terms of the spin-$\frac{1}{2}$ XY model or the Kitaev chain. Such a relatively simple phase diagram allows us to exhaustively explore the consequences of the quench protocol we will define below.

A Fourier transformation $c_n = (1/\sqrt{N}) \sum _k e^{ikn} c_k$ leads to the diagonal Hamiltonian
\begin{equation}
    {\cal H} = \sum_{k > 0} {\cal {C}}^\dag  H_k {\cal {C}}
\end{equation}
where ${\cal {C}}^\dag= (c_k^\dag ,c_{ - k})$ and $ H_k = \vec{d}_k.\vec\sigma  $ is the Bogoliubov-de Gennes (BdG) Hamiltonian with $\vec{d}_k \equiv (d_x(k),d_y(k),d_z(k))$ given by
\begin{equation}
    \vec{d}_k =(0,-2J \delta \sin(k),-2J\cos(k)-2h).
\end{equation}
This can be straightforwardly diagonalized to give 
\begin{eqnarray}\label{eq3}
{\cal H} = \sum_k \varepsilon _k  (\beta _k^\dag \beta _k -1/2)
\end{eqnarray}
with energy spectrum $\pm\varepsilon_k =\pm|\vec{d}_k| $, and eigenvectors 
\begin{equation}
\left|\chi _k^+\right\rangle 
=\left(\begin{array}{c}
\cos(\theta _k)\\
i\sin (\theta _k)
\end{array}\right)
\textrm{ and }
\left|\chi _k^-\right\rangle 
=\left(\begin{array}{c}
i\sin(\theta_k)\\\cos(\theta_k)
\end{array}\right) 
\end{equation}
for the positive and negative energy states. The angle is defined by $\tan (2 \theta _k) =- d_y(k)/d_z(k) $, with the Bogoliubov transformation $c_k = \cos (\theta _k){\beta _k} + i\sin (\theta_k)\beta _{ - k}^\dag $. The summation in Eq.~\eqref{eq3} runs over $k = 2\pi m/N$, with $m = 0,\pm 1,... ,\pm (N -1)/2$ for odd $N$  or $m = 0,\pm 1,...,\pm (N/2 -1), N/2$ for even $N$.

Throughout this paper we keep $J$ fixed as the energy scale, $\delta=1$, and quench the parameter $h$. The initial state is the ground state of the Hamiltonian with $h=h_i$, the time-evolving Hamiltonian following the first quench at time $t=0$ has $h=h_m$, and the time-evolving Hamiltonian following the first quench at time $t=0$ has $h=h_f$. It is therefore convenient to introduce three angles for each $h_j$, $\theta_k^j$, and spectra $\varepsilon_k^j$, where j can be $i$, $m$, or $f$. Rotations between the eigenbases of each Hamiltonian involve three angles and we define $\theta_k^{mi}\equiv\theta_k^m-\theta_k^i$, $\theta_k^{fm}\equiv\theta_k^f-\theta_k^m$, and $\theta_k^{fi}\equiv\theta_k^f-\theta_k^i$.

%%%%%%%%%%%%%%%%%%%%%%%%%%%%%%%%%%%%%%%%%%%%%%%%
\section{Quenches and dynamical quantum phase transitions}\label{sec:quenches}
%%%%%%%%%%%%%%%%%%%%%%%%%%%%%%%%%%%%%%%%%%%%%%%%

Although here we will focus entirely on double quench dynamics, the idea can naturally be easily extended to multiple quenches. We first fix the initial state $\left|\Psi^i\right\rangle$ of the system as the ground state of the initial Hamiltonian ${\cal H}(h_i)$ in a phase $A$, then, a single quench at $t=0$ will be done meaning the state is evolved by Hamiltonian ${\cal H}(h_m)$ which could be in the same phase A or a different phase B. The system is then time evolve until a given time $t=T$ when the time evolving Hamiltonian is suddenly switched to ${\cal H}(h_f)$ which could be in either phase $A$ or phase $B$. This is the second quench, naturally, the state is continuously evolving. I.e.~the time evolved state is
\begin{equation}
\left|\Psi(t) \right\rangle=
\begin{cases}
  e^{-i{\cal H}(h_m)t}
\left|\Psi^i\right\rangle  & \textrm{ if } 0\leq t\leq T \\
  e^{-i{\cal H}(h_f)t}e^{-i{\cal H}(h_m)T}
\left|\Psi^i\right\rangle & \textrm{ if } t\geq T.
\end{cases}
\end{equation}
The critical times will emerge at the times when the rate function is non-analytic which defines as
\begin{equation}
R(t)=-\frac{1}{N}\log| l(t)|^2
\end{equation}
where
\begin{equation}
l(t)=\left\langle \Psi (t)|\Psi^i \right\rangle
\end{equation}
is the Loschmidt echo, and $N$ is the system size.

Phase transitions can be associated with divergences in suitable physical susceptibilities when continuous control parameters are changed, such as temperature or an external field \cite{Sachdev2011,Carr2011}. Thermodynamically they are described by non-analyticities in the free energy density, which can be pictured using Fisher zeros in complex temperature  \cite{Fisher1965} or, for example, Lee-Yang zeros in complex magnetic plane \cite{Yang1952}. Additionally, at zero temperature the ground state wave function of a many-body system can undergo an abrupt change in its qualitative structure, giving rise to quantum phase transitions~\cite{Zanardi2006,Zanardi2007b}. In the theory of DQPTs~\cite{Heyl2018}, non-analyticities in the free energy as a function of a control parameter become non-analyticities as a function of time in the rate function.  The critical times at which DQPTs occur are when the Fisher zeros, defined by $l(-iz)=0$, cross the imaginary axis, corresponding to real times.

%%%%%%%%%%%%%%%%%%%%%%%%%%%%%%%%%%%%%%%%%%%%%%%%
\subsection{The first quench}
%%%%%%%%%%%%%%%%%%%%%%%%%%%%%%%%%%%%%%%%%%%%%%%%

To reiterate, the first quench starts with the system at $t=0$ in the ground state of the initial Hamiltonian,  $|\Psi_{h_i}\rangle =\prod_{k>0} |\chi _k^-(h_i)\rangle$, and then doing a quench as $h_i$ to $h_m$. Consequently, the Loschmidt echo is
\begin{equation}
l(t) = \prod_{k>0} {\cal L}_k(t)
\end{equation}
where
\begin{equation}\label{lo1}
{\cal L}_k(t)=\cos ^2(\theta_k^{mi})e^{it \varepsilon_k^m} + \sin ^2(\theta_k^{mi})e^{ - it\varepsilon_k^m}.
\end{equation}
The Fisher zeros for $t\leq T$ are
\begin{equation}
z_n(k) = \frac{1}{2\varepsilon_k^m}\left[ \ln ( \tan ^2(\theta_k^{mi}) ) + i\pi (2n + 1) \right]
\end{equation}
Now, critical times appear periodically at 
\begin{equation}
t_n^* = t^*\left[2n + 1 \right]\textrm{ where }n\in\mathbb{W}
\end{equation}
with $t^*=\pi/(2\varepsilon_{k^*}^m)$ under the condition $\cos(2\theta_{k^*}^{mi})=0$, which leads to
\begin{equation} \label{hm}
\left(J \cos(k)+h_m\right) \left(J \cos(k)+h_i\right)+[J\delta \sin(k)]^2=0.
\end{equation}
This can be solved to give the critical momenta as
\begin{equation} \label{hms}
\cos(k^*)=-\frac{h_i+h_m\pm \Delta_k}{2J(1-\delta^2)},
\end{equation}
with $\Delta_k=\sqrt{(h_i+h_m)^2-4(1-\delta^2)(J^2\delta^2+h_ih_m)}$, from which we find the critical times $t^*_c$ for the DQPTs following the first quench. In the following, we focus on the case of $\delta=1$ where the model simplifies to the Ising model. In this case, the critical momentum become $\cos(k^*)=-(J^2+h_ih_m)/(J(h_i+h_m))$ so that
\begin{equation}
t^*=\frac{\pi}{4} \sqrt{\frac{h_i+h_m}{(h_i-h_m)(J^2-h_m^2)}}
\end{equation}
gives the critical times.

\begin{figure*}[t]
\includegraphics[width=0.24\linewidth]{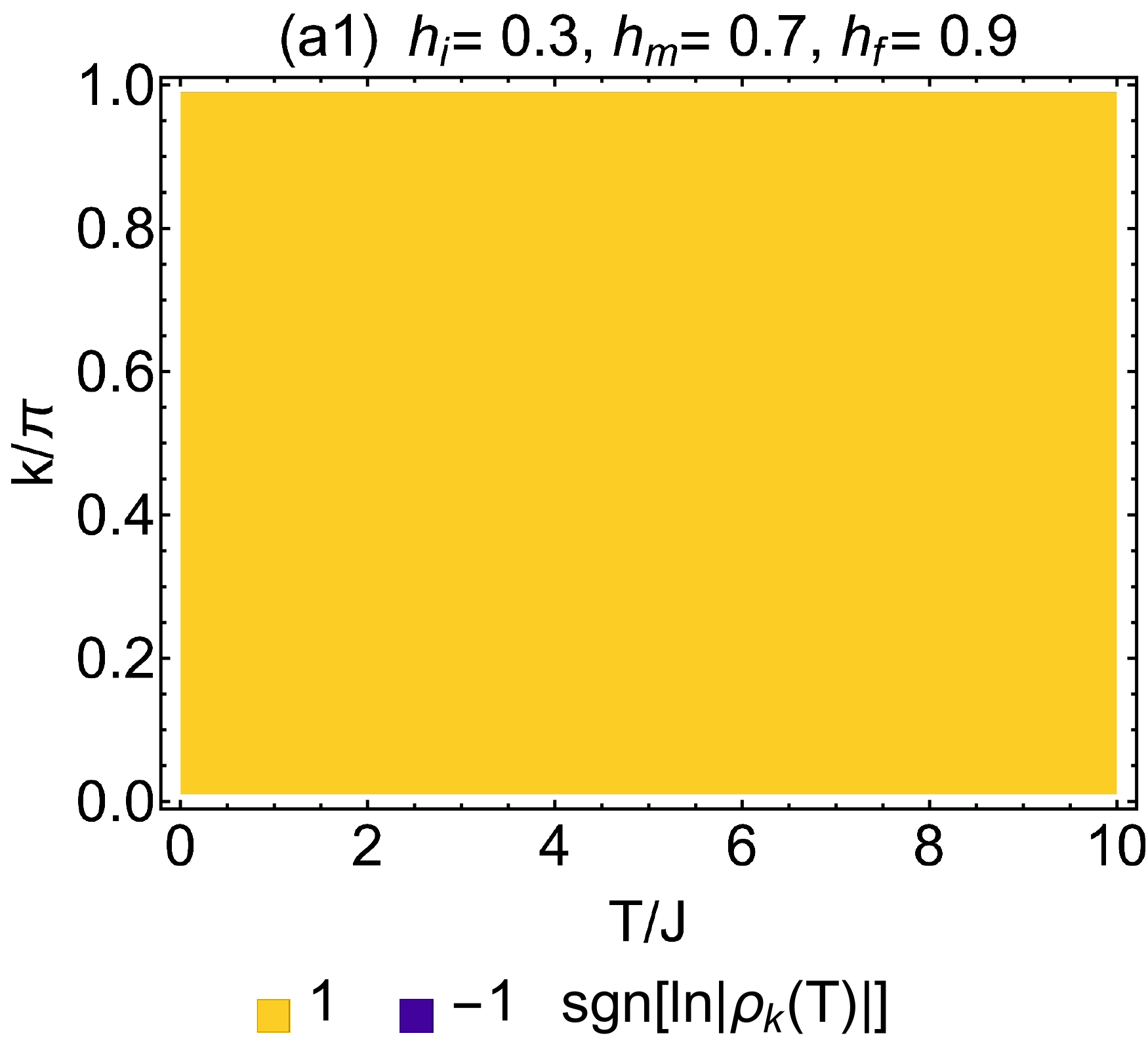}
\includegraphics[width=0.24\linewidth]{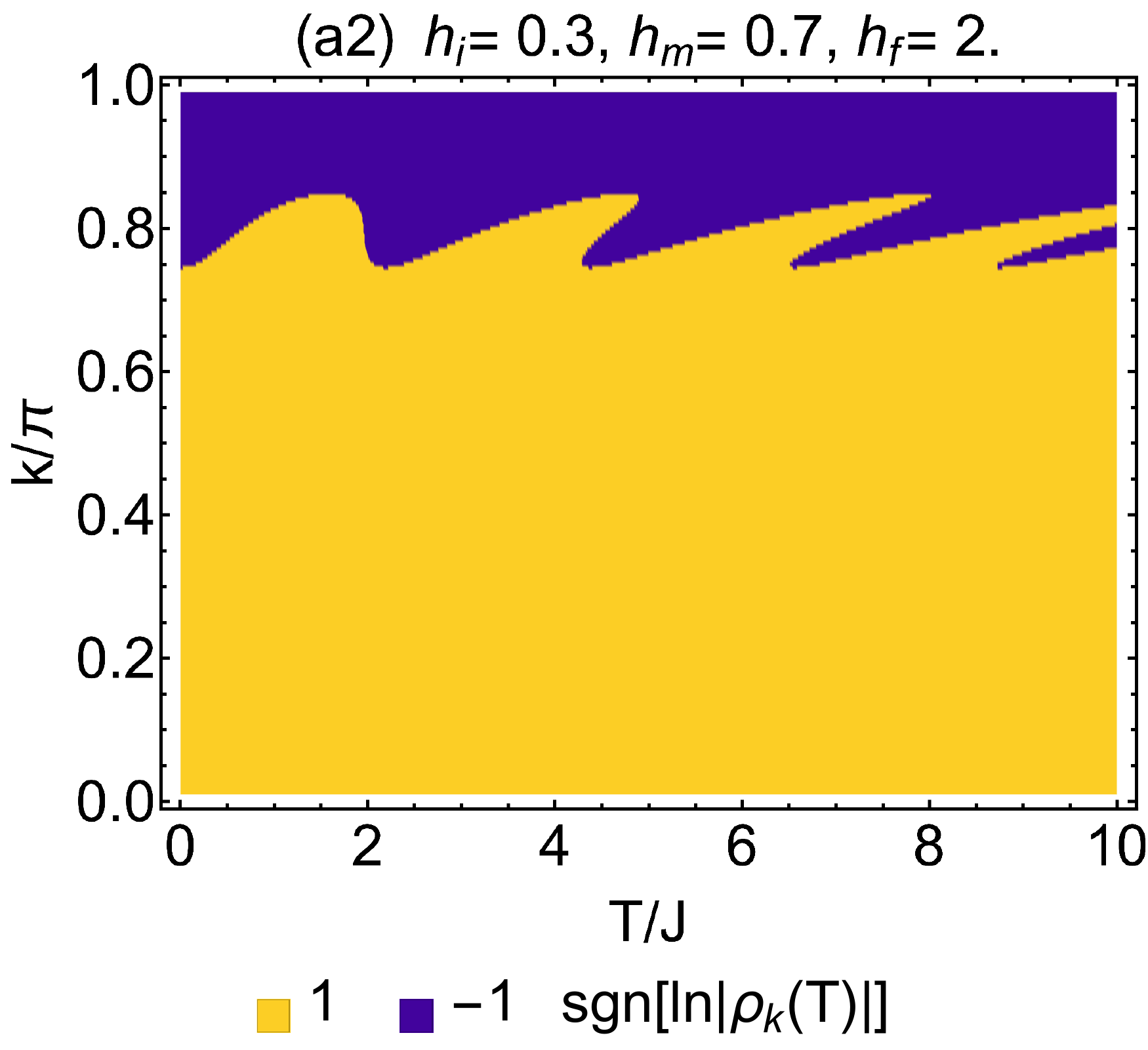}
\includegraphics[width=0.24\linewidth]{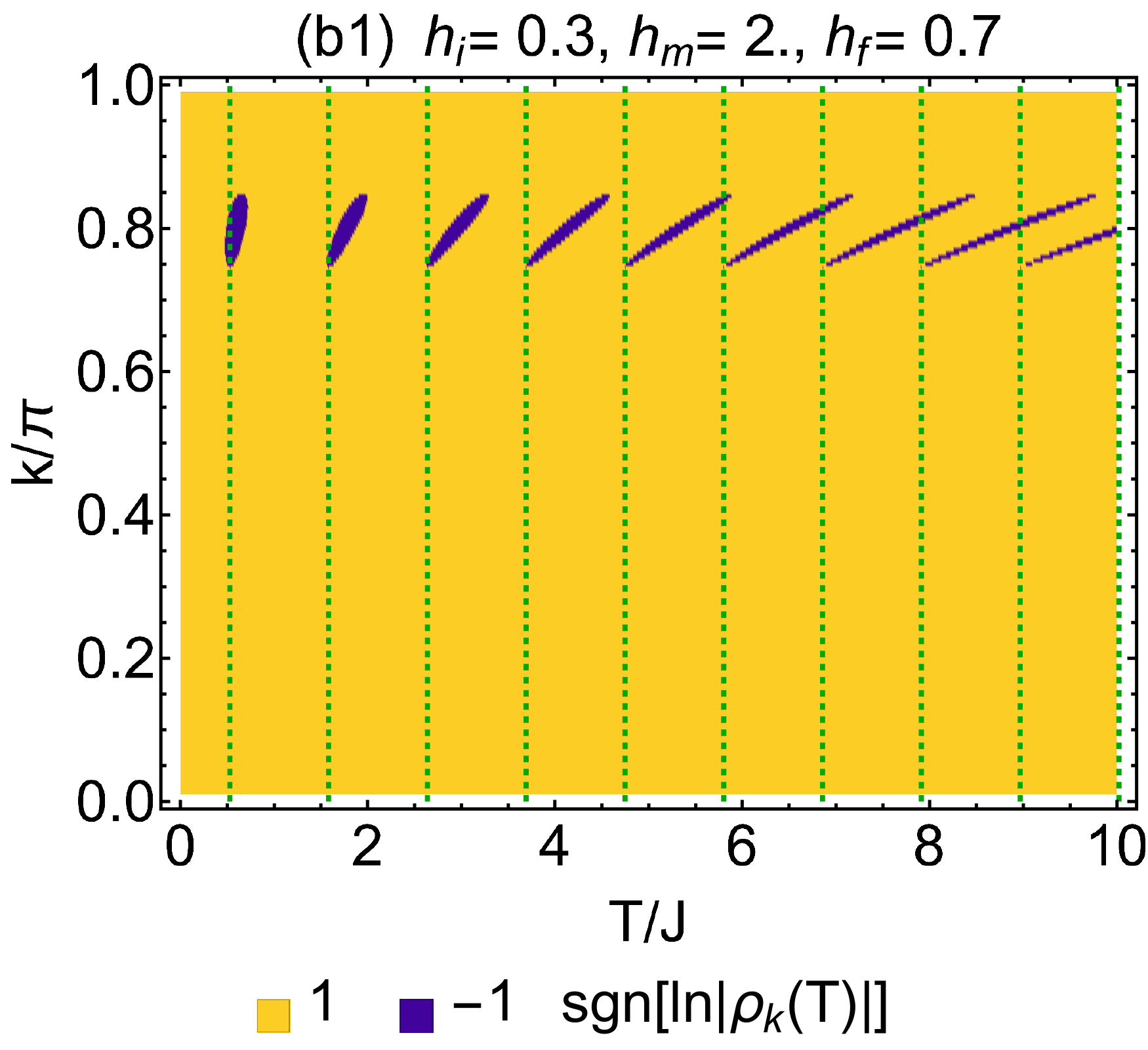}
\includegraphics[width=0.24\linewidth]{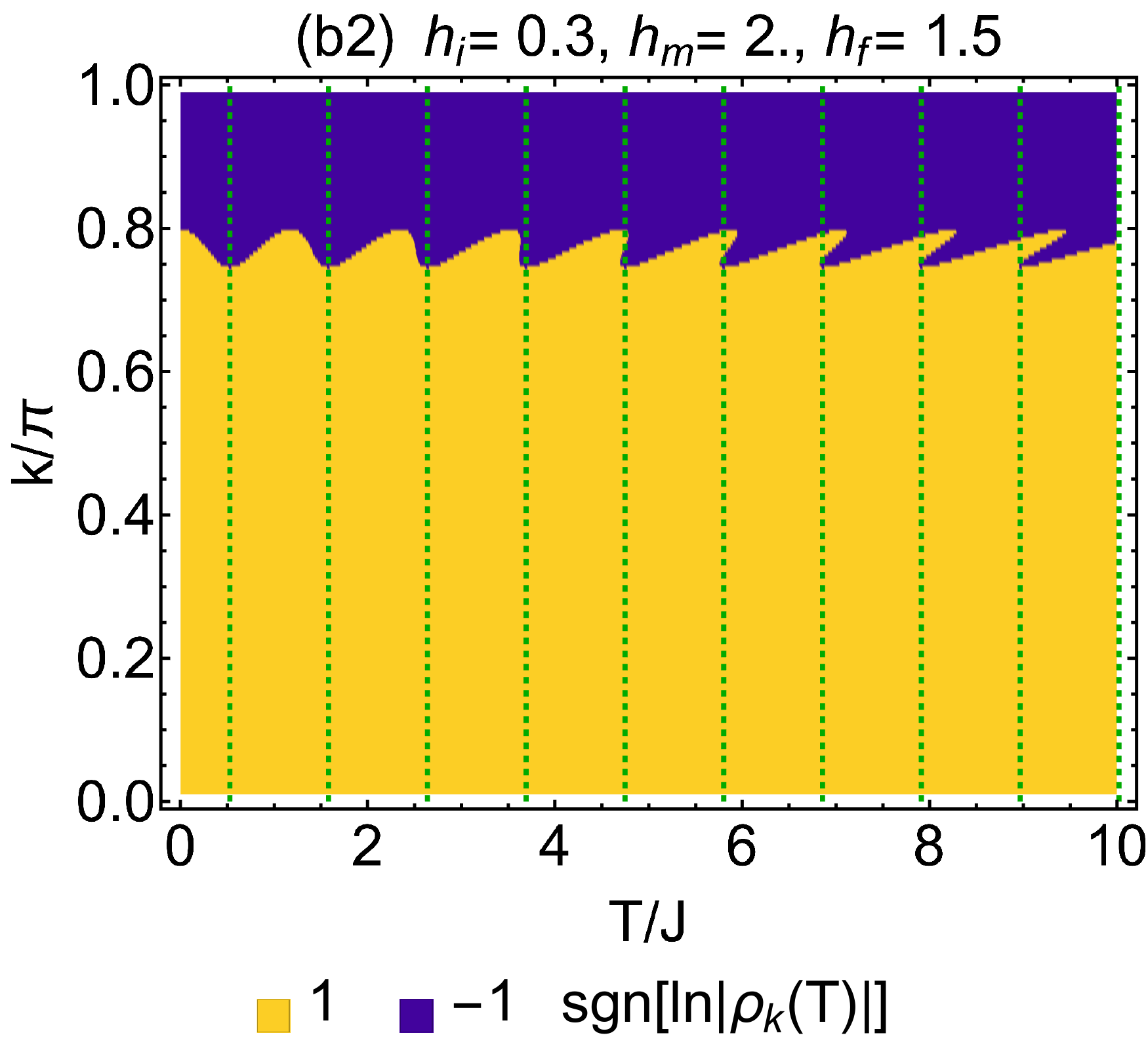}\\
\includegraphics[width=0.24\linewidth]{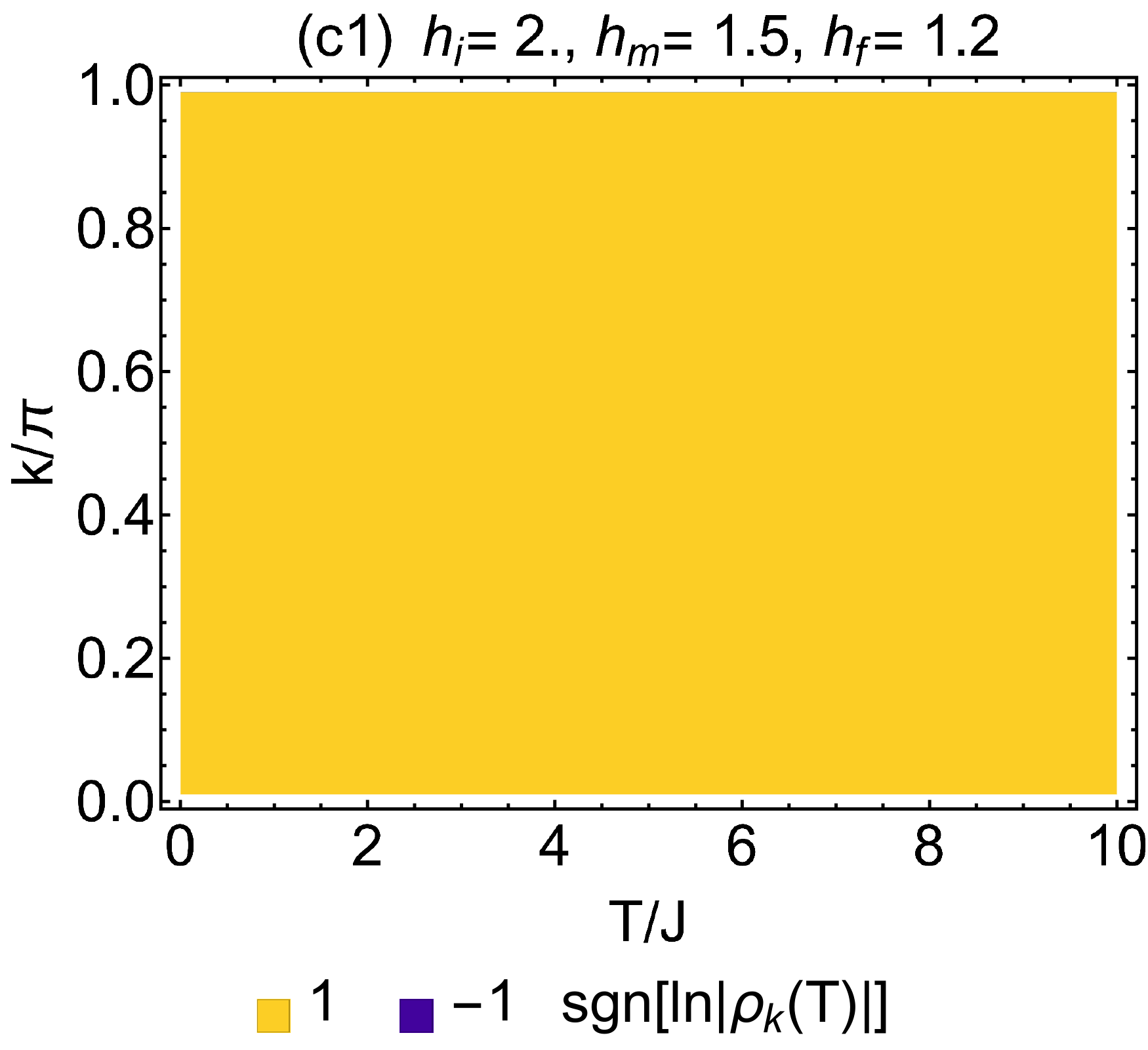}
\includegraphics[width=0.24\linewidth]{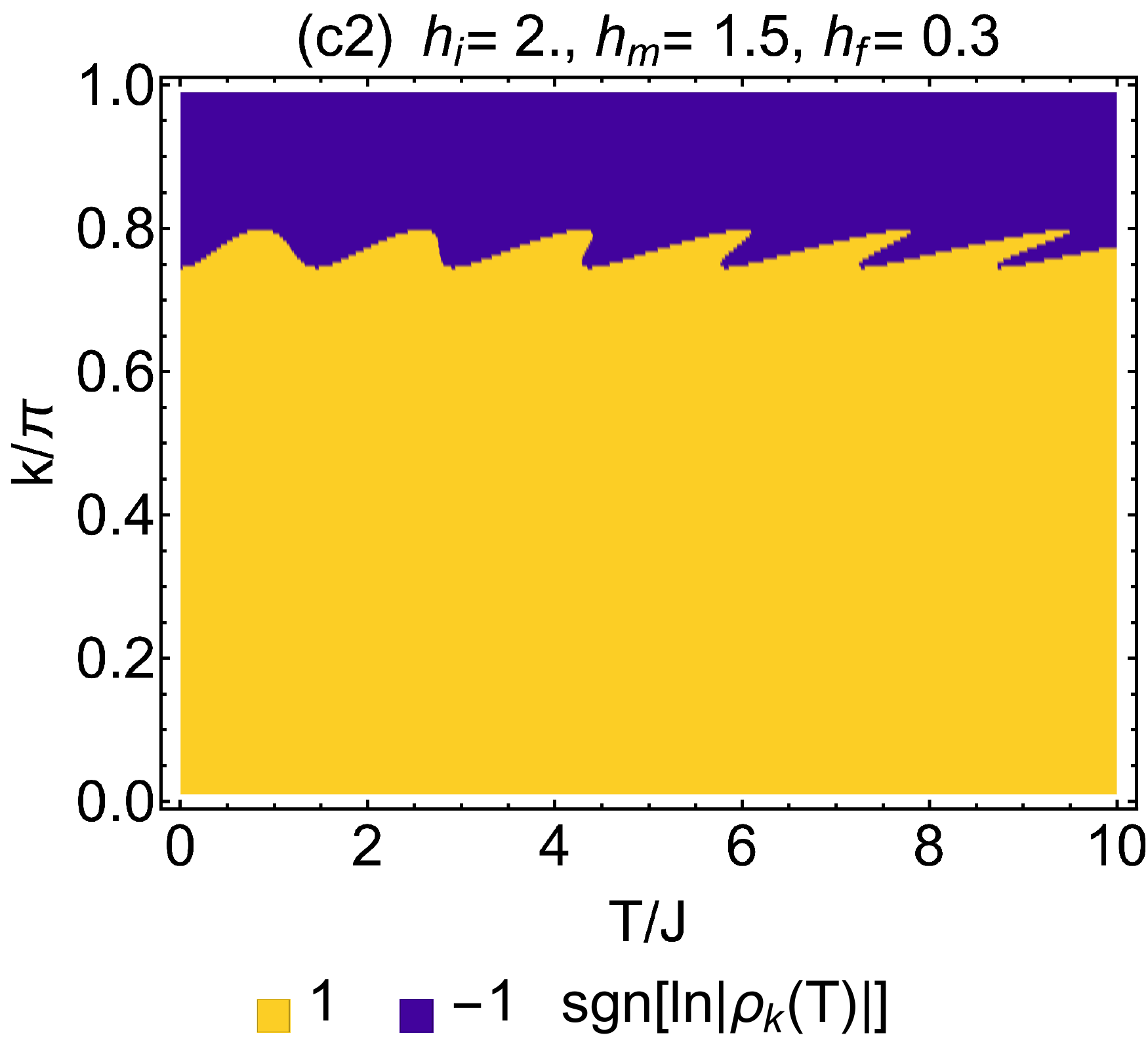}
\includegraphics[width=0.24\linewidth]{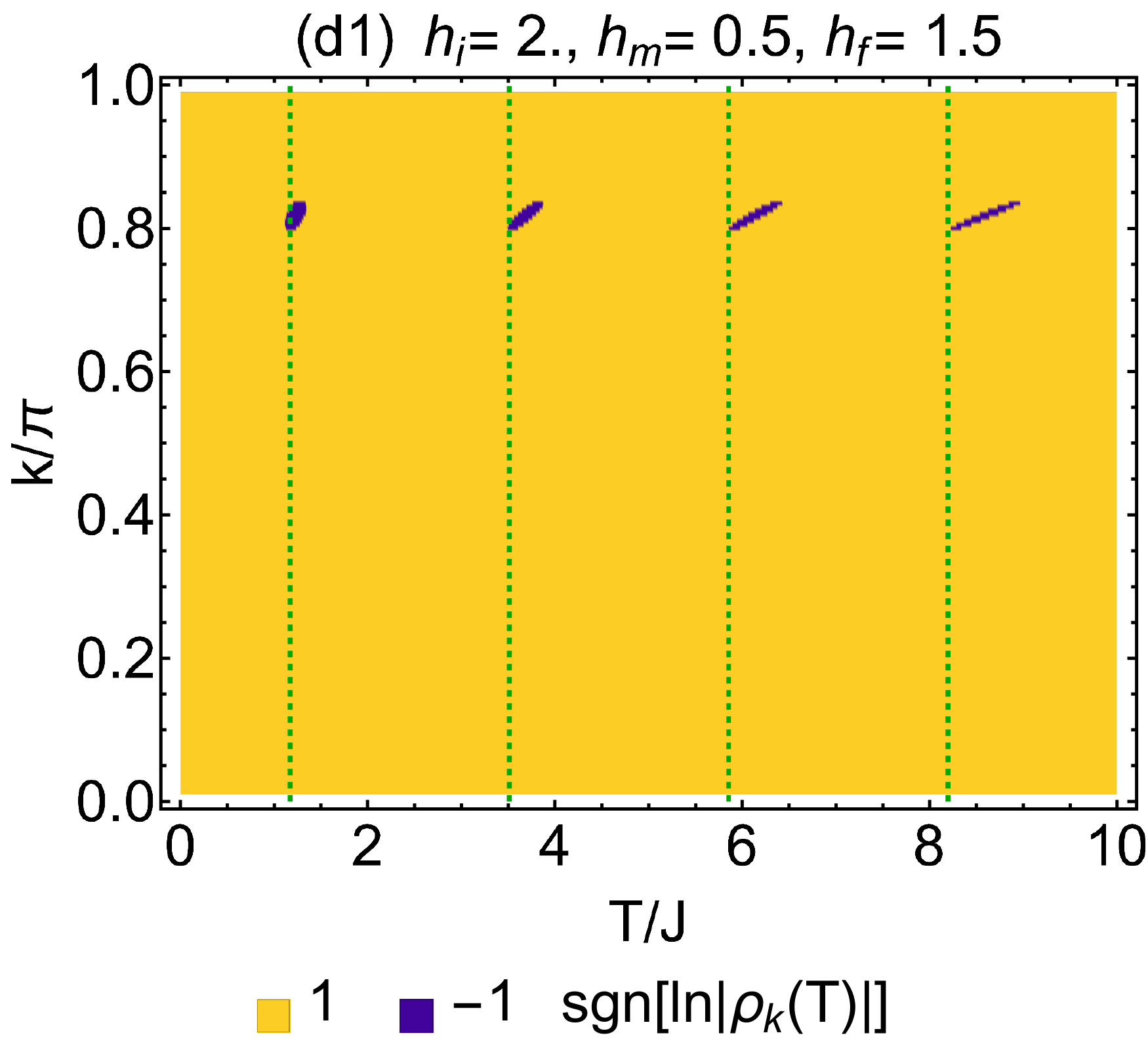}
\includegraphics[width=0.24\linewidth]{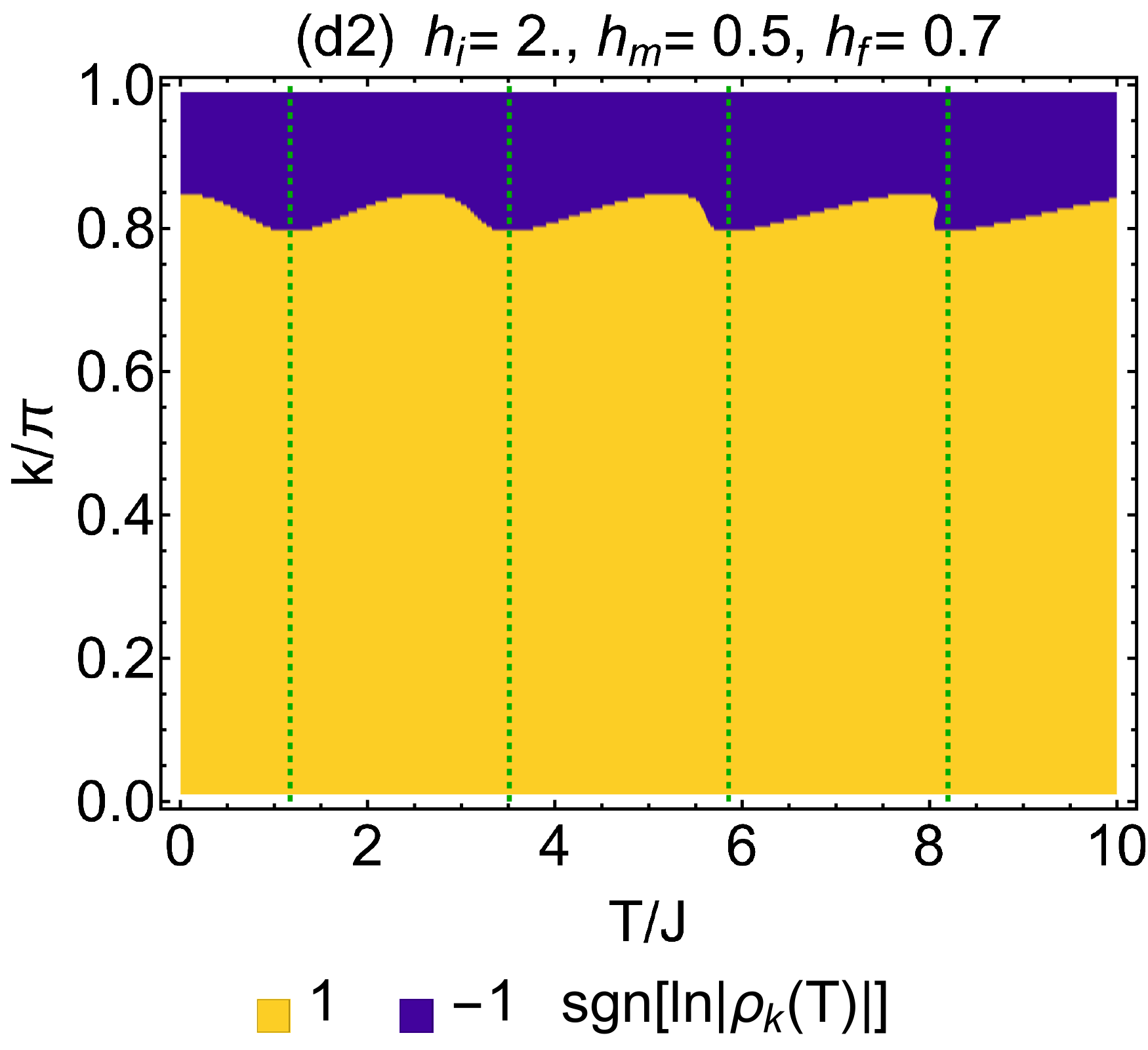}
\caption{Density plots of $\textrm{sgn}\,[\ln\rho_k]$, values of $T$ at which $\ln\rho_k$ changes sign will have DQPTs. The quench protocols are fully described in Table \ref{quenchtable}. Where appropriate the critical times for DQPTs following the first quench have been marked in green dashed lines.}\label{fig:manyrho}
\end{figure*}

%%%%%%%%%%%%%%%%%%%%%%%%%%%%%%%%%%%%%%%%%%%%%%%%
\subsection{The second quench}
%%%%%%%%%%%%%%%%%%%%%%%%%%%%%%%%%%%%%%%%%%%%%%%%

The second quench is performed at time $t=T$ as $h_m$ to $h_f$. In this case for the Loschmidt echo we have
\begin{equation}\label{lo2}
{\cal L}_k(t)=
e^{iT[\varepsilon_k^m-\varepsilon_k^f]}
\left[{\cal A}_ke^{it\varepsilon_k^f} +
{\cal B}_k e^{-it\varepsilon_k^f}\right] 
\end{equation}
where
\begin{align}
{\cal A}_k =& \cos (\theta_k^{fi})\bigg[
\cos (\theta_k^{mi})\cos (\theta_k^{fm})\nonumber\\
&- \sin (\theta_k^{mi})\sin (\theta_k^{fm}){e^{ - 2iT\varepsilon_k^m}}\bigg]
\end{align}
and
\begin{align}
{\cal B}_k =& \sin (\theta_k^{fi})\bigg[ \cos (\theta_k^{mi})\sin (\theta_k^{fm})  \nonumber  \\
&+\sin (\theta_k^{mi})\cos (\theta_k^{fm})e^{ - 2iT\varepsilon_k^m} \bigg]e^{2iT\varepsilon_k^f}.
\end{align}
It can be checked explicitly that for $t=T$ equations \eqref{lo1} and \eqref{lo2} agree as required. By defining $\rho _ke^{i\varphi _k} = {\cal A}_k/{\cal B}_k$ for real $\rho_k$ and $\varphi_k$ the Fisher zeros for $t\geq T$ can be obtained as
\begin{eqnarray}
z_n(k) = \frac{1}{2\varepsilon_k^f}\left[i\pi (2n + 1)-\ln\rho_k -i\varphi_k \right].
\end{eqnarray}
Hence the critical points appear under the condition that $\rho _{k^*}=1$, in the form of
\begin{eqnarray}
t_n^* = \frac{1}{2\varepsilon_{k^*}^f}\left[ \pi (2n + 1) - \varphi _{k^*} \right]~;~n=0,1,2,...
\end{eqnarray}
On the other hand, seeking the condition $\rho _{k^*}=1$ leads one to $h_f=H(k^*,T)$ where
\begin{equation}
H_k(T)\equiv -J \cos(k) - \frac{J \delta \sin(k)}{\tan(\Omega_k(T))}
\end{equation}
with $2\beta_{kT} = 1-\cos (2T \varepsilon_k^m)$ and
\begin{align}
\Omega_k(T)=& \cos^{-1}\left[\beta_{kT}\sin(2\theta_k^{mi}) \sin(2\theta_k^{fm}))\right]\nonumber\\
&-\tan^{-1}\left[\frac{J\delta\sin(k)}{J\cos(k)+h_i}\right].
\end{align}
It is easy to check at $T=0$, this equation reduces to Eq.~(\ref{hm}). In addition, by defining $a_k = J\delta \sin (k)$ and $b_k = J\cos (k)$ a lengthy calculation yields  $H_k(T) = -\Upsilon_k(T) /\Theta _k(T)$ with
\begin{align}\label{upstheta}
\Upsilon_k(T)=&
\beta_{kT}a_k^2h_m(h_i-h_m)+\left(a_k^2 + b_k^2\right)^2
+2b_k^2{h_i}{h_m} \nonumber\\
 &+(a_k^2{b_k}+b_k^3){h_i}
 +2(a_k^2{b_k} + b_k^3){h_m}\nonumber\\
 &+(a_k^2+ 2b_k^2)h_m^2
 +{b_k}{h_i}h_m^2 \textrm{ and}\nonumber \\  
\Theta _k(T) =& a_k^2{b_k} + b_k^3 + (a_k^2 - \beta_{kT}a_k^2 + b_k^2){h_i} 
+ {h_i}h_m^2\nonumber\\
 &+(\beta_{kT}a_k^2 + 2b_k^2){h_m}
 +2{b_k}{h_i}{h_m} + {b_k}h_m^2.
\end{align}
In the following section we will use these results to determine the conditions under which DQPTs occur following a second quench.

%%%%%%%%%%%%%%%%%%%%%%%%%%%%%%%%%%%%%%%%%%%%%%%%
\section{Double Quenches}\label{sec:dqpt}
%%%%%%%%%%%%%%%%%%%%%%%%%%%%%%%%%%%%%%%%%%%%%%%%

To simplify our analysis we will consider $\delta=1$ throughout. We then wish to consider each possible scenario of quenches, see table \ref{quenchtable}. We can start in either the topologically trivial phase (TrP) or topologically non-trivial  phase (ToP). For both the first and second quench we can then either cross the phase boundary or not, giving a total of eight distinct possibilities. As this model is relatively simple we discount the possibility that we can see qualitatively different behaviour except for quenches which cross different equilibrium phase boundaries~\cite{Vajna2015,Maslowski2020}.

\begin{table}
    \centering
    \begin{tabular}{|l|c|c|c|c|}\hline
       Fig.~\ref{fig:manyrho} & Initial & Second & Final & Type \\
       Label &  &  & &  \\\hline
       (a1) & ToP $h_i=0.3$ & ToP $h_m=0.7$ & ToP $h_f=0.9$ & AAA\\
       (a2) & ToP $h_i=0.3$ & Top $h_m=0.7$ & TrP $h_f=2.0$ & AAB\\
       (b1) & ToP $h_i=0.3$ & TrP $h_m=2.0$ & ToP $h_f=0.7$ & ABA\\
       (b2) & ToP $h_i=0.3$ & TrP $h_m=2.0$ & TrP $h_f=1.5$ & ABB\\
       (c1) & TrP $h_i=2.0$ & TrP $h_m=1.5$ & TrP $h_f=1.2$ & AAA\\
       (c2) & TrP $h_i=2.0$ & TrP $h_m=1.5$ & ToP $h_f=0.3$ & AAB\\
       (d1) & TrP $h_i=2.0$ & ToP $h_m=0.5$ & TrP $h_f=1.5$ & ABA\\
       (d2) & TrP $h_i=2.0$ & ToP $h_m=0.5$ & ToP $h_f=0.7$ & ABB\\\hline
    \end{tabular}
    \caption{A list of all possible quench scenarios, with the exemplary values of $h$ used in the calculations here and whether the phase is a topologically trivial phase (TrP) or a topologically non-trivial  phase (ToP). All values are given in units of $J$. The final column gives a shorthand to show whether a phase boundary is crossed, A is by definition the phase of the initial state.}
    \label{quenchtable}
\end{table}

%%%%%%%%%%%%%%%%%%%%%%%%%%%%%%%%%%%%%%%%%%%%%%%%
\subsection{Analysis of different quenches and existence criteria of DQPTs}
%%%%%%%%%%%%%%%%%%%%%%%%%%%%%%%%%%%%%%%%%%%%%%%%

In order to unambiguously determine when there are dynamical quantum phase transitions following the second quench we consider $\ln\rho_k(T)$. We recall that a DQPT will occur when $\rho_k(T)=1$. Therefore we look for times $T$ at which $\ln\rho_k(T)$ crosses zero as a function of $k$, which will ensure the existence of DQPTs. This is plotted in Fig.~\ref{fig:manyrho} for an example of each quench scenario in Table \ref{quenchtable}. Several distinct interesting cases can be seen. Perhaps as would be expected in this simple model, if we never quench across an equilibrium phase transition then there are no DQPTs following either quench. We then have a set of cases in which there are always DQPTs for any $T$, and cases for which DQPTs occur only for specific times $T$.

As is clear from Fig.~\ref{fig:manyrho}, DQPTs occur for all times $T$ if one of the quenches crosses an equilibrium topological phase boundary relative to the initial state, a fact we will prove below. This demonstrates that properties of the initial state persist for all times~\cite{Kennes2018}. These results also show that as $T$ increases multiple critical momenta, and therefore multiple critical times appear (see Fig.~\ref{fig:return3}, upper panel). On the other side, if one quenches across a phase boundary and then back so that the final Hamiltonian and the initial Hamiltonian belong to the same equilibrium phase, then DQPTs occur only at a certain $T$. As $T$ gets larger then it appears that DQPTs occur for longer ranges of $T$, with multiple critical times appearing again (see Fig.~\ref{fig:return3}, lower panel). The importance of the critical times for when we quench the second time are clear from Fig.~\ref{fig:manyrho}.

\begin{figure}
\includegraphics[width=0.49 \columnwidth]{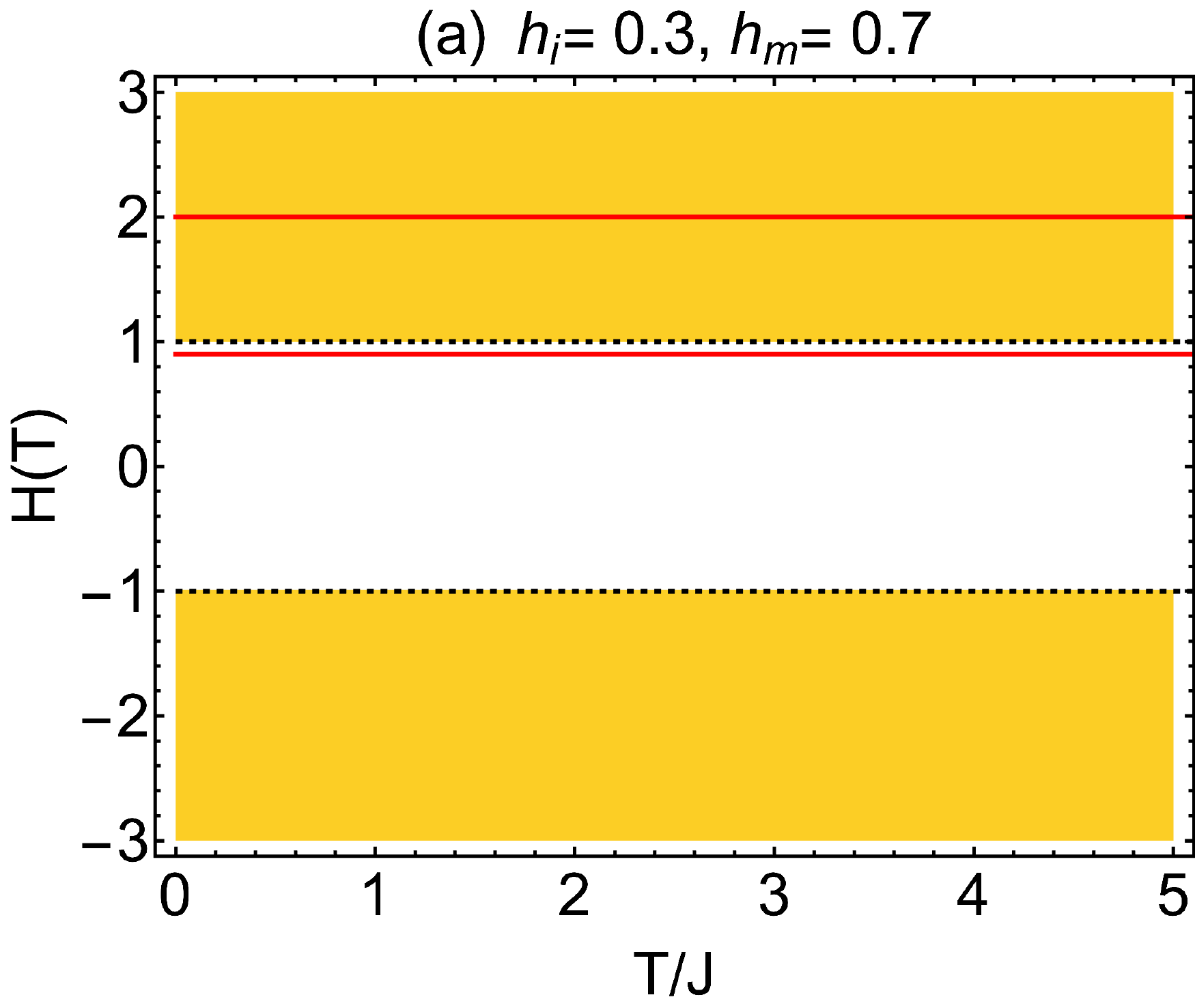}
\includegraphics[width=0.49 \columnwidth]{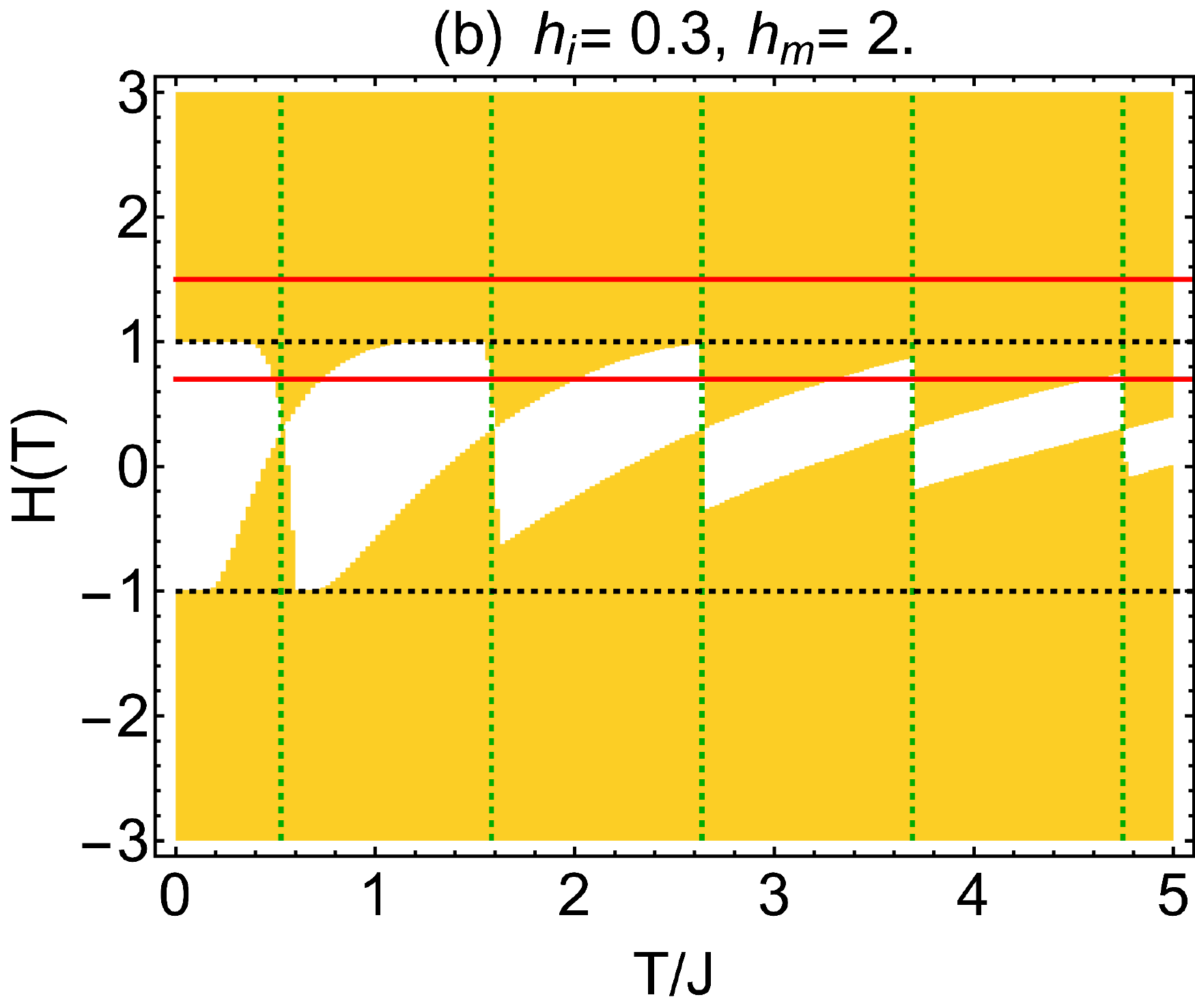}\\
\includegraphics[width=0.49 \columnwidth]{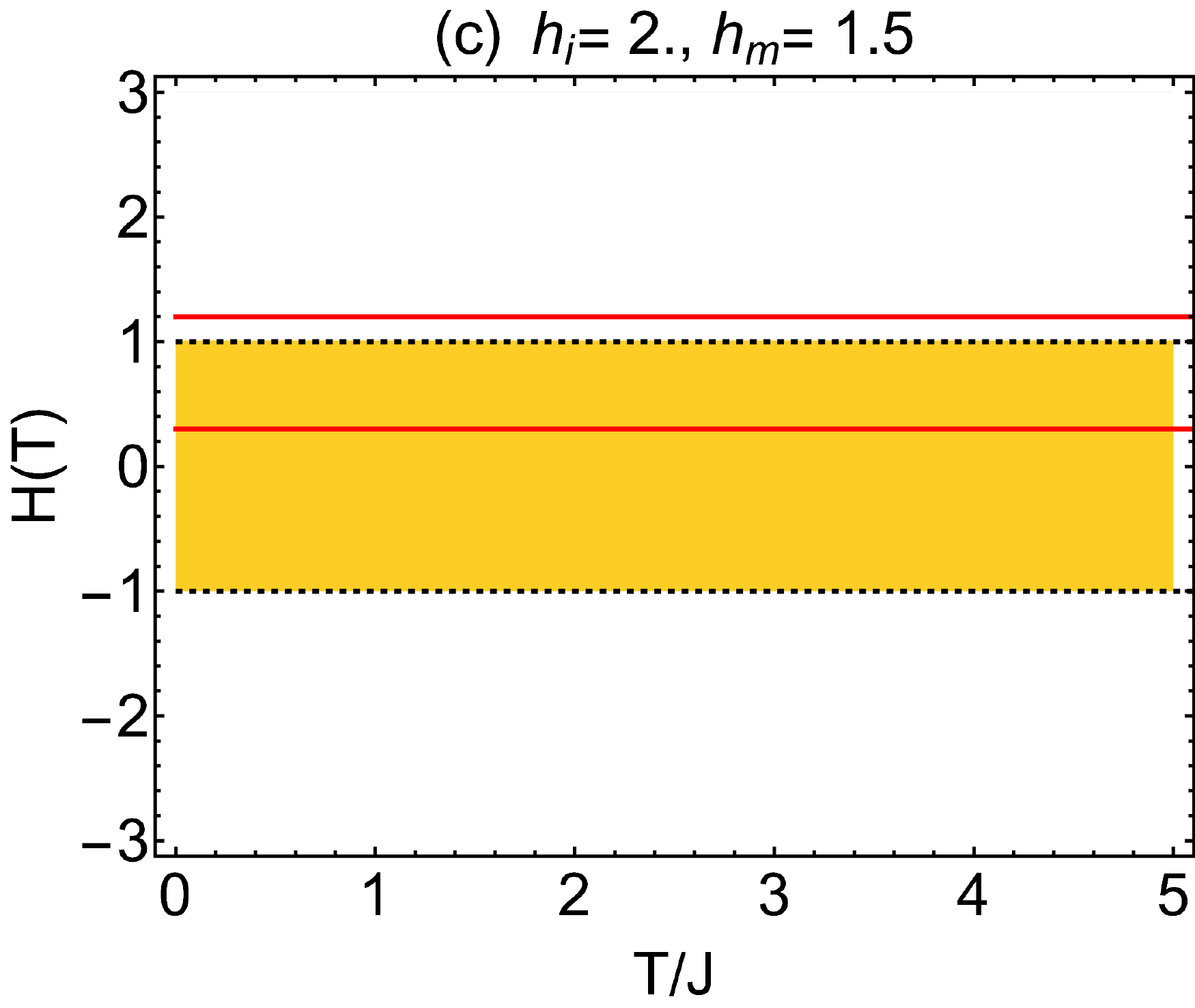}
\includegraphics[width=0.49 \columnwidth]{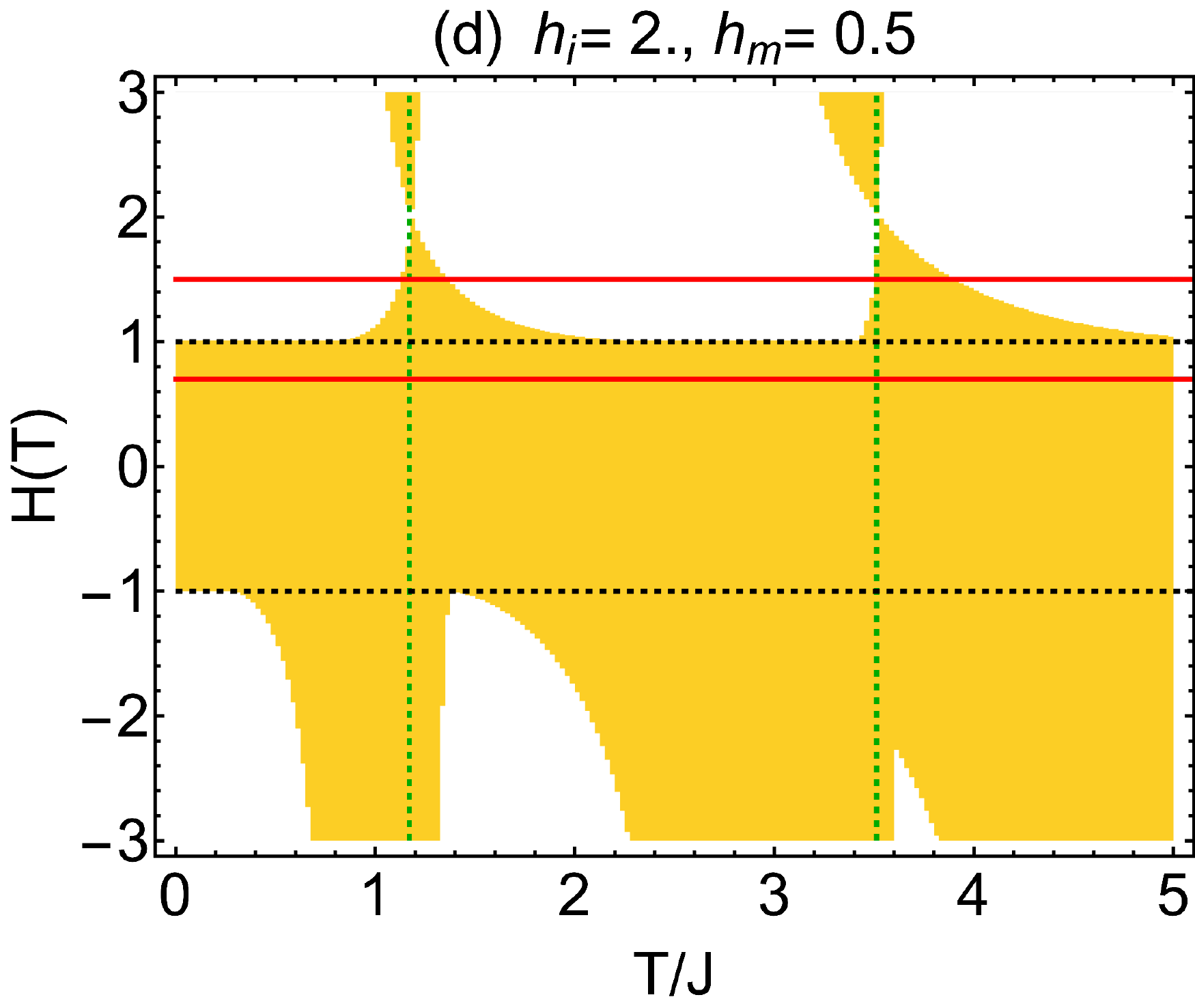}
  \caption{Here we show a plot of all values $H(T)$ where $H(T)=\{H_k(T):k\in[0,2\pi]\}$. At any particular second quench time $T$ if there exists a momentum $k$ for which $h_f=H_k(T)$, i.e.~$h_f\in\{H(T)\}$, then there will be DQPTs. That is provided $\{h_f,T\}$ is in the yellow region there will be DQPTs. Also plotted are the values for $h_f$ used in the quenches in table \ref{quenchtable} as red lines, and the dashed black lines show the phase boundaries for the equilibrium phase diagram. Where they exist critical times for the first quench are shown as green dashed lines. The importance of these lines for the DQPTs following the second quench is clearly visible.}\label{fig:hft}
\end{figure}

To understand these results we can use Eq.~\eqref{upstheta} to consider, following a particular first quench, for which $\{h_f,T\}$ DQPTs will then occur following the second quench. The condition for the critical momenta for the DQPTs after the second quench is $h_f=H_k(T)$, and if there are solutions to this then there will be critical times and DQPTs. Solving this numerically we can plot for any time $T$ the possible values $H(T)$ which $H_k(T)$ can obtain for all $k$. These results are shown in Fig.~\ref{fig:hft}. Consistently with Fig.~\ref{fig:manyrho} we can see under which conditions the DQPTs depend on the value of $T$ and when they do not. The extrema of the curves of $H_k(T)$ give the critical value of $h_f$ for which DQPTs exist for $T$. In Fig.~\ref{fig:hft}(a) for example for all $T$ the critical $h_f$ is equal to $1$, the equilibrium phase boundary. However for (b) the critical $h_f$ becomes $T$-dependent.

\begin{figure}
\includegraphics[width=0.49\columnwidth]{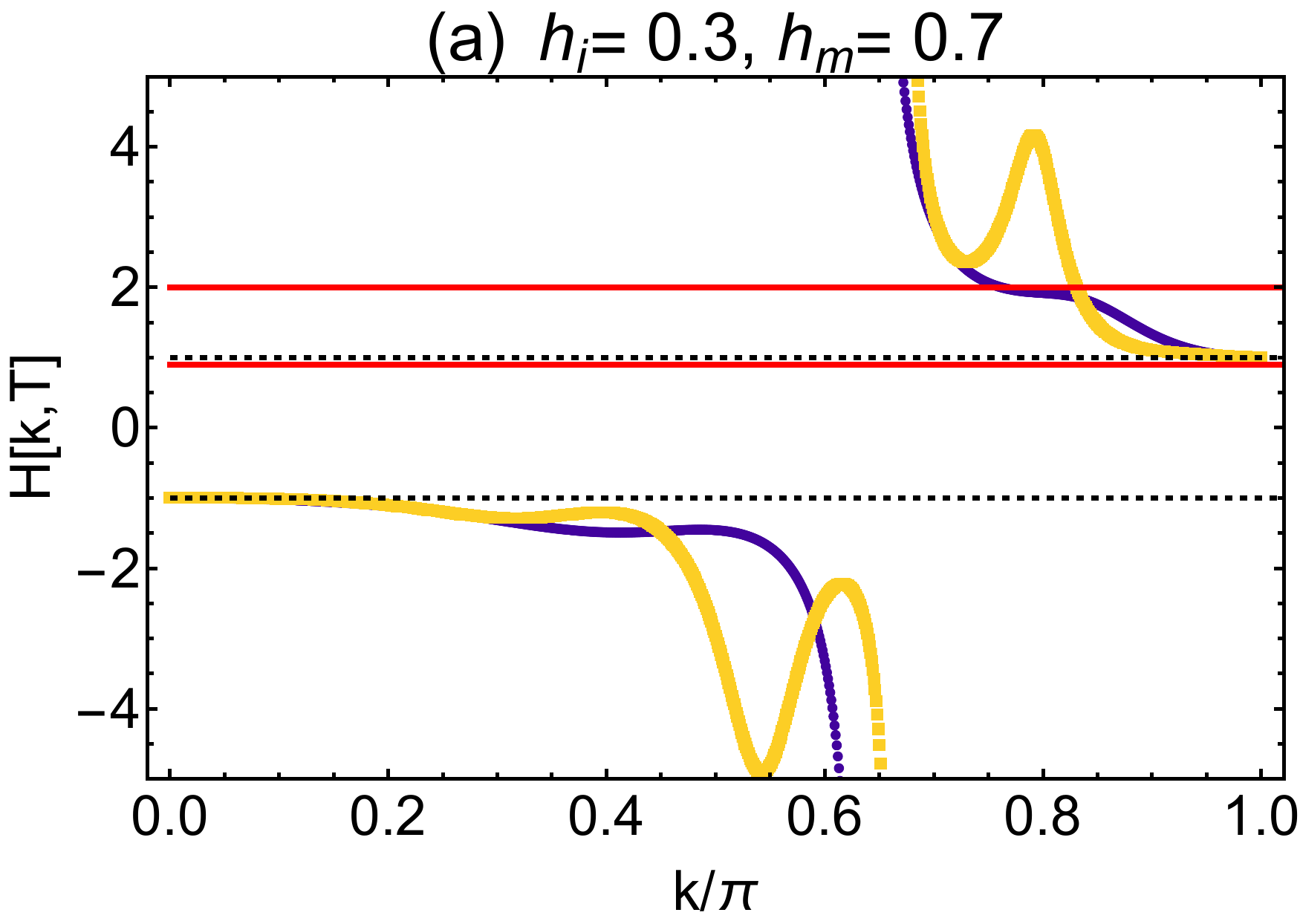}
\includegraphics[width=0.49\columnwidth]{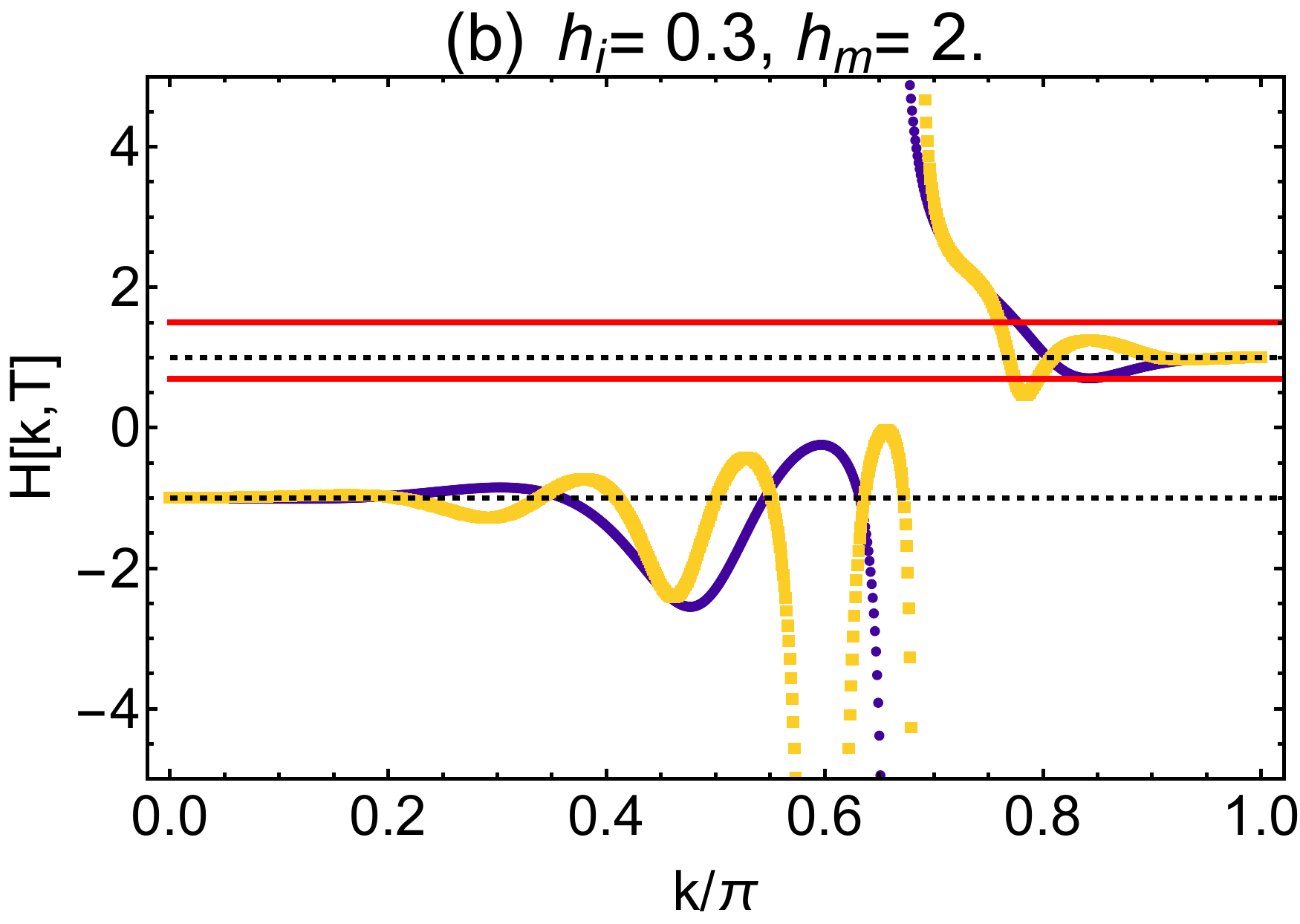}\\
\includegraphics[width=0.49\columnwidth]{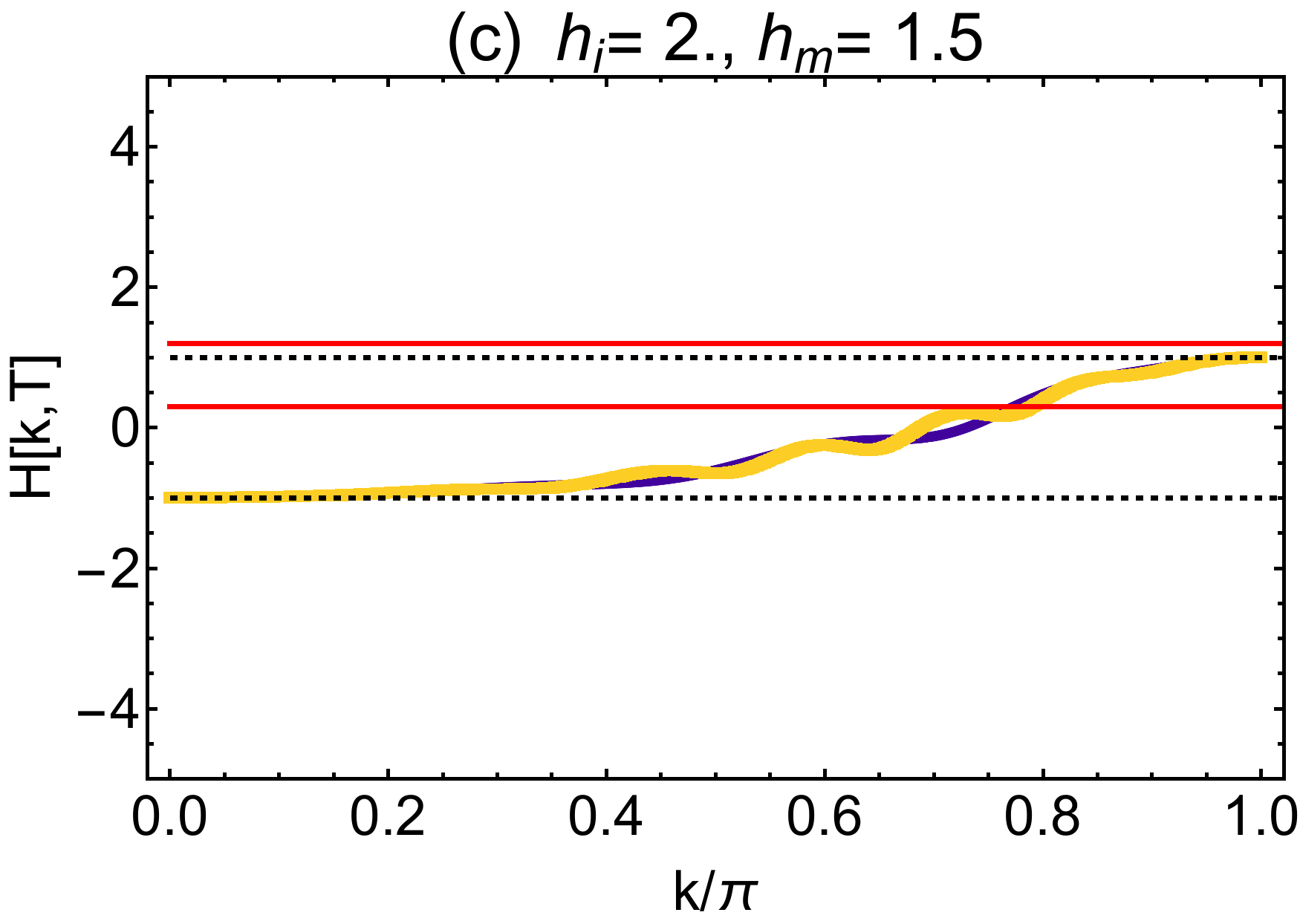}
\includegraphics[width=0.49\columnwidth]{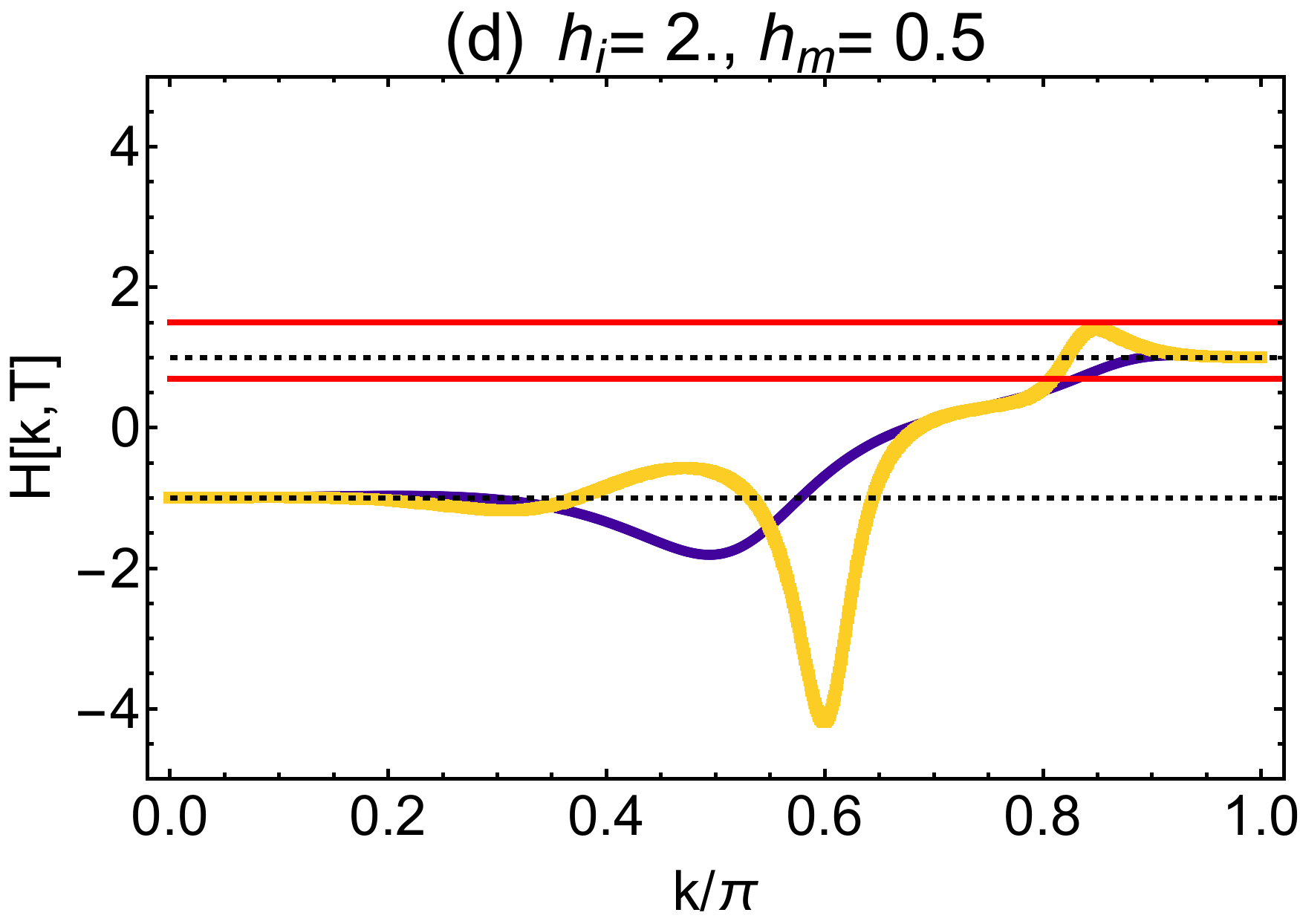}
  \caption{$H_k(T)$ versus $k$ for $T=2$ (purple circles) and $T=4$ (yellow squares). Also plotted are the values for $h_f$ used in the quenches in table \ref{quenchtable} as red lines, and the dashed black lines show the phase boundaries for the equilibrium phase diagram. Critical times for the first quench are shown as green dashed lines. The condition for DQPTs following the second quench is if f $h_f$ crosses $H_k(T)$. Compare with figure \ref{fig:hft}.}\label{fig:hfk}
\end{figure}

We note that $H_k(T)$ is symmetric and hence it suffices to consider $k\in[0,\pi]$. One can easily show that $H(0)=-1$ and $H(\pi)=1$, following form this it means that between $k=0$ and $k=\pi$, $H_k(T)$ must either pass through zero or diverge. Equivalently either $\Theta(k)$ or $\Upsilon(k)$ has a zero respectively, see Eq.~\eqref{upstheta}. In turn, these two possibilities ensure DQPTs for certain values of $h_f$. If $|h_i|\geq1$ then $\Theta(k)$ must have a zero and if  $|h_i|\leq1$ $\Upsilon(k)$ must have a zero. It follows that if $|h_i|\geq1$ then for $|h_f|\leq1$
there is certainly a DQPT after the second quench. Similarly, if $|h_i|\leq1$ then for $|h_f|\geq1$ there is also certainly a DQPT after the second quench. See Fig.~\ref{fig:hfk} for examples at particular times $T$. This behavior is also clearly reflected in Fig.~\ref{fig:hft}.  Additional DQPTs are also possible as $H_k(T)$ can in principle vary between its limits. As can be seen in Fig.~\ref{fig:hft} these are related to the critical times of the first quench.

\begin{figure}
\includegraphics[width=0.49\columnwidth]{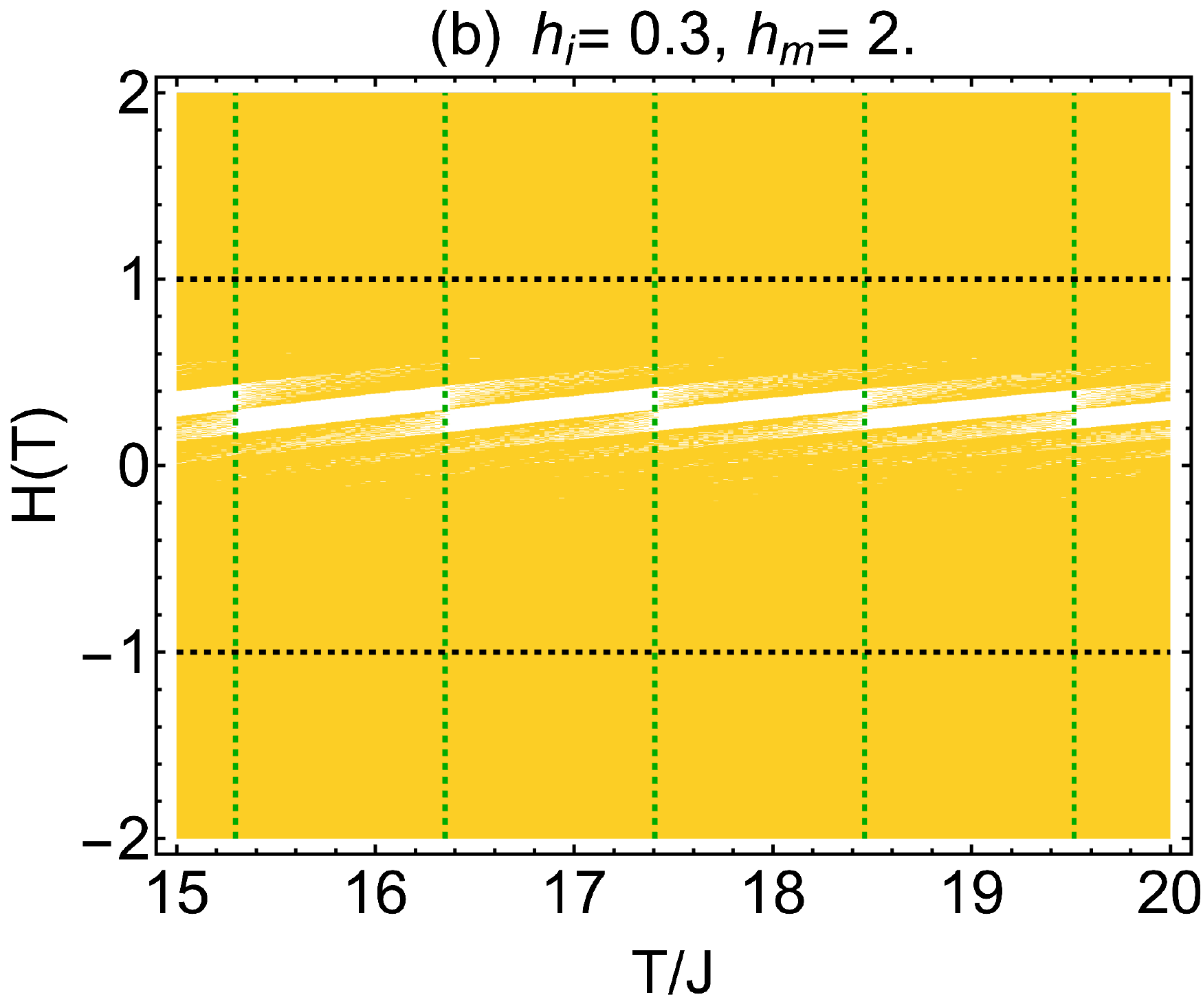}
\includegraphics[width=0.49\columnwidth]{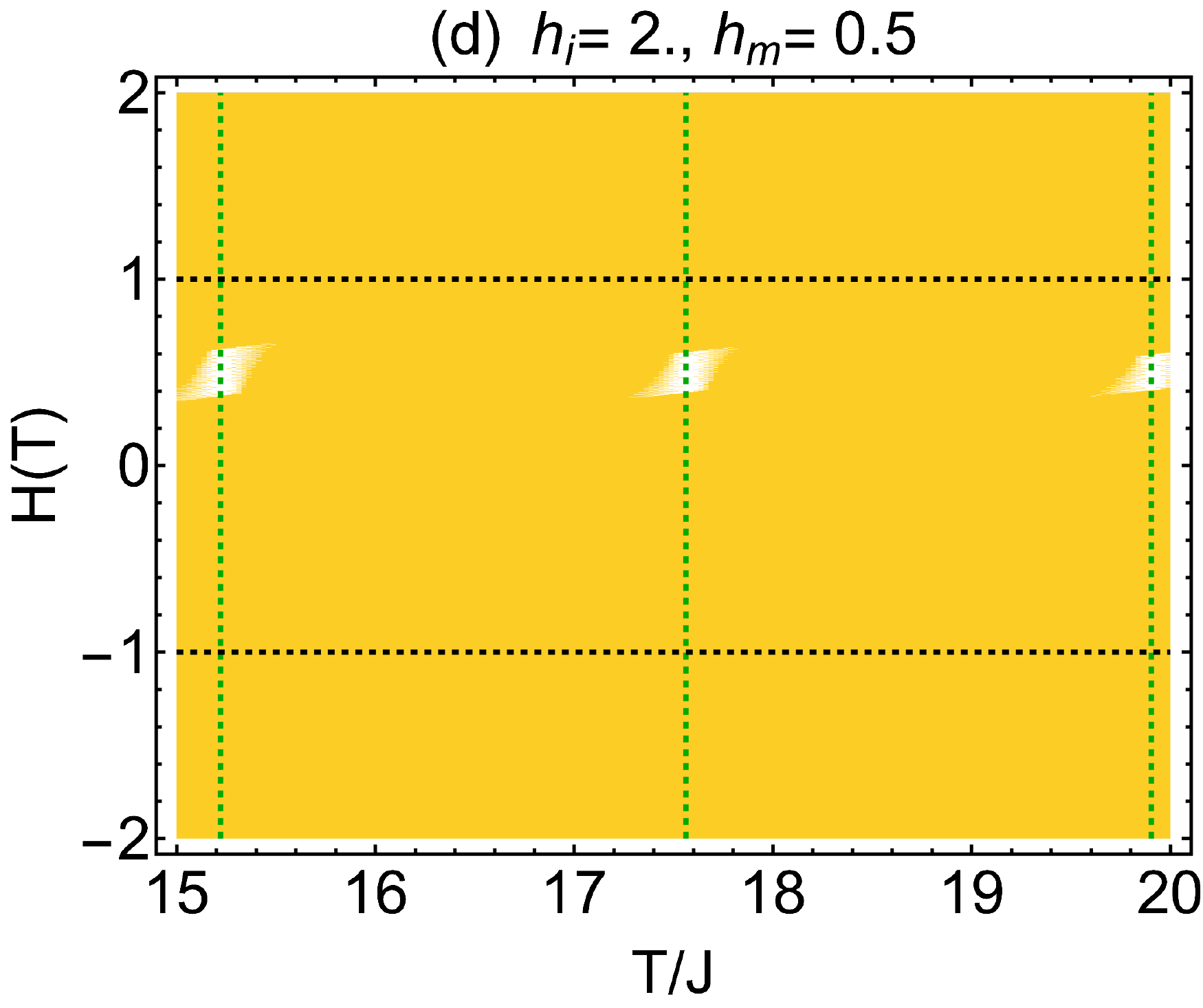}
  \caption{Here we show a plot of all values $H(T)$ see Fig.~\ref{fig:hft} and main text for details. Also plotted are the values for $h_f$ used in the quenches in table \ref{quenchtable} as red lines, and the dashed black lines show the phase boundaries for the equilibrium phase diagram. Where they exist critical times for the first quench are shown as green dashed lines. At large times $T$ one can see that DQPTs become inevitable.}\label{fig:hftlongt}
\end{figure}

At long times for the quenches of the form ABA, i.e.~(b1) and (d1) in table \ref{quenchtable}, DQPTs become inevitable, see Fig.~\ref{fig:hftlongt}. The additional solutions for the critical momenta and times become elongated as a function of $T$, ending solutions for all $T$ provided it is large enough.

%%%%%%%%%%%%%%%%%%%%%%%%%%%%%%%%%%%%%%%%%%%%%%%%
\subsection{DQPTs in the rate function}
%%%%%%%%%%%%%%%%%%%%%%%%%%%%%%%%%%%%%%%%%%%%%%%%

\begin{figure}
\includegraphics[width=0.8\columnwidth]{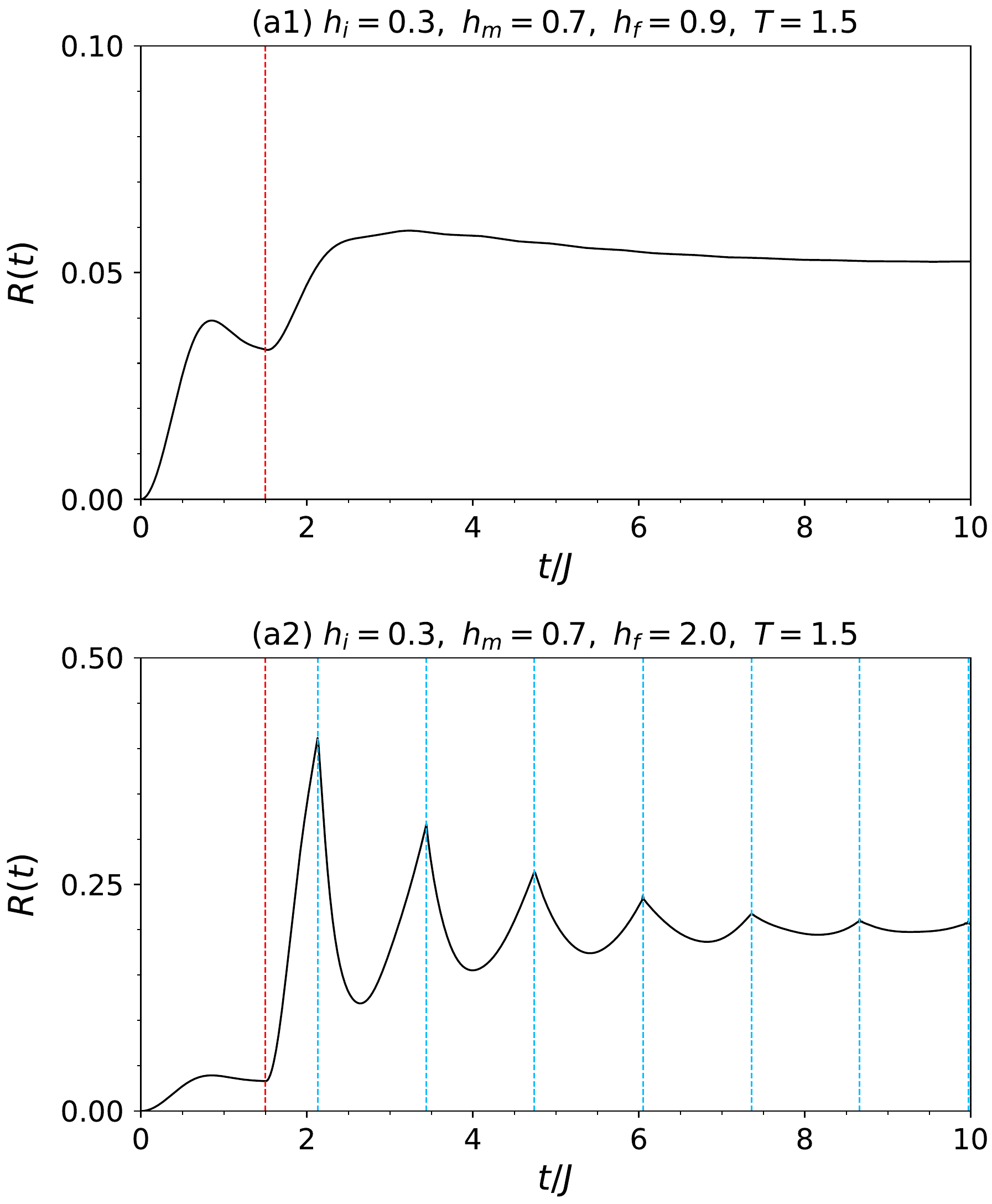}
  \caption{Time evolution of the rate function for a double quench for cases (a1) and (a2). The vertical red line is the time $T$ of the second quench, and the blue lines show the DQPTs following the second quench. In both of these cases there are no DQPTs following the first quench.}\label{fig:return1}
\end{figure}

\begin{figure}
\includegraphics[width=0.8\columnwidth]{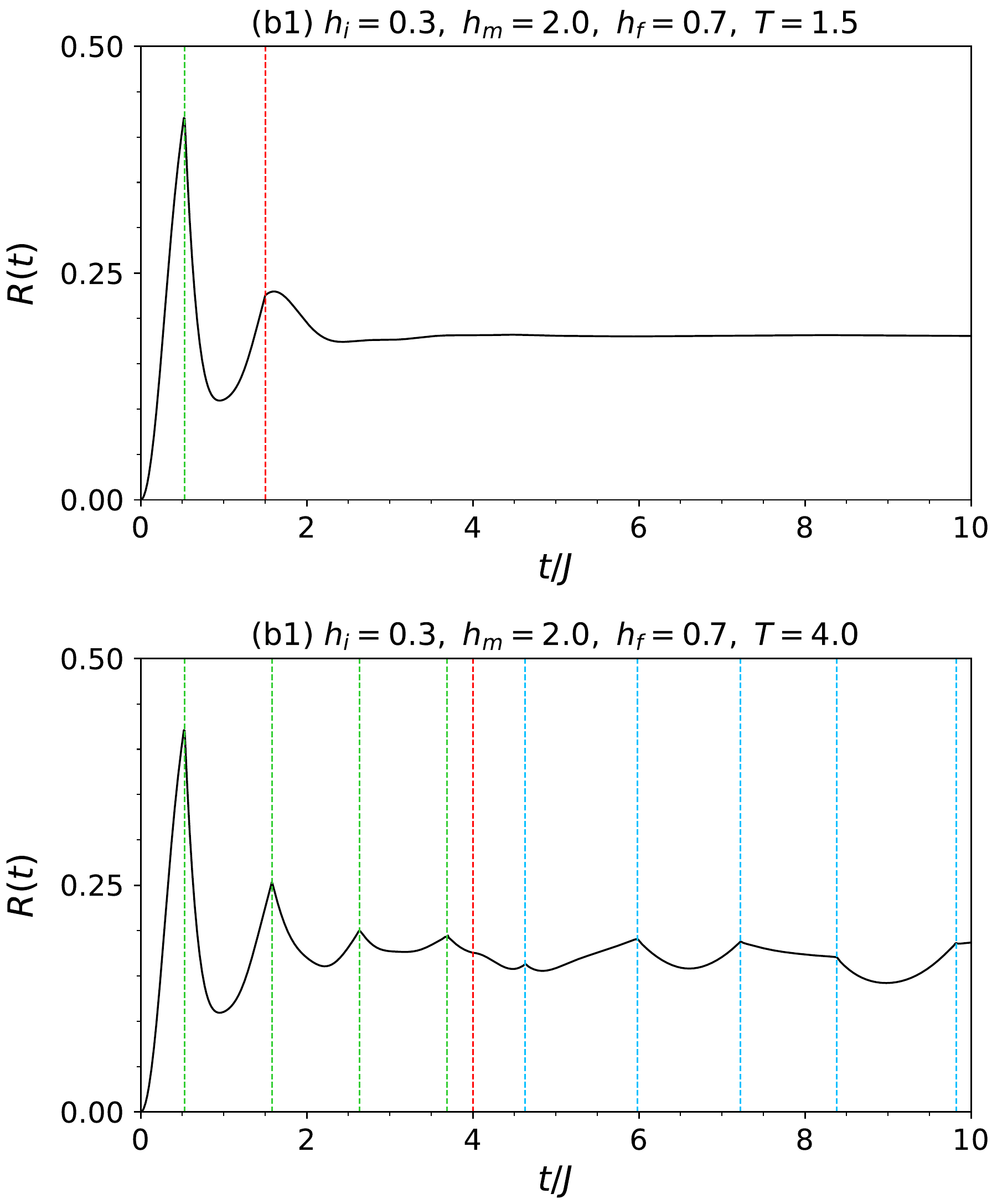}
  \caption{Time evolution of the rate function for a double quench for case (b1). The vertical red line is the time $T$ of the second quench, the green lines show the DQPTs following the first quench, and the blue lines show the DQPTs following the second quench.}\label{fig:return2}
\end{figure}

\begin{figure}
\includegraphics[width=0.8\columnwidth]{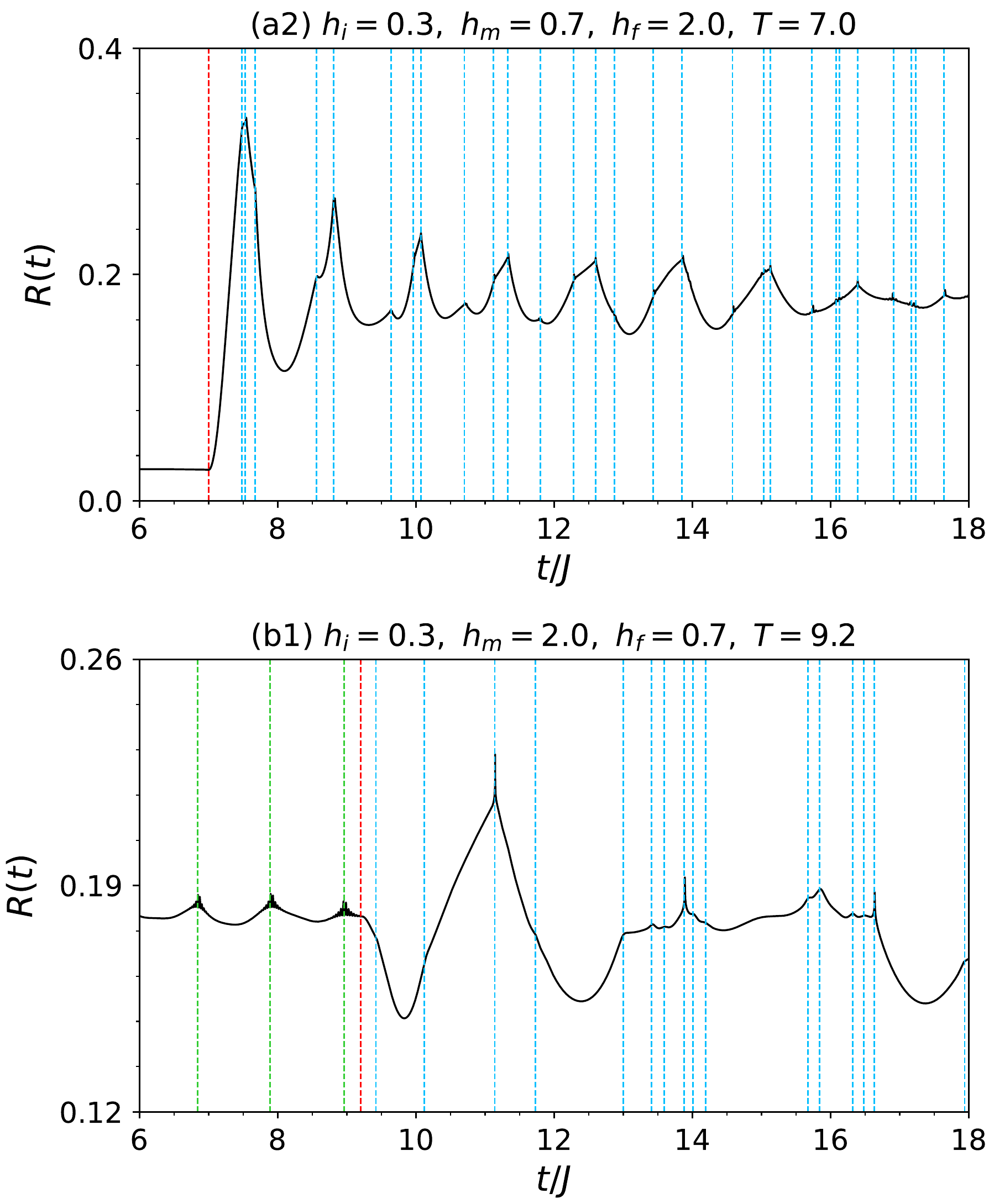}
  \caption{Time evolution of the rate function for a double quench for cases (a2) and (b1) where multiple critical times appear. The vertical red line is the time $T$ of the second quench, the green lines show the DQPTs following the first quench, and the blue lines show the DQPTs following the second quench.}\label{fig:return3}
\end{figure}

In Figs.~\ref{fig:return1} to \ref{fig:return3} we show explicitly the DQPTs in the rate function. $T$ is the time of the second quench, and hence if $T<t^*$ there are no DQPTs from the first quench. The results are in agreement with the preceding section. For (a1) and (a2), see Fig.~\ref{fig:return1}, DQPTs occur only in case (a2), which is an example of an AAB quench. For AAA (a1) no DQPTs are seen. We then focus on the ABA quench (b1). Depending on $T$ either no DQPTs are seen, see Fig.~\ref{fig:return2} upper panel, DQPTs occur with a single discernible critical time, see Fig.~\ref{fig:return2} lower panel, or DQPTs occur with multiple discernible critical times, see Fig.~\ref{fig:return3}

%%%%%%%%%%%%%%%%%%%%%%%%%%%%%%%%%%%%%%%%%%%%%%%%
\section{Conclusions}\label{sec:conc}
%%%%%%%%%%%%%%%%%%%%%%%%%%%%%%%%%%%%%%%%%%%%%%%%

In this article, we have explored the consequences of double quenches on the existence of DQPTs. Additionally, we have been partially motivated by to what extent one can gain information about ``dynamical phases'' from this consideration. We have considered an exemplary one-dimensional model which maps between a topological insulator or superconductor and an integrable spin model, and have exhaustively considered all the possible quench scenarios. Due to the relative simplicity of the model we were able to directly prove some of the results for the existence of the DQPTs.

To summarise, we can group the general behavior by the type of quench, we note that whether A is a topologically trivial or non-trivial phase plays no role here. Following the first quench we have as expected only DQPTs for quenches AB, provided $T>t^*$. We now focus on the DQPTs following the second quench. For quenches of the form AAA (a1,c1) there are never DQPTs. For quenches of the form AAB (a2,c2) there are always DQPTs for all $T$. This remains true for quenches ABB (b2,d2). This shows that there is a memory of the initial state which persists for all time. However, for quenches of the form ABA (b1,d1) we find that if $T$ is relatively small there are DQPTs that occur near $t_n^*$. This suggests that near critical times the time-evolved state changes its properties in a way that could be cautiously compared to a dynamical phase. For longer times these regions expand until DQPTs become inevitable for large $T$. This is in contrast to Kennes et al.~\cite{Kennes2018} who find that for ABA type double quenches that as $T$ increases DQPTs can be avoided.

Extensions to more complicated models where DQPTs do not follow the equilibrium phase diagram, to higher dimensions, to boundary effects, and to multiple quenches would be interesting for further work.

%%%%%%%%%%%%%%%%%%%%%%%%%%%%%%%%%%%%%%%%%%%%%%%%
\acknowledgments
This work was supported by the National Science Centre (NCN, Poland) by the grant 2019/35/B/ST3/03625.
%%%%%%%%%%%%%%%%%%%%%%%%%%%%%%%%%%%%%%%%%%%%%%%%

%%%%%%%%%%%%%%%%%%%%%%%%%%%%%%%%%%%%%%%%%%%%%%%%
% \appendix

%%%%%%%%%%%%%%%%%%%%%%%%%%%%%%%%%%%%%%%%%%%%%%%%
%%%%%%%%%%%%%%%%%%%%%%%%%%%%%%%%%%%%%%%%%%%%%%%%

%\bibliography{library}

\begin{thebibliography}{113}%
\makeatletter
\providecommand \@ifxundefined [1]{%
 \@ifx{#1\undefined}
}%
\providecommand \@ifnum [1]{%
 \ifnum #1\expandafter \@firstoftwo
 \else \expandafter \@secondoftwo
 \fi
}%
\providecommand \@ifx [1]{%
 \ifx #1\expandafter \@firstoftwo
 \else \expandafter \@secondoftwo
 \fi
}%
\providecommand \natexlab [1]{#1}%
\providecommand \enquote  [1]{``#1''}%
\providecommand \bibnamefont  [1]{#1}%
\providecommand \bibfnamefont [1]{#1}%
\providecommand \citenamefont [1]{#1}%
\providecommand \href@noop [0]{\@secondoftwo}%
\providecommand \href [0]{\begingroup \@sanitize@url \@href}%
\providecommand \@href[1]{\@@startlink{#1}\@@href}%
\providecommand \@@href[1]{\endgroup#1\@@endlink}%
\providecommand \@sanitize@url [0]{\catcode `\\12\catcode `\$12\catcode
  `\&12\catcode `\#12\catcode `\^12\catcode `\_12\catcode `\%12\relax}%
\providecommand \@@startlink[1]{}%
\providecommand \@@endlink[0]{}%
\providecommand \url  [0]{\begingroup\@sanitize@url \@url }%
\providecommand \@url [1]{\endgroup\@href {#1}{\urlprefix }}%
\providecommand \urlprefix  [0]{URL }%
\providecommand \Eprint [0]{\href }%
\providecommand \doibase [0]{https://doi.org/}%
\providecommand \selectlanguage [0]{\@gobble}%
\providecommand \bibinfo  [0]{\@secondoftwo}%
\providecommand \bibfield  [0]{\@secondoftwo}%
\providecommand \translation [1]{[#1]}%
\providecommand \BibitemOpen [0]{}%
\providecommand \bibitemStop [0]{}%
\providecommand \bibitemNoStop [0]{.\EOS\space}%
\providecommand \EOS [0]{\spacefactor3000\relax}%
\providecommand \BibitemShut  [1]{\csname bibitem#1\endcsname}%
\let\auto@bib@innerbib\@empty
%</preamble>
\bibitem [{\citenamefont {Streltsov}\ \emph {et~al.}(2017)\citenamefont
  {Streltsov}, \citenamefont {Adesso},\ and\ \citenamefont
  {Plenio}}]{Streltsov2017}%
  \BibitemOpen
  \bibfield  {author} {\bibinfo {author} {\bibfnamefont {A.}~\bibnamefont
  {Streltsov}}, \bibinfo {author} {\bibfnamefont {G.}~\bibnamefont {Adesso}},\
  and\ \bibinfo {author} {\bibfnamefont {M.~B.}\ \bibnamefont {Plenio}},\
  }\bibfield  {title} {\bibinfo {title} {{\emph{Colloquium}} : {{Quantum}}
  coherence as a resource},\ }\href
  {https://doi.org/10.1103/RevModPhys.89.041003} {\bibfield  {journal}
  {\bibinfo  {journal} {Reviews of Modern Physics}\ }\textbf {\bibinfo {volume}
  {89}},\ \bibinfo {pages} {041003} (\bibinfo {year} {2017})}\BibitemShut
  {NoStop}%
\bibitem [{\citenamefont {Basov}\ \emph {et~al.}(2017)\citenamefont {Basov},
  \citenamefont {Averitt},\ and\ \citenamefont {Hsieh}}]{Basov2017}%
  \BibitemOpen
  \bibfield  {author} {\bibinfo {author} {\bibfnamefont {D.~N.}\ \bibnamefont
  {Basov}}, \bibinfo {author} {\bibfnamefont {R.~D.}\ \bibnamefont {Averitt}},\
  and\ \bibinfo {author} {\bibfnamefont {D.}~\bibnamefont {Hsieh}},\ }\bibfield
   {title} {\bibinfo {title} {Towards properties on demand in quantum
  materials},\ }\href {https://doi.org/10.1038/nmat5017} {\bibfield  {journal}
  {\bibinfo  {journal} {Nature Materials}\ }\textbf {\bibinfo {volume} {16}},\
  \bibinfo {pages} {1077} (\bibinfo {year} {2017})}\BibitemShut {NoStop}%
\bibitem [{\citenamefont {Martinez}\ \emph {et~al.}(2016)\citenamefont
  {Martinez}, \citenamefont {Muschik}, \citenamefont {Schindler}, \citenamefont
  {Nigg}, \citenamefont {Erhard}, \citenamefont {Heyl}, \citenamefont {Hauke},
  \citenamefont {Dalmonte}, \citenamefont {Monz}, \citenamefont {Zoller},\ and\
  \citenamefont {Blatt}}]{Martinez2016}%
  \BibitemOpen
  \bibfield  {author} {\bibinfo {author} {\bibfnamefont {E.~A.}\ \bibnamefont
  {Martinez}}, \bibinfo {author} {\bibfnamefont {C.~A.}\ \bibnamefont
  {Muschik}}, \bibinfo {author} {\bibfnamefont {P.}~\bibnamefont {Schindler}},
  \bibinfo {author} {\bibfnamefont {D.}~\bibnamefont {Nigg}}, \bibinfo {author}
  {\bibfnamefont {A.}~\bibnamefont {Erhard}}, \bibinfo {author} {\bibfnamefont
  {M.}~\bibnamefont {Heyl}}, \bibinfo {author} {\bibfnamefont {P.}~\bibnamefont
  {Hauke}}, \bibinfo {author} {\bibfnamefont {M.}~\bibnamefont {Dalmonte}},
  \bibinfo {author} {\bibfnamefont {T.}~\bibnamefont {Monz}}, \bibinfo {author}
  {\bibfnamefont {P.}~\bibnamefont {Zoller}},\ and\ \bibinfo {author}
  {\bibfnamefont {R.}~\bibnamefont {Blatt}},\ }\bibfield  {title} {\bibinfo
  {title} {Real-time dynamics of lattice gauge theories with a few-qubit
  quantum computer},\ }\href {https://doi.org/10.1038/nature18318} {\bibfield
  {journal} {\bibinfo  {journal} {Nature}\ }\textbf {\bibinfo {volume} {534}},\
  \bibinfo {pages} {516} (\bibinfo {year} {2016})}\BibitemShut {NoStop}%
\bibitem [{\citenamefont {Halimeh}\ and\ \citenamefont
  {{Zauner-Stauber}}(2017)}]{Halimeh2017}%
  \BibitemOpen
  \bibfield  {author} {\bibinfo {author} {\bibfnamefont {J.~C.}\ \bibnamefont
  {Halimeh}}\ and\ \bibinfo {author} {\bibfnamefont {V.}~\bibnamefont
  {{Zauner-Stauber}}},\ }\bibfield  {title} {\bibinfo {title} {Dynamical phase
  diagram of quantum spin chains with long-range interactions},\ }\href
  {https://doi.org/10.1103/PhysRevB.96.134427} {\bibfield  {journal} {\bibinfo
  {journal} {Physical Review B}\ }\textbf {\bibinfo {volume} {96}},\ \bibinfo
  {pages} {134427} (\bibinfo {year} {2017})}\BibitemShut {NoStop}%
\bibitem [{\citenamefont {Heyl}\ \emph {et~al.}(2013)\citenamefont {Heyl},
  \citenamefont {Polkovnikov},\ and\ \citenamefont {Kehrein}}]{Heyl2013}%
  \BibitemOpen
  \bibfield  {author} {\bibinfo {author} {\bibfnamefont {M.}~\bibnamefont
  {Heyl}}, \bibinfo {author} {\bibfnamefont {A.}~\bibnamefont {Polkovnikov}},\
  and\ \bibinfo {author} {\bibfnamefont {S.}~\bibnamefont {Kehrein}},\
  }\bibfield  {title} {\bibinfo {title} {Dynamical quantum phase transitions in
  the transverse-field ising model},\ }\href
  {https://doi.org/10.1103/PhysRevLett.110.135704} {\bibfield  {journal}
  {\bibinfo  {journal} {Physical Review Letters}\ }\textbf {\bibinfo {volume}
  {110}},\ \bibinfo {pages} {135704} (\bibinfo {year} {2013})}\BibitemShut
  {NoStop}%
\bibitem [{\citenamefont {Heyl}(2018)}]{Heyl2018a}%
  \BibitemOpen
  \bibfield  {author} {\bibinfo {author} {\bibfnamefont {M.}~\bibnamefont
  {Heyl}},\ }\bibfield  {title} {\bibinfo {title} {Dynamical quantum phase
  transitions: A review},\ }\href {https://doi.org/10.1088/1361-6633/aaaf9a}
  {\bibfield  {journal} {\bibinfo  {journal} {Reports on Progress in Physics}\
  }\textbf {\bibinfo {volume} {81}},\ \bibinfo {pages} {054001} (\bibinfo
  {year} {2018})}\BibitemShut {NoStop}%
\bibitem [{\citenamefont {Sedlmayr}(2019)}]{Sedlmayr2019a}%
  \BibitemOpen
  \bibfield  {author} {\bibinfo {author} {\bibfnamefont {N.}~\bibnamefont
  {Sedlmayr}},\ }\bibfield  {title} {\bibinfo {title} {Dynamical {{Phase
  Transitions}} in {{Topological Insulators}}},\ }\href
  {https://doi.org/10.12693/APhysPolA.135.1191} {\bibfield  {journal} {\bibinfo
   {journal} {Acta Physica Polonica A}\ }\textbf {\bibinfo {volume} {135}},\
  \bibinfo {pages} {1191} (\bibinfo {year} {2019})}\BibitemShut {NoStop}%
\bibitem [{\citenamefont {Heyl}(2015)}]{Heyl2015}%
  \BibitemOpen
  \bibfield  {author} {\bibinfo {author} {\bibfnamefont {M.}~\bibnamefont
  {Heyl}},\ }\bibfield  {title} {\bibinfo {title} {Scaling and {{Universality}}
  at {{Dynamical Quantum Phase Transitions}}},\ }\href
  {https://doi.org/10.1103/PhysRevLett.115.140602} {\bibfield  {journal}
  {\bibinfo  {journal} {Physical Review Letters}\ }\textbf {\bibinfo {volume}
  {115}},\ \bibinfo {pages} {140602} (\bibinfo {year} {2015})}\BibitemShut
  {NoStop}%
\bibitem [{\citenamefont {Sharma}\ \emph {et~al.}(2015)\citenamefont {Sharma},
  \citenamefont {Suzuki},\ and\ \citenamefont {Dutta}}]{Sharma2015}%
  \BibitemOpen
  \bibfield  {author} {\bibinfo {author} {\bibfnamefont {S.}~\bibnamefont
  {Sharma}}, \bibinfo {author} {\bibfnamefont {S.}~\bibnamefont {Suzuki}},\
  and\ \bibinfo {author} {\bibfnamefont {A.}~\bibnamefont {Dutta}},\ }\bibfield
   {title} {\bibinfo {title} {Quenches and dynamical phase transitions in a
  nonintegrable quantum {{Ising}} model},\ }\href
  {https://doi.org/10.1103/PhysRevB.92.104306} {\bibfield  {journal} {\bibinfo
  {journal} {Physical Review B}\ }\textbf {\bibinfo {volume} {92}},\ \bibinfo
  {pages} {104306} (\bibinfo {year} {2015})}\BibitemShut {NoStop}%
\bibitem [{\citenamefont {Homrighausen}\ \emph {et~al.}(2017)\citenamefont
  {Homrighausen}, \citenamefont {Abeling}, \citenamefont {{Zauner-Stauber}},\
  and\ \citenamefont {Halimeh}}]{Homrighausen2017}%
  \BibitemOpen
  \bibfield  {author} {\bibinfo {author} {\bibfnamefont {I.}~\bibnamefont
  {Homrighausen}}, \bibinfo {author} {\bibfnamefont {N.~O.}\ \bibnamefont
  {Abeling}}, \bibinfo {author} {\bibfnamefont {V.}~\bibnamefont
  {{Zauner-Stauber}}},\ and\ \bibinfo {author} {\bibfnamefont {J.~C.}\
  \bibnamefont {Halimeh}},\ }\bibfield  {title} {\bibinfo {title} {Anomalous
  dynamical phase in quantum spin chains with long-range interactions},\ }\href
  {https://doi.org/10.1103/PhysRevB.96.104436} {\bibfield  {journal} {\bibinfo
  {journal} {Physical Review B}\ }\textbf {\bibinfo {volume} {96}},\ \bibinfo
  {pages} {104436} (\bibinfo {year} {2017})}\BibitemShut {NoStop}%
\bibitem [{\citenamefont {Halimeh}\ \emph {et~al.}(2020)\citenamefont
  {Halimeh}, \citenamefont {Van~Damme}, \citenamefont {{Zauner-Stauber}},\ and\
  \citenamefont {Vanderstraeten}}]{Halimeh2018}%
  \BibitemOpen
  \bibfield  {author} {\bibinfo {author} {\bibfnamefont {J.~C.}\ \bibnamefont
  {Halimeh}}, \bibinfo {author} {\bibfnamefont {M.}~\bibnamefont {Van~Damme}},
  \bibinfo {author} {\bibfnamefont {V.}~\bibnamefont {{Zauner-Stauber}}},\ and\
  \bibinfo {author} {\bibfnamefont {L.}~\bibnamefont {Vanderstraeten}},\
  }\bibfield  {title} {\bibinfo {title} {Quasiparticle {{Origin}} of
  {{Dynamical Quantum Phase Transitions}}},\ }\href
  {https://doi.org/10.1103/PhysRevResearch.2.033111} {\bibfield  {journal}
  {\bibinfo  {journal} {Physical Review Research}\ }\textbf {\bibinfo {volume}
  {2}},\ \bibinfo {pages} {033111} (\bibinfo {year} {2020})}\BibitemShut
  {NoStop}%
\bibitem [{\citenamefont {Shpielberg}\ \emph {et~al.}(2018)\citenamefont
  {Shpielberg}, \citenamefont {Nemoto},\ and\ \citenamefont
  {Caetano}}]{Shpielberg2018}%
  \BibitemOpen
  \bibfield  {author} {\bibinfo {author} {\bibfnamefont {O.}~\bibnamefont
  {Shpielberg}}, \bibinfo {author} {\bibfnamefont {T.}~\bibnamefont {Nemoto}},\
  and\ \bibinfo {author} {\bibfnamefont {J.}~\bibnamefont {Caetano}},\
  }\bibfield  {title} {\bibinfo {title} {Universality in dynamical phase
  transitions of diffusive systems},\ }\href
  {https://doi.org/10.1103/PhysRevE.98.052116} {\bibfield  {journal} {\bibinfo
  {journal} {Physical Review E}\ }\textbf {\bibinfo {volume} {98}},\ \bibinfo
  {pages} {052116} (\bibinfo {year} {2018})}\BibitemShut {NoStop}%
\bibitem [{\citenamefont {Zunkovic}\ \emph {et~al.}(2018)\citenamefont
  {Zunkovic}, \citenamefont {Heyl}, \citenamefont {Knap},\ and\ \citenamefont
  {Silva}}]{Zunkovic2018}%
  \BibitemOpen
  \bibfield  {author} {\bibinfo {author} {\bibfnamefont {B.}~\bibnamefont
  {Zunkovic}}, \bibinfo {author} {\bibfnamefont {M.}~\bibnamefont {Heyl}},
  \bibinfo {author} {\bibfnamefont {M.}~\bibnamefont {Knap}},\ and\ \bibinfo
  {author} {\bibfnamefont {A.}~\bibnamefont {Silva}},\ }\bibfield  {title}
  {\bibinfo {title} {Dynamical {{Quantum Phase Transitions}} in {{Spin Chains}}
  with {{Long-Range Interactions}}: {{Merging}} different concepts of
  non-equilibrium criticality},\ }\href
  {https://doi.org/10.1103/PhysRevLett.120.130601} {\bibfield  {journal}
  {\bibinfo  {journal} {Physical Review Letters}\ }\textbf {\bibinfo {volume}
  {120}},\ \bibinfo {pages} {130601} (\bibinfo {year} {2018})}\BibitemShut
  {NoStop}%
\bibitem [{\citenamefont {Srivastav}\ \emph {et~al.}(2019)\citenamefont
  {Srivastav}, \citenamefont {Bhattacharya},\ and\ \citenamefont
  {Dutta}}]{Srivastav2019}%
  \BibitemOpen
  \bibfield  {author} {\bibinfo {author} {\bibfnamefont {V.}~\bibnamefont
  {Srivastav}}, \bibinfo {author} {\bibfnamefont {U.}~\bibnamefont
  {Bhattacharya}},\ and\ \bibinfo {author} {\bibfnamefont {A.}~\bibnamefont
  {Dutta}},\ }\bibfield  {title} {\bibinfo {title} {Dynamical quantum phase
  transitions in extended toric-code models},\ }\href
  {https://doi.org/10.1103/PhysRevB.100.144203} {\bibfield  {journal} {\bibinfo
   {journal} {Physical Review B}\ }\textbf {\bibinfo {volume} {100}},\ \bibinfo
  {pages} {144203} (\bibinfo {year} {2019})}\BibitemShut {NoStop}%
\bibitem [{\citenamefont {Hagym{\'a}si}\ \emph {et~al.}(2019)\citenamefont
  {Hagym{\'a}si}, \citenamefont {Hubig}, \citenamefont {Legeza},\ and\
  \citenamefont {Schollw{\"o}ck}}]{Hagymasi2019}%
  \BibitemOpen
  \bibfield  {author} {\bibinfo {author} {\bibfnamefont {I.}~\bibnamefont
  {Hagym{\'a}si}}, \bibinfo {author} {\bibfnamefont {C.}~\bibnamefont {Hubig}},
  \bibinfo {author} {\bibfnamefont {{\"O}.}~\bibnamefont {Legeza}},\ and\
  \bibinfo {author} {\bibfnamefont {U.}~\bibnamefont {Schollw{\"o}ck}},\
  }\bibfield  {title} {\bibinfo {title} {Dynamical {{Topological Quantum Phase
  Transitions}} in {{Nonintegrable Models}}},\ }\href
  {https://doi.org/10.1103/PhysRevLett.122.250601} {\bibfield  {journal}
  {\bibinfo  {journal} {Physical Review Letters}\ }\textbf {\bibinfo {volume}
  {122}},\ \bibinfo {pages} {250601} (\bibinfo {year} {2019})}\BibitemShut
  {NoStop}%
\bibitem [{\citenamefont {Huang}\ \emph {et~al.}(2019)\citenamefont {Huang},
  \citenamefont {Banerjee},\ and\ \citenamefont {Heyl}}]{Huang2019}%
  \BibitemOpen
  \bibfield  {author} {\bibinfo {author} {\bibfnamefont {Y.-P.}\ \bibnamefont
  {Huang}}, \bibinfo {author} {\bibfnamefont {D.}~\bibnamefont {Banerjee}},\
  and\ \bibinfo {author} {\bibfnamefont {M.}~\bibnamefont {Heyl}},\ }\bibfield
  {title} {\bibinfo {title} {Dynamical {{Quantum Phase Transitions}} in
  {{U}}(1) {{Quantum Link Models}}},\ }\href
  {https://doi.org/10.1103/PhysRevLett.122.250401} {\bibfield  {journal}
  {\bibinfo  {journal} {Physical Review Letters}\ }\textbf {\bibinfo {volume}
  {122}},\ \bibinfo {pages} {250401} (\bibinfo {year} {2019})}\BibitemShut
  {NoStop}%
\bibitem [{\citenamefont {Gurarie}(2019)}]{Gurarie2019}%
  \BibitemOpen
  \bibfield  {author} {\bibinfo {author} {\bibfnamefont {V.}~\bibnamefont
  {Gurarie}},\ }\bibfield  {title} {\bibinfo {title} {Dynamical quantum phase
  transitions in the random field {{Ising}} model},\ }\href
  {https://doi.org/10.1103/PhysRevA.100.031601} {\bibfield  {journal} {\bibinfo
   {journal} {Physical Review A}\ }\textbf {\bibinfo {volume} {100}},\ \bibinfo
  {pages} {031601} (\bibinfo {year} {2019})}\BibitemShut {NoStop}%
\bibitem [{\citenamefont {Abdi}(2019)}]{Abdi2019}%
  \BibitemOpen
  \bibfield  {author} {\bibinfo {author} {\bibfnamefont {M.}~\bibnamefont
  {Abdi}},\ }\bibfield  {title} {\bibinfo {title} {Dynamical quantum phase
  transition in {{Bose-Einstein}} condensates},\ }\href
  {https://doi.org/10.1103/PhysRevB.100.184310} {\bibfield  {journal} {\bibinfo
   {journal} {Physical Review B}\ }\textbf {\bibinfo {volume} {100}},\ \bibinfo
  {pages} {184310} (\bibinfo {year} {2019})}\BibitemShut {NoStop}%
\bibitem [{\citenamefont {Cao}\ \emph {et~al.}(2020)\citenamefont {Cao},
  \citenamefont {Li}, \citenamefont {Zhong},\ and\ \citenamefont
  {Tong}}]{Cao2020}%
  \BibitemOpen
  \bibfield  {author} {\bibinfo {author} {\bibfnamefont {K.}~\bibnamefont
  {Cao}}, \bibinfo {author} {\bibfnamefont {W.}~\bibnamefont {Li}}, \bibinfo
  {author} {\bibfnamefont {M.}~\bibnamefont {Zhong}},\ and\ \bibinfo {author}
  {\bibfnamefont {P.}~\bibnamefont {Tong}},\ }\bibfield  {title} {\bibinfo
  {title} {Influence of weak disorder on the dynamical quantum phase
  transitions in the anisotropic {{XY}} chain},\ }\href
  {https://doi.org/10.1103/PhysRevB.102.014207} {\bibfield  {journal} {\bibinfo
   {journal} {Physical Review B}\ }\textbf {\bibinfo {volume} {102}},\ \bibinfo
  {pages} {014207} (\bibinfo {year} {2020})}\BibitemShut {NoStop}%
\bibitem [{\citenamefont {Puebla}(2020)}]{Puebla2020}%
  \BibitemOpen
  \bibfield  {author} {\bibinfo {author} {\bibfnamefont {R.}~\bibnamefont
  {Puebla}},\ }\bibfield  {title} {\bibinfo {title} {Finite-component dynamical
  quantum phase transitions},\ }\href
  {https://doi.org/10.1103/PhysRevB.102.220302} {\bibfield  {journal} {\bibinfo
   {journal} {Physical Review B}\ }\textbf {\bibinfo {volume} {102}},\ \bibinfo
  {pages} {220302} (\bibinfo {year} {2020})}\BibitemShut {NoStop}%
\bibitem [{\citenamefont {Soriente}\ \emph {et~al.}(2020)\citenamefont
  {Soriente}, \citenamefont {Chitra},\ and\ \citenamefont
  {Zilberberg}}]{Soriente2020}%
  \BibitemOpen
  \bibfield  {author} {\bibinfo {author} {\bibfnamefont {M.}~\bibnamefont
  {Soriente}}, \bibinfo {author} {\bibfnamefont {R.}~\bibnamefont {Chitra}},\
  and\ \bibinfo {author} {\bibfnamefont {O.}~\bibnamefont {Zilberberg}},\
  }\bibfield  {title} {\bibinfo {title} {Distinguishing phases using the
  dynamical response of driven-dissipative light-matter systems},\ }\href
  {https://doi.org/10.1103/PhysRevA.101.023823} {\bibfield  {journal} {\bibinfo
   {journal} {Physical Review A}\ }\textbf {\bibinfo {volume} {101}},\ \bibinfo
  {pages} {023823} (\bibinfo {year} {2020})}\BibitemShut {NoStop}%
\bibitem [{\citenamefont {Porta}\ \emph {et~al.}(2020)\citenamefont {Porta},
  \citenamefont {Cavaliere}, \citenamefont {Sassetti},\ and\ \citenamefont
  {Traverso~Ziani}}]{Porta2020}%
  \BibitemOpen
  \bibfield  {author} {\bibinfo {author} {\bibfnamefont {S.}~\bibnamefont
  {Porta}}, \bibinfo {author} {\bibfnamefont {F.}~\bibnamefont {Cavaliere}},
  \bibinfo {author} {\bibfnamefont {M.}~\bibnamefont {Sassetti}},\ and\
  \bibinfo {author} {\bibfnamefont {N.}~\bibnamefont {Traverso~Ziani}},\
  }\bibfield  {title} {\bibinfo {title} {Topological classification of
  dynamical quantum phase transitions in the xy chain},\ }\href
  {https://doi.org/10.1038/s41598-020-69621-8} {\bibfield  {journal} {\bibinfo
  {journal} {Scientific Reports}\ }\textbf {\bibinfo {volume} {10}},\ \bibinfo
  {pages} {12766} (\bibinfo {year} {2020})}\BibitemShut {NoStop}%
\bibitem [{\citenamefont {Mishra}\ \emph {et~al.}(2020)\citenamefont {Mishra},
  \citenamefont {Jafari},\ and\ \citenamefont {Akbari}}]{Mishra2020}%
  \BibitemOpen
  \bibfield  {author} {\bibinfo {author} {\bibfnamefont {U.}~\bibnamefont
  {Mishra}}, \bibinfo {author} {\bibfnamefont {R.}~\bibnamefont {Jafari}},\
  and\ \bibinfo {author} {\bibfnamefont {A.}~\bibnamefont {Akbari}},\
  }\bibfield  {title} {\bibinfo {title} {Disordered {{Kitaev}} chain with
  long-range pairing: {{Loschmidt}} echo revivals and dynamical phase
  transitions},\ }\href {https://doi.org/10.1088/1751-8121/ab97de} {\bibfield
  {journal} {\bibinfo  {journal} {Journal of Physics A: Mathematical and
  Theoretical}\ }\textbf {\bibinfo {volume} {53}},\ \bibinfo {pages} {375301}
  (\bibinfo {year} {2020})}\BibitemShut {NoStop}%
\bibitem [{\citenamefont {Link}\ and\ \citenamefont {Strunz}(2020)}]{Link2020}%
  \BibitemOpen
  \bibfield  {author} {\bibinfo {author} {\bibfnamefont {V.}~\bibnamefont
  {Link}}\ and\ \bibinfo {author} {\bibfnamefont {W.~T.}\ \bibnamefont
  {Strunz}},\ }\bibfield  {title} {\bibinfo {title} {Dynamical {{Phase
  Transitions}} in {{Dissipative Quantum Dynamics}} with {{Quantum Optical
  Realization}}},\ }\href {https://doi.org/10.1103/PhysRevLett.125.143602}
  {\bibfield  {journal} {\bibinfo  {journal} {Physical Review Letters}\
  }\textbf {\bibinfo {volume} {125}},\ \bibinfo {pages} {143602} (\bibinfo
  {year} {2020})}\BibitemShut {NoStop}%
\bibitem [{\citenamefont {Sun}\ and\ \citenamefont {Wei}(2020)}]{Sun2020}%
  \BibitemOpen
  \bibfield  {author} {\bibinfo {author} {\bibfnamefont {G.}~\bibnamefont
  {Sun}}\ and\ \bibinfo {author} {\bibfnamefont {B.-B.}\ \bibnamefont {Wei}},\
  }\bibfield  {title} {\bibinfo {title} {Dynamical quantum phase transitions in
  a spin chain with deconfined quantum critical points},\ }\href
  {https://doi.org/10.1103/PhysRevB.102.094302} {\bibfield  {journal} {\bibinfo
   {journal} {Physical Review B}\ }\textbf {\bibinfo {volume} {102}},\ \bibinfo
  {pages} {094302} (\bibinfo {year} {2020})}\BibitemShut {NoStop}%
\bibitem [{\citenamefont {Rylands}\ and\ \citenamefont
  {Galitski}(2020)}]{Rylands2020}%
  \BibitemOpen
  \bibfield  {author} {\bibinfo {author} {\bibfnamefont {C.}~\bibnamefont
  {Rylands}}\ and\ \bibinfo {author} {\bibfnamefont {V.}~\bibnamefont
  {Galitski}},\ }\bibfield  {title} {\bibinfo {title} {Dynamical {{Quantum
  Phase}} transitions and {{Recurrences}} in the {{Non-Equilibrium BCS}}
  model},\ }\href@noop {} {\bibfield  {journal} {\bibinfo  {journal}
  {arXiv:2001.10084 [cond-mat]}\ } (\bibinfo {year} {2020})}\BibitemShut
  {NoStop}%
\bibitem [{\citenamefont {Uhrich}\ \emph {et~al.}(2020)\citenamefont {Uhrich},
  \citenamefont {Defenu}, \citenamefont {Jafari},\ and\ \citenamefont
  {Halimeh}}]{Uhrich2020}%
  \BibitemOpen
  \bibfield  {author} {\bibinfo {author} {\bibfnamefont {P.}~\bibnamefont
  {Uhrich}}, \bibinfo {author} {\bibfnamefont {N.}~\bibnamefont {Defenu}},
  \bibinfo {author} {\bibfnamefont {R.}~\bibnamefont {Jafari}},\ and\ \bibinfo
  {author} {\bibfnamefont {J.~C.}\ \bibnamefont {Halimeh}},\ }\bibfield
  {title} {\bibinfo {title} {Out-of-equilibrium phase diagram of long-range
  superconductors},\ }\href {https://doi.org/10.1103/PhysRevB.101.245148}
  {\bibfield  {journal} {\bibinfo  {journal} {Physical Review B}\ }\textbf
  {\bibinfo {volume} {101}},\ \bibinfo {pages} {245148} (\bibinfo {year}
  {2020})}\BibitemShut {NoStop}%
\bibitem [{\citenamefont {Syed}\ \emph {et~al.}(2021)\citenamefont {Syed},
  \citenamefont {Enss},\ and\ \citenamefont {Defenu}}]{Syed2021}%
  \BibitemOpen
  \bibfield  {author} {\bibinfo {author} {\bibfnamefont {M.}~\bibnamefont
  {Syed}}, \bibinfo {author} {\bibfnamefont {T.}~\bibnamefont {Enss}},\ and\
  \bibinfo {author} {\bibfnamefont {N.}~\bibnamefont {Defenu}},\ }\bibfield
  {title} {\bibinfo {title} {Dynamical quantum phase transition in a bosonic
  system with long-range interactions},\ }\href
  {https://doi.org/10.1103/PhysRevB.103.064306} {\bibfield  {journal} {\bibinfo
   {journal} {Physical Review B}\ }\textbf {\bibinfo {volume} {103}},\ \bibinfo
  {pages} {064306} (\bibinfo {year} {2021})}\BibitemShut {NoStop}%
\bibitem [{\citenamefont {Trapin}\ \emph {et~al.}(2021)\citenamefont {Trapin},
  \citenamefont {Halimeh},\ and\ \citenamefont {Heyl}}]{Trapin2021}%
  \BibitemOpen
  \bibfield  {author} {\bibinfo {author} {\bibfnamefont {D.}~\bibnamefont
  {Trapin}}, \bibinfo {author} {\bibfnamefont {J.~C.}\ \bibnamefont
  {Halimeh}},\ and\ \bibinfo {author} {\bibfnamefont {M.}~\bibnamefont
  {Heyl}},\ }\bibfield  {title} {\bibinfo {title} {Unconventional critical
  exponents at dynamical quantum phase transitions in a random {{Ising}}
  chain},\ }\href {https://doi.org/10.1103/PhysRevB.104.115159} {\bibfield
  {journal} {\bibinfo  {journal} {Physical Review B}\ }\textbf {\bibinfo
  {volume} {104}},\ \bibinfo {pages} {115159} (\bibinfo {year}
  {2021})}\BibitemShut {NoStop}%
\bibitem [{\citenamefont {Yu}\ \emph {et~al.}(2021)\citenamefont {Yu},
  \citenamefont {Sacramento}, \citenamefont {Li},\ and\ \citenamefont
  {Lin}}]{Yu2021}%
  \BibitemOpen
  \bibfield  {author} {\bibinfo {author} {\bibfnamefont {W.~C.}\ \bibnamefont
  {Yu}}, \bibinfo {author} {\bibfnamefont {P.~D.}\ \bibnamefont {Sacramento}},
  \bibinfo {author} {\bibfnamefont {Y.~C.}\ \bibnamefont {Li}},\ and\ \bibinfo
  {author} {\bibfnamefont {H.-Q.}\ \bibnamefont {Lin}},\ }\bibfield  {title}
  {\bibinfo {title} {Correlations and dynamical quantum phase transitions in an
  interacting topological insulator},\ }\href
  {https://doi.org/10.1103/PhysRevB.104.085104} {\bibfield  {journal} {\bibinfo
   {journal} {Physical Review B}\ }\textbf {\bibinfo {volume} {104}},\ \bibinfo
  {pages} {085104} (\bibinfo {year} {2021})}\BibitemShut {NoStop}%
\bibitem [{\citenamefont {Zhou}\ \emph {et~al.}(2021)\citenamefont {Zhou},
  \citenamefont {Zeng},\ and\ \citenamefont {Chen}}]{Zhou2021c}%
  \BibitemOpen
  \bibfield  {author} {\bibinfo {author} {\bibfnamefont {B.}~\bibnamefont
  {Zhou}}, \bibinfo {author} {\bibfnamefont {Y.}~\bibnamefont {Zeng}},\ and\
  \bibinfo {author} {\bibfnamefont {S.}~\bibnamefont {Chen}},\ }\bibfield
  {title} {\bibinfo {title} {Exact zeros of the {{Loschmidt}} echo and quantum
  speed limit time for the dynamical quantum phase transition in finite-size
  systems},\ }\href {https://doi.org/10.1103/PhysRevB.104.094311} {\bibfield
  {journal} {\bibinfo  {journal} {Physical Review B}\ }\textbf {\bibinfo
  {volume} {104}},\ \bibinfo {pages} {094311} (\bibinfo {year}
  {2021})}\BibitemShut {NoStop}%
\bibitem [{\citenamefont {Sadrzadeh}\ \emph {et~al.}(2021)\citenamefont
  {Sadrzadeh}, \citenamefont {Jafari},\ and\ \citenamefont
  {Langari}}]{Sadrzadeh2021}%
  \BibitemOpen
  \bibfield  {author} {\bibinfo {author} {\bibfnamefont {M.}~\bibnamefont
  {Sadrzadeh}}, \bibinfo {author} {\bibfnamefont {R.}~\bibnamefont {Jafari}},\
  and\ \bibinfo {author} {\bibfnamefont {A.}~\bibnamefont {Langari}},\
  }\bibfield  {title} {\bibinfo {title} {Dynamical topological quantum phase
  transitions at criticality},\ }\href
  {https://doi.org/10.1103/PhysRevB.103.144305} {\bibfield  {journal} {\bibinfo
   {journal} {Physical Review B}\ }\textbf {\bibinfo {volume} {103}},\ \bibinfo
  {pages} {144305} (\bibinfo {year} {2021})}\BibitemShut {NoStop}%
\bibitem [{\citenamefont {Halimeh}\ \emph
  {et~al.}(2021{\natexlab{a}})\citenamefont {Halimeh}, \citenamefont
  {Van~Damme}, \citenamefont {Guo}, \citenamefont {Lang},\ and\ \citenamefont
  {Hauke}}]{Halimeh2021}%
  \BibitemOpen
  \bibfield  {author} {\bibinfo {author} {\bibfnamefont {J.~C.}\ \bibnamefont
  {Halimeh}}, \bibinfo {author} {\bibfnamefont {M.}~\bibnamefont {Van~Damme}},
  \bibinfo {author} {\bibfnamefont {L.}~\bibnamefont {Guo}}, \bibinfo {author}
  {\bibfnamefont {J.}~\bibnamefont {Lang}},\ and\ \bibinfo {author}
  {\bibfnamefont {P.}~\bibnamefont {Hauke}},\ }\bibfield  {title} {\bibinfo
  {title} {Dynamical phase transitions in quantum spin models with
  antiferromagnetic long-range interactions},\ }\href
  {https://doi.org/10.1103/PhysRevB.104.115133} {\bibfield  {journal} {\bibinfo
   {journal} {Physical Review B}\ }\textbf {\bibinfo {volume} {104}},\ \bibinfo
  {pages} {115133} (\bibinfo {year} {2021}{\natexlab{a}})}\BibitemShut
  {NoStop}%
\bibitem [{\citenamefont {Halimeh}\ \emph
  {et~al.}(2021{\natexlab{b}})\citenamefont {Halimeh}, \citenamefont {Trapin},
  \citenamefont {Van~Damme},\ and\ \citenamefont {Heyl}}]{Halimeh2021a}%
  \BibitemOpen
  \bibfield  {author} {\bibinfo {author} {\bibfnamefont {J.~C.}\ \bibnamefont
  {Halimeh}}, \bibinfo {author} {\bibfnamefont {D.}~\bibnamefont {Trapin}},
  \bibinfo {author} {\bibfnamefont {M.}~\bibnamefont {Van~Damme}},\ and\
  \bibinfo {author} {\bibfnamefont {M.}~\bibnamefont {Heyl}},\ }\bibfield
  {title} {\bibinfo {title} {Local measures of dynamical quantum phase
  transitions},\ }\href {https://doi.org/10.1103/PhysRevB.104.075130}
  {\bibfield  {journal} {\bibinfo  {journal} {Physical Review B}\ }\textbf
  {\bibinfo {volume} {104}},\ \bibinfo {pages} {075130} (\bibinfo {year}
  {2021}{\natexlab{b}})}\BibitemShut {NoStop}%
\bibitem [{\citenamefont {De~Nicola}\ \emph {et~al.}(2021)\citenamefont
  {De~Nicola}, \citenamefont {Michailidis},\ and\ \citenamefont
  {Serbyn}}]{DeNicola2021}%
  \BibitemOpen
  \bibfield  {author} {\bibinfo {author} {\bibfnamefont {S.}~\bibnamefont
  {De~Nicola}}, \bibinfo {author} {\bibfnamefont {A.~A.}\ \bibnamefont
  {Michailidis}},\ and\ \bibinfo {author} {\bibfnamefont {M.}~\bibnamefont
  {Serbyn}},\ }\bibfield  {title} {\bibinfo {title} {Entanglement {{View}} of
  {{Dynamical Quantum Phase Transitions}}},\ }\href
  {https://doi.org/10.1103/PhysRevLett.126.040602} {\bibfield  {journal}
  {\bibinfo  {journal} {Physical Review Letters}\ }\textbf {\bibinfo {volume}
  {126}},\ \bibinfo {pages} {040602} (\bibinfo {year} {2021})}\BibitemShut
  {NoStop}%
\bibitem [{\citenamefont {Rylands}\ \emph {et~al.}(2021)\citenamefont
  {Rylands}, \citenamefont {Yuzbashyan}, \citenamefont {Gurarie}, \citenamefont
  {Zabalo},\ and\ \citenamefont {Galitski}}]{Rylands2021}%
  \BibitemOpen
  \bibfield  {author} {\bibinfo {author} {\bibfnamefont {C.}~\bibnamefont
  {Rylands}}, \bibinfo {author} {\bibfnamefont {E.~A.}\ \bibnamefont
  {Yuzbashyan}}, \bibinfo {author} {\bibfnamefont {V.}~\bibnamefont {Gurarie}},
  \bibinfo {author} {\bibfnamefont {A.}~\bibnamefont {Zabalo}},\ and\ \bibinfo
  {author} {\bibfnamefont {V.}~\bibnamefont {Galitski}},\ }\bibfield  {title}
  {\bibinfo {title} {Loschmidt echo of far-from-equilibrium fermionic
  superfluids},\ }\href {https://doi.org/10.1016/j.aop.2021.168554} {\bibfield
  {journal} {\bibinfo  {journal} {Annals of Physics}\ }\textbf {\bibinfo
  {volume} {435}},\ \bibinfo {pages} {168554} (\bibinfo {year}
  {2021})}\BibitemShut {NoStop}%
\bibitem [{\citenamefont {Peotta}\ \emph {et~al.}(2021)\citenamefont {Peotta},
  \citenamefont {Brange}, \citenamefont {Deger}, \citenamefont {Ojanen},\ and\
  \citenamefont {Flindt}}]{Peotta2021}%
  \BibitemOpen
  \bibfield  {author} {\bibinfo {author} {\bibfnamefont {S.}~\bibnamefont
  {Peotta}}, \bibinfo {author} {\bibfnamefont {F.}~\bibnamefont {Brange}},
  \bibinfo {author} {\bibfnamefont {A.}~\bibnamefont {Deger}}, \bibinfo
  {author} {\bibfnamefont {T.}~\bibnamefont {Ojanen}},\ and\ \bibinfo {author}
  {\bibfnamefont {C.}~\bibnamefont {Flindt}},\ }\bibfield  {title} {\bibinfo
  {title} {Determination of {{Dynamical Quantum Phase Transitions}} in
  {{Strongly Correlated Many-Body Systems Using Loschmidt Cumulants}}},\ }\href
  {https://doi.org/10.1103/PhysRevX.11.041018} {\bibfield  {journal} {\bibinfo
  {journal} {Physical Review X}\ }\textbf {\bibinfo {volume} {11}},\ \bibinfo
  {pages} {041018} (\bibinfo {year} {2021})}\BibitemShut {NoStop}%
\bibitem [{\citenamefont {Modak}\ and\ \citenamefont
  {Rakshit}(2021)}]{Modak2021}%
  \BibitemOpen
  \bibfield  {author} {\bibinfo {author} {\bibfnamefont {R.}~\bibnamefont
  {Modak}}\ and\ \bibinfo {author} {\bibfnamefont {D.}~\bibnamefont
  {Rakshit}},\ }\bibfield  {title} {\bibinfo {title} {Many-body dynamical phase
  transition in a quasiperiodic potential},\ }\href
  {https://doi.org/10.1103/PhysRevB.103.224310} {\bibfield  {journal} {\bibinfo
   {journal} {Physical Review B}\ }\textbf {\bibinfo {volume} {103}},\ \bibinfo
  {pages} {224310} (\bibinfo {year} {2021})}\BibitemShut {NoStop}%
\bibitem [{\citenamefont {Cheraghi}\ and\ \citenamefont
  {Mahdavifar}(2021)}]{Cheraghi2021}%
  \BibitemOpen
  \bibfield  {author} {\bibinfo {author} {\bibfnamefont {H.}~\bibnamefont
  {Cheraghi}}\ and\ \bibinfo {author} {\bibfnamefont {S.}~\bibnamefont
  {Mahdavifar}},\ }\bibfield  {title} {\bibinfo {title} {Dynamical {{Quantum
  Phase Transitions}} in the {{1D Nonintegrable Spin}}-1/2 {{Transverse Field
  XZZ Model}}},\ }\href {https://doi.org/10.1002/andp.202000542} {\bibfield
  {journal} {\bibinfo  {journal} {Annalen der Physik}\ ,\ \bibinfo {pages}
  {2000542}} (\bibinfo {year} {2021})}\BibitemShut {NoStop}%
\bibitem [{\citenamefont {Cao}\ \emph {et~al.}(2021)\citenamefont {Cao},
  \citenamefont {Ming},\ and\ \citenamefont {Tong}}]{Cao2021}%
  \BibitemOpen
  \bibfield  {author} {\bibinfo {author} {\bibfnamefont {K.}~\bibnamefont
  {Cao}}, \bibinfo {author} {\bibfnamefont {Z.}~\bibnamefont {Ming}},\ and\
  \bibinfo {author} {\bibfnamefont {P.}~\bibnamefont {Tong}},\ }\bibfield
  {title} {\bibinfo {title} {Dynamical quantum phase transition in quantum spin
  chains with gapless phases},\ }\href@noop {} {\bibfield  {journal} {\bibinfo
  {journal} {arXiv:2106.00191 [cond-mat]}\ } (\bibinfo {year}
  {2021})}\BibitemShut {NoStop}%
\bibitem [{\citenamefont {Bandyopadhyay}\ \emph {et~al.}(2021)\citenamefont
  {Bandyopadhyay}, \citenamefont {Polkovnikov},\ and\ \citenamefont
  {Dutta}}]{Bandyopadhyay2021}%
  \BibitemOpen
  \bibfield  {author} {\bibinfo {author} {\bibfnamefont {S.}~\bibnamefont
  {Bandyopadhyay}}, \bibinfo {author} {\bibfnamefont {A.}~\bibnamefont
  {Polkovnikov}},\ and\ \bibinfo {author} {\bibfnamefont {A.}~\bibnamefont
  {Dutta}},\ }\bibfield  {title} {\bibinfo {title} {Observing {{Dynamical
  Quantum Phase Transitions}} through {{Quasilocal String Operators}}},\ }\href
  {https://doi.org/10.1103/PhysRevLett.126.200602} {\bibfield  {journal}
  {\bibinfo  {journal} {Physical Review Letters}\ }\textbf {\bibinfo {volume}
  {126}},\ \bibinfo {pages} {200602} (\bibinfo {year} {2021})}\BibitemShut
  {NoStop}%
\bibitem [{\citenamefont {Vajna}\ and\ \citenamefont
  {D{\'o}ra}(2015)}]{Vajna2015}%
  \BibitemOpen
  \bibfield  {author} {\bibinfo {author} {\bibfnamefont {S.}~\bibnamefont
  {Vajna}}\ and\ \bibinfo {author} {\bibfnamefont {B.}~\bibnamefont
  {D{\'o}ra}},\ }\bibfield  {title} {\bibinfo {title} {Topological
  classification of dynamical phase transitions},\ }\href
  {https://doi.org/10.1103/PhysRevB.91.155127} {\bibfield  {journal} {\bibinfo
  {journal} {Physical Review B}\ }\textbf {\bibinfo {volume} {91}},\ \bibinfo
  {pages} {155127} (\bibinfo {year} {2015})}\BibitemShut {NoStop}%
\bibitem [{\citenamefont {Schmitt}\ and\ \citenamefont
  {Kehrein}(2015)}]{Schmitt2015}%
  \BibitemOpen
  \bibfield  {author} {\bibinfo {author} {\bibfnamefont {M.}~\bibnamefont
  {Schmitt}}\ and\ \bibinfo {author} {\bibfnamefont {S.}~\bibnamefont
  {Kehrein}},\ }\bibfield  {title} {\bibinfo {title} {Dynamical quantum phase
  transitions in the {{Kitaev}} honeycomb model},\ }\href
  {https://doi.org/10.1103/PhysRevB.92.075114} {\bibfield  {journal} {\bibinfo
  {journal} {Physical Review B}\ }\textbf {\bibinfo {volume} {92}},\ \bibinfo
  {pages} {075114} (\bibinfo {year} {2015})}\BibitemShut {NoStop}%
\bibitem [{\citenamefont {Jafari}(2016)}]{Jafari2016}%
  \BibitemOpen
  \bibfield  {author} {\bibinfo {author} {\bibfnamefont {R.}~\bibnamefont
  {Jafari}},\ }\bibfield  {title} {\bibinfo {title} {Quench dynamics and ground
  state fidelity of the one-dimensional extended quantum compass model in a
  transverse field},\ }\href {https://doi.org/10.1088/1751-8113/49/18/185004}
  {\bibfield  {journal} {\bibinfo  {journal} {Journal of Physics A:
  Mathematical and Theoretical}\ }\textbf {\bibinfo {volume} {49}},\ \bibinfo
  {pages} {185004} (\bibinfo {year} {2016})}\BibitemShut {NoStop}%
\bibitem [{\citenamefont {Jafari}\ and\ \citenamefont
  {Johannesson}(2017{\natexlab{a}})}]{Jafari2017a}%
  \BibitemOpen
  \bibfield  {author} {\bibinfo {author} {\bibfnamefont {R.}~\bibnamefont
  {Jafari}}\ and\ \bibinfo {author} {\bibfnamefont {H.}~\bibnamefont
  {Johannesson}},\ }\bibfield  {title} {\bibinfo {title} {Loschmidt {{Echo
  Revivals}}: {{Critical}} and {{Noncritical}}},\ }\href
  {https://doi.org/10.1103/PhysRevLett.118.015701} {\bibfield  {journal}
  {\bibinfo  {journal} {Physical Review Letters}\ }\textbf {\bibinfo {volume}
  {118}},\ \bibinfo {pages} {015701} (\bibinfo {year}
  {2017}{\natexlab{a}})}\BibitemShut {NoStop}%
\bibitem [{\citenamefont {Sedlmayr}\ \emph
  {et~al.}(2018{\natexlab{a}})\citenamefont {Sedlmayr}, \citenamefont
  {J{\"a}ger}, \citenamefont {Maiti},\ and\ \citenamefont
  {Sirker}}]{Sedlmayr2018}%
  \BibitemOpen
  \bibfield  {author} {\bibinfo {author} {\bibfnamefont {N.}~\bibnamefont
  {Sedlmayr}}, \bibinfo {author} {\bibfnamefont {P.}~\bibnamefont {J{\"a}ger}},
  \bibinfo {author} {\bibfnamefont {M.}~\bibnamefont {Maiti}},\ and\ \bibinfo
  {author} {\bibfnamefont {J.}~\bibnamefont {Sirker}},\ }\bibfield  {title}
  {\bibinfo {title} {Bulk-boundary correspondence for dynamical phase
  transitions in one-dimensional topological insulators and superconductors},\
  }\href {https://doi.org/10.1103/PhysRevB.97.064304} {\bibfield  {journal}
  {\bibinfo  {journal} {Physical Review B}\ }\textbf {\bibinfo {volume} {97}},\
  \bibinfo {pages} {064304} (\bibinfo {year} {2018}{\natexlab{a}})}\BibitemShut
  {NoStop}%
\bibitem [{\citenamefont {Jafari}\ \emph {et~al.}(2018)\citenamefont {Jafari},
  \citenamefont {Johannesson}, \citenamefont {Langari},\ and\ \citenamefont
  {{Martin-Delgado}}}]{Jafari2018}%
  \BibitemOpen
  \bibfield  {author} {\bibinfo {author} {\bibfnamefont {R.}~\bibnamefont
  {Jafari}}, \bibinfo {author} {\bibfnamefont {H.}~\bibnamefont {Johannesson}},
  \bibinfo {author} {\bibfnamefont {A.}~\bibnamefont {Langari}},\ and\ \bibinfo
  {author} {\bibfnamefont {M.~A.}\ \bibnamefont {{Martin-Delgado}}},\
  }\bibfield  {title} {\bibinfo {title} {Quench dynamics and zero-energy modes:
  The case of the {{Creutz}} model},\ }\href
  {https://doi.org/10.1103/PhysRevB.99.054302} {\bibfield  {journal} {\bibinfo
  {journal} {Physical Review B}\ }\textbf {\bibinfo {volume} {99}},\ \bibinfo
  {pages} {054302} (\bibinfo {year} {2018})}\BibitemShut {NoStop}%
\bibitem [{\citenamefont {Zache}\ \emph {et~al.}(2019)\citenamefont {Zache},
  \citenamefont {Mueller}, \citenamefont {Schneider}, \citenamefont
  {Jendrzejewski}, \citenamefont {Berges},\ and\ \citenamefont
  {Hauke}}]{Zache2019}%
  \BibitemOpen
  \bibfield  {author} {\bibinfo {author} {\bibfnamefont {T.~V.}\ \bibnamefont
  {Zache}}, \bibinfo {author} {\bibfnamefont {N.}~\bibnamefont {Mueller}},
  \bibinfo {author} {\bibfnamefont {J.~T.}\ \bibnamefont {Schneider}}, \bibinfo
  {author} {\bibfnamefont {F.}~\bibnamefont {Jendrzejewski}}, \bibinfo {author}
  {\bibfnamefont {J.}~\bibnamefont {Berges}},\ and\ \bibinfo {author}
  {\bibfnamefont {P.}~\bibnamefont {Hauke}},\ }\bibfield  {title} {\bibinfo
  {title} {Dynamical {{Topological Transitions}} in the {{Massive Schwinger
  Model}} with a \$\textbackslash theta\$ {{Term}}},\ }\href
  {https://doi.org/10.1103/PhysRevLett.122.050403} {\bibfield  {journal}
  {\bibinfo  {journal} {Physical Review Letters}\ }\textbf {\bibinfo {volume}
  {122}},\ \bibinfo {pages} {050403} (\bibinfo {year} {2019})}\BibitemShut
  {NoStop}%
\bibitem [{\citenamefont {Mas{\l}owski}\ and\ \citenamefont
  {Sedlmayr}(2020)}]{Maslowski2020}%
  \BibitemOpen
  \bibfield  {author} {\bibinfo {author} {\bibfnamefont {T.}~\bibnamefont
  {Mas{\l}owski}}\ and\ \bibinfo {author} {\bibfnamefont {N.}~\bibnamefont
  {Sedlmayr}},\ }\bibfield  {title} {\bibinfo {title} {Quasiperiodic dynamical
  quantum phase transitions in multiband topological insulators and connections
  with entanglement entropy and fidelity susceptibility},\ }\href
  {https://doi.org/10.1103/PhysRevB.101.014301} {\bibfield  {journal} {\bibinfo
   {journal} {Physical Review B}\ }\textbf {\bibinfo {volume} {101}},\ \bibinfo
  {pages} {014301} (\bibinfo {year} {2020})}\BibitemShut {NoStop}%
\bibitem [{\citenamefont {Okugawa}\ \emph {et~al.}(2021)\citenamefont
  {Okugawa}, \citenamefont {Oshiyama},\ and\ \citenamefont
  {Ohzeki}}]{Okugawa2021}%
  \BibitemOpen
  \bibfield  {author} {\bibinfo {author} {\bibfnamefont {R.}~\bibnamefont
  {Okugawa}}, \bibinfo {author} {\bibfnamefont {H.}~\bibnamefont {Oshiyama}},\
  and\ \bibinfo {author} {\bibfnamefont {M.}~\bibnamefont {Ohzeki}},\
  }\bibfield  {title} {\bibinfo {title} {Mirror-symmetry-protected dynamical
  quantum phase transitions in topological crystalline insulators},\ }\href
  {https://doi.org/10.1103/PhysRevResearch.3.043064} {\bibfield  {journal}
  {\bibinfo  {journal} {Physical Review Research}\ }\textbf {\bibinfo {volume}
  {3}},\ \bibinfo {pages} {043064} (\bibinfo {year} {2021})}\BibitemShut
  {NoStop}%
\bibitem [{\citenamefont {Zhou}\ \emph {et~al.}(2018)\citenamefont {Zhou},
  \citenamefont {Wang}, \citenamefont {Wang},\ and\ \citenamefont
  {Gong}}]{Zhou2018}%
  \BibitemOpen
  \bibfield  {author} {\bibinfo {author} {\bibfnamefont {L.}~\bibnamefont
  {Zhou}}, \bibinfo {author} {\bibfnamefont {Q.-h.}\ \bibnamefont {Wang}},
  \bibinfo {author} {\bibfnamefont {H.}~\bibnamefont {Wang}},\ and\ \bibinfo
  {author} {\bibfnamefont {J.}~\bibnamefont {Gong}},\ }\bibfield  {title}
  {\bibinfo {title} {Dynamical quantum phase transitions in non-{{Hermitian}}
  lattices},\ }\href {https://doi.org/10.1103/PhysRevA.98.022129} {\bibfield
  {journal} {\bibinfo  {journal} {Physical Review A}\ }\textbf {\bibinfo
  {volume} {98}},\ \bibinfo {pages} {022129} (\bibinfo {year}
  {2018})}\BibitemShut {NoStop}%
\bibitem [{\citenamefont {Zhou}\ and\ \citenamefont
  {Du}(2021{\natexlab{a}})}]{Zhou2021}%
  \BibitemOpen
  \bibfield  {author} {\bibinfo {author} {\bibfnamefont {L.}~\bibnamefont
  {Zhou}}\ and\ \bibinfo {author} {\bibfnamefont {Q.}~\bibnamefont {Du}},\
  }\bibfield  {title} {\bibinfo {title} {Non-{{Hermitian}} topological phases
  and dynamical quantum phase transitions: A generic connection},\ }\href
  {https://doi.org/10.1088/1367-2630/ac0574} {\bibfield  {journal} {\bibinfo
  {journal} {New Journal of Physics}\ }\textbf {\bibinfo {volume} {23}},\
  \bibinfo {pages} {063041} (\bibinfo {year} {2021}{\natexlab{a}})}\BibitemShut
  {NoStop}%
\bibitem [{\citenamefont {Mondal}\ and\ \citenamefont
  {Nag}(2022)}]{Mondal2022}%
  \BibitemOpen
  \bibfield  {author} {\bibinfo {author} {\bibfnamefont {D.}~\bibnamefont
  {Mondal}}\ and\ \bibinfo {author} {\bibfnamefont {T.}~\bibnamefont {Nag}},\
  }\href@noop {} {\bibinfo {title} {Finite temperature dynamical quantum phase
  transition in a non-{{Hermitian}} system}} (\bibinfo {year} {2022}),\ \Eprint
  {https://arxiv.org/abs/2212.05839} {arXiv:2212.05839 [cond-mat.stat-mech]}
  \BibitemShut {NoStop}%
\bibitem [{\citenamefont {Sharma}\ \emph {et~al.}(2014)\citenamefont {Sharma},
  \citenamefont {Russomanno}, \citenamefont {Santoro},\ and\ \citenamefont
  {Dutta}}]{Sharma2014}%
  \BibitemOpen
  \bibfield  {author} {\bibinfo {author} {\bibfnamefont {S.}~\bibnamefont
  {Sharma}}, \bibinfo {author} {\bibfnamefont {A.}~\bibnamefont {Russomanno}},
  \bibinfo {author} {\bibfnamefont {G.~E.}\ \bibnamefont {Santoro}},\ and\
  \bibinfo {author} {\bibfnamefont {A.}~\bibnamefont {Dutta}},\ }\bibfield
  {title} {\bibinfo {title} {Loschmidt echo and dynamical fidelity in
  periodically driven quantum systems},\ }\href
  {https://doi.org/10.1209/0295-5075/106/67003} {\bibfield  {journal} {\bibinfo
   {journal} {Europhysics Letters}\ }\textbf {\bibinfo {volume} {106}},\
  \bibinfo {pages} {67003} (\bibinfo {year} {2014})}\BibitemShut {NoStop}%
\bibitem [{\citenamefont {Yang}\ \emph {et~al.}(2019)\citenamefont {Yang},
  \citenamefont {Zhou}, \citenamefont {Ma}, \citenamefont {Kong}, \citenamefont
  {Wang}, \citenamefont {Qin}, \citenamefont {Rong}, \citenamefont {Wang},
  \citenamefont {Shi}, \citenamefont {Gong},\ and\ \citenamefont
  {Du}}]{Yang2019}%
  \BibitemOpen
  \bibfield  {author} {\bibinfo {author} {\bibfnamefont {K.}~\bibnamefont
  {Yang}}, \bibinfo {author} {\bibfnamefont {L.}~\bibnamefont {Zhou}}, \bibinfo
  {author} {\bibfnamefont {W.}~\bibnamefont {Ma}}, \bibinfo {author}
  {\bibfnamefont {X.}~\bibnamefont {Kong}}, \bibinfo {author} {\bibfnamefont
  {P.}~\bibnamefont {Wang}}, \bibinfo {author} {\bibfnamefont {X.}~\bibnamefont
  {Qin}}, \bibinfo {author} {\bibfnamefont {X.}~\bibnamefont {Rong}}, \bibinfo
  {author} {\bibfnamefont {Y.}~\bibnamefont {Wang}}, \bibinfo {author}
  {\bibfnamefont {F.}~\bibnamefont {Shi}}, \bibinfo {author} {\bibfnamefont
  {J.}~\bibnamefont {Gong}},\ and\ \bibinfo {author} {\bibfnamefont
  {J.}~\bibnamefont {Du}},\ }\bibfield  {title} {\bibinfo {title} {Floquet
  dynamical quantum phase transitions},\ }\href
  {https://doi.org/10.1103/PhysRevB.100.085308} {\bibfield  {journal} {\bibinfo
   {journal} {Physical Review B}\ }\textbf {\bibinfo {volume} {100}},\ \bibinfo
  {pages} {085308} (\bibinfo {year} {2019})}\BibitemShut {NoStop}%
\bibitem [{\citenamefont {Zamani}\ \emph {et~al.}(2020)\citenamefont {Zamani},
  \citenamefont {Jafari},\ and\ \citenamefont {Langari}}]{Zamani2020}%
  \BibitemOpen
  \bibfield  {author} {\bibinfo {author} {\bibfnamefont {S.}~\bibnamefont
  {Zamani}}, \bibinfo {author} {\bibfnamefont {R.}~\bibnamefont {Jafari}},\
  and\ \bibinfo {author} {\bibfnamefont {A.}~\bibnamefont {Langari}},\
  }\bibfield  {title} {\bibinfo {title} {Floquet dynamical quantum phase
  transition in the extended {{XY}} model: {{Nonadiabatic}} to adiabatic
  topological transition},\ }\href
  {https://doi.org/10.1103/PhysRevB.102.144306} {\bibfield  {journal} {\bibinfo
   {journal} {Physical Review B}\ }\textbf {\bibinfo {volume} {102}},\ \bibinfo
  {pages} {144306} (\bibinfo {year} {2020})}\BibitemShut {NoStop}%
\bibitem [{\citenamefont {Zhou}\ and\ \citenamefont
  {Du}(2021{\natexlab{b}})}]{Zhou2021a}%
  \BibitemOpen
  \bibfield  {author} {\bibinfo {author} {\bibfnamefont {L.}~\bibnamefont
  {Zhou}}\ and\ \bibinfo {author} {\bibfnamefont {Q.}~\bibnamefont {Du}},\
  }\bibfield  {title} {\bibinfo {title} {Floquet dynamical quantum phase
  transitions in periodically quenched systems},\ }\href
  {https://doi.org/10.1088/1361-648X/ac0b60} {\bibfield  {journal} {\bibinfo
  {journal} {Journal of Physics: Condensed Matter}\ }\textbf {\bibinfo {volume}
  {33}},\ \bibinfo {pages} {345403} (\bibinfo {year}
  {2021}{\natexlab{b}})}\BibitemShut {NoStop}%
\bibitem [{\citenamefont {Jafari}\ and\ \citenamefont
  {Akbari}(2021)}]{Jafari2021}%
  \BibitemOpen
  \bibfield  {author} {\bibinfo {author} {\bibfnamefont {R.}~\bibnamefont
  {Jafari}}\ and\ \bibinfo {author} {\bibfnamefont {A.}~\bibnamefont
  {Akbari}},\ }\bibfield  {title} {\bibinfo {title} {Floquet dynamical phase
  transition and entanglement spectrum},\ }\href
  {https://doi.org/10.1103/PhysRevA.103.012204} {\bibfield  {journal} {\bibinfo
   {journal} {Physical Review A}\ }\textbf {\bibinfo {volume} {103}},\ \bibinfo
  {pages} {012204} (\bibinfo {year} {2021})}\BibitemShut {NoStop}%
\bibitem [{\citenamefont {Hamazaki}(2021)}]{Hamazaki2021}%
  \BibitemOpen
  \bibfield  {author} {\bibinfo {author} {\bibfnamefont {R.}~\bibnamefont
  {Hamazaki}},\ }\bibfield  {title} {\bibinfo {title} {Exceptional dynamical
  quantum phase transitions in periodically driven systems},\ }\href
  {https://doi.org/10.1038/s41467-021-25355-3} {\bibfield  {journal} {\bibinfo
  {journal} {Nature Communications}\ }\textbf {\bibinfo {volume} {12}},\
  \bibinfo {pages} {5108} (\bibinfo {year} {2021})}\BibitemShut {NoStop}%
\bibitem [{\citenamefont {Zamani}\ \emph {et~al.}(2022)\citenamefont {Zamani},
  \citenamefont {Jafari},\ and\ \citenamefont {Langari}}]{Zamani2022}%
  \BibitemOpen
  \bibfield  {author} {\bibinfo {author} {\bibfnamefont {S.}~\bibnamefont
  {Zamani}}, \bibinfo {author} {\bibfnamefont {R.}~\bibnamefont {Jafari}},\
  and\ \bibinfo {author} {\bibfnamefont {A.}~\bibnamefont {Langari}},\
  }\bibfield  {title} {\bibinfo {title} {Out-of-time-order correlations and
  {{Floquet}} dynamical quantum phase transition},\ }\href
  {https://doi.org/10.1103/PhysRevB.105.094304} {\bibfield  {journal} {\bibinfo
   {journal} {Physical Review B}\ }\textbf {\bibinfo {volume} {105}},\ \bibinfo
  {pages} {094304} (\bibinfo {year} {2022})}\BibitemShut {NoStop}%
\bibitem [{\citenamefont {Luan}\ \emph {et~al.}(2022)\citenamefont {Luan},
  \citenamefont {Zhang},\ and\ \citenamefont {Wang}}]{Luan2022}%
  \BibitemOpen
  \bibfield  {author} {\bibinfo {author} {\bibfnamefont {L.-N.}\ \bibnamefont
  {Luan}}, \bibinfo {author} {\bibfnamefont {M.-Y.}\ \bibnamefont {Zhang}},\
  and\ \bibinfo {author} {\bibfnamefont {L.}~\bibnamefont {Wang}},\ }\bibfield
  {title} {\bibinfo {title} {Floquet dynamical quantum phase transitions of the
  {{XY}} spin-chain under periodic quenching},\ }\href
  {https://doi.org/10.1016/j.physa.2022.127866} {\bibfield  {journal} {\bibinfo
   {journal} {Physica A: Statistical Mechanics and its Applications}\ }\textbf
  {\bibinfo {volume} {604}},\ \bibinfo {pages} {127866} (\bibinfo {year}
  {2022})}\BibitemShut {NoStop}%
\bibitem [{\citenamefont {Jafari}\ \emph {et~al.}(2022)\citenamefont {Jafari},
  \citenamefont {Akbari}, \citenamefont {Mishra},\ and\ \citenamefont
  {Johannesson}}]{Jafari2022}%
  \BibitemOpen
  \bibfield  {author} {\bibinfo {author} {\bibfnamefont {R.}~\bibnamefont
  {Jafari}}, \bibinfo {author} {\bibfnamefont {A.}~\bibnamefont {Akbari}},
  \bibinfo {author} {\bibfnamefont {U.}~\bibnamefont {Mishra}},\ and\ \bibinfo
  {author} {\bibfnamefont {H.}~\bibnamefont {Johannesson}},\ }\bibfield
  {title} {\bibinfo {title} {Floquet dynamical quantum phase transitions under
  synchronized periodic driving},\ }\href
  {https://doi.org/10.1103/PhysRevB.105.094311} {\bibfield  {journal} {\bibinfo
   {journal} {Physical Review B}\ }\textbf {\bibinfo {volume} {105}},\ \bibinfo
  {pages} {094311} (\bibinfo {year} {2022})}\BibitemShut {NoStop}%
\bibitem [{\citenamefont {Hasan}\ and\ \citenamefont {Kane}(2010)}]{Hasan2010}%
  \BibitemOpen
  \bibfield  {author} {\bibinfo {author} {\bibfnamefont {M.~Z.}\ \bibnamefont
  {Hasan}}\ and\ \bibinfo {author} {\bibfnamefont {C.~L.}\ \bibnamefont
  {Kane}},\ }\bibfield  {title} {\bibinfo {title} {Colloquium: {{Topological}}
  insulators},\ }\href {https://doi.org/10.1103/RevModPhys.82.3045} {\bibfield
  {journal} {\bibinfo  {journal} {Reviews of Modern Physics}\ }\textbf
  {\bibinfo {volume} {82}},\ \bibinfo {pages} {3045} (\bibinfo {year}
  {2010})}\BibitemShut {NoStop}%
\bibitem [{\citenamefont {Teo}\ and\ \citenamefont {Kane}(2010)}]{Teo2010}%
  \BibitemOpen
  \bibfield  {author} {\bibinfo {author} {\bibfnamefont {J.~C.~Y.}\
  \bibnamefont {Teo}}\ and\ \bibinfo {author} {\bibfnamefont {C.~L.}\
  \bibnamefont {Kane}},\ }\bibfield  {title} {\bibinfo {title} {Topological
  defects and gapless modes in insulators and superconductors},\ }\href
  {https://doi.org/10.1103/PhysRevB.82.115120} {\bibfield  {journal} {\bibinfo
  {journal} {Physical Review B}\ }\textbf {\bibinfo {volume} {82}},\ \bibinfo
  {pages} {115120} (\bibinfo {year} {2010})}\BibitemShut {NoStop}%
\bibitem [{\citenamefont {Budich}\ and\ \citenamefont
  {Heyl}(2016)}]{Budich2016}%
  \BibitemOpen
  \bibfield  {author} {\bibinfo {author} {\bibfnamefont {J.~C.}\ \bibnamefont
  {Budich}}\ and\ \bibinfo {author} {\bibfnamefont {M.}~\bibnamefont {Heyl}},\
  }\bibfield  {title} {\bibinfo {title} {Dynamical topological order parameters
  far from equilibrium},\ }\href {https://doi.org/10.1103/PhysRevB.93.085416}
  {\bibfield  {journal} {\bibinfo  {journal} {Physical Review B}\ }\textbf
  {\bibinfo {volume} {93}},\ \bibinfo {pages} {85416} (\bibinfo {year}
  {2016})}\BibitemShut {NoStop}%
\bibitem [{\citenamefont {Heyl}\ and\ \citenamefont {Budich}(2017)}]{Heyl2017}%
  \BibitemOpen
  \bibfield  {author} {\bibinfo {author} {\bibfnamefont {M.}~\bibnamefont
  {Heyl}}\ and\ \bibinfo {author} {\bibfnamefont {J.~C.}\ \bibnamefont
  {Budich}},\ }\bibfield  {title} {\bibinfo {title} {Dynamical topological
  quantum phase transitions for mixed states},\ }\href
  {https://doi.org/10.1103/PhysRevB.96.180304} {\bibfield  {journal} {\bibinfo
  {journal} {Physical Review B}\ }\textbf {\bibinfo {volume} {96}},\ \bibinfo
  {pages} {180304} (\bibinfo {year} {2017})}\BibitemShut {NoStop}%
\bibitem [{\citenamefont {Bhattacharya}\ \emph {et~al.}(2017)\citenamefont
  {Bhattacharya}, \citenamefont {Bandyopadhyay},\ and\ \citenamefont
  {Dutta}}]{Bhattacharya2017a}%
  \BibitemOpen
  \bibfield  {author} {\bibinfo {author} {\bibfnamefont {U.}~\bibnamefont
  {Bhattacharya}}, \bibinfo {author} {\bibfnamefont {S.}~\bibnamefont
  {Bandyopadhyay}},\ and\ \bibinfo {author} {\bibfnamefont {A.}~\bibnamefont
  {Dutta}},\ }\bibfield  {title} {\bibinfo {title} {Mixed state dynamical
  quantum phase transitions},\ }\href
  {https://doi.org/10.1103/PhysRevB.96.180303} {\bibfield  {journal} {\bibinfo
  {journal} {Physical Review B}\ }\textbf {\bibinfo {volume} {96}},\ \bibinfo
  {pages} {180303} (\bibinfo {year} {2017})}\BibitemShut {NoStop}%
\bibitem [{\citenamefont {Huang}\ and\ \citenamefont
  {Balatsky}(2016)}]{Huang2016}%
  \BibitemOpen
  \bibfield  {author} {\bibinfo {author} {\bibfnamefont {Z.}~\bibnamefont
  {Huang}}\ and\ \bibinfo {author} {\bibfnamefont {A.~V.}\ \bibnamefont
  {Balatsky}},\ }\bibfield  {title} {\bibinfo {title} {Dynamical {{Quantum
  Phase Transitions}}: {{Role}} of {{Topological Nodes}} in {{Wave Function
  Overlaps}}},\ }\href {https://doi.org/10.1103/PhysRevLett.117.086802}
  {\bibfield  {journal} {\bibinfo  {journal} {Physical Review Letters}\
  }\textbf {\bibinfo {volume} {117}},\ \bibinfo {pages} {086802} (\bibinfo
  {year} {2016})}\BibitemShut {NoStop}%
\bibitem [{\citenamefont {Jafari}(2019)}]{Jafari2019}%
  \BibitemOpen
  \bibfield  {author} {\bibinfo {author} {\bibfnamefont {R.}~\bibnamefont
  {Jafari}},\ }\bibfield  {title} {\bibinfo {title} {Dynamical {{Quantum Phase
  Transition}} and {{Quasi Particle Excitation}}},\ }\href
  {https://doi.org/10.1038/s41598-019-39595-3} {\bibfield  {journal} {\bibinfo
  {journal} {Scientific Reports}\ }\textbf {\bibinfo {volume} {9}},\ \bibinfo
  {pages} {2871} (\bibinfo {year} {2019})}\BibitemShut {NoStop}%
\bibitem [{\citenamefont {Mendl}\ and\ \citenamefont
  {Budich}(2019)}]{Mendl2019}%
  \BibitemOpen
  \bibfield  {author} {\bibinfo {author} {\bibfnamefont {C.~B.}\ \bibnamefont
  {Mendl}}\ and\ \bibinfo {author} {\bibfnamefont {J.~C.}\ \bibnamefont
  {Budich}},\ }\bibfield  {title} {\bibinfo {title} {Stability of dynamical
  quantum phase transitions in quenched topological insulators: {{From}}
  multiband to disordered systems},\ }\href
  {https://doi.org/10.1103/PhysRevB.100.224307} {\bibfield  {journal} {\bibinfo
   {journal} {Physical Review B}\ }\textbf {\bibinfo {volume} {100}},\ \bibinfo
  {pages} {224307} (\bibinfo {year} {2019})}\BibitemShut {NoStop}%
\bibitem [{\citenamefont {De~Nicola}\ \emph {et~al.}(2022)\citenamefont
  {De~Nicola}, \citenamefont {Michailidis},\ and\ \citenamefont
  {Serbyn}}]{DeNicola2022}%
  \BibitemOpen
  \bibfield  {author} {\bibinfo {author} {\bibfnamefont {S.}~\bibnamefont
  {De~Nicola}}, \bibinfo {author} {\bibfnamefont {A.~A.}\ \bibnamefont
  {Michailidis}},\ and\ \bibinfo {author} {\bibfnamefont {M.}~\bibnamefont
  {Serbyn}},\ }\bibfield  {title} {\bibinfo {title} {Entanglement and
  precession in two-dimensional dynamical quantum phase transitions},\ }\href
  {https://doi.org/10.1103/PhysRevB.105.165149} {\bibfield  {journal} {\bibinfo
   {journal} {Physical Review B}\ }\textbf {\bibinfo {volume} {105}},\ \bibinfo
  {pages} {165149} (\bibinfo {year} {2022})}\BibitemShut {NoStop}%
\bibitem [{\citenamefont {Hashizume}\ \emph {et~al.}(2022)\citenamefont
  {Hashizume}, \citenamefont {McCulloch},\ and\ \citenamefont
  {Halimeh}}]{Hashizume2022}%
  \BibitemOpen
  \bibfield  {author} {\bibinfo {author} {\bibfnamefont {T.}~\bibnamefont
  {Hashizume}}, \bibinfo {author} {\bibfnamefont {I.~P.}\ \bibnamefont
  {McCulloch}},\ and\ \bibinfo {author} {\bibfnamefont {J.~C.}\ \bibnamefont
  {Halimeh}},\ }\bibfield  {title} {\bibinfo {title} {Dynamical phase
  transitions in the two-dimensional transverse-field {{Ising}} model},\ }\href
  {https://doi.org/10.1103/PhysRevResearch.4.013250} {\bibfield  {journal}
  {\bibinfo  {journal} {Physical Review Research}\ }\textbf {\bibinfo {volume}
  {4}},\ \bibinfo {pages} {013250} (\bibinfo {year} {2022})}\BibitemShut
  {NoStop}%
\bibitem [{\citenamefont {Brange}\ \emph {et~al.}(2022)\citenamefont {Brange},
  \citenamefont {Peotta}, \citenamefont {Flindt},\ and\ \citenamefont
  {Ojanen}}]{Brange2022}%
  \BibitemOpen
  \bibfield  {author} {\bibinfo {author} {\bibfnamefont {F.}~\bibnamefont
  {Brange}}, \bibinfo {author} {\bibfnamefont {S.}~\bibnamefont {Peotta}},
  \bibinfo {author} {\bibfnamefont {C.}~\bibnamefont {Flindt}},\ and\ \bibinfo
  {author} {\bibfnamefont {T.}~\bibnamefont {Ojanen}},\ }\bibfield  {title}
  {\bibinfo {title} {Dynamical quantum phase transitions in strongly correlated
  two-dimensional spin lattices following a quench},\ }\href
  {https://doi.org/10.1103/PhysRevResearch.4.033032} {\bibfield  {journal}
  {\bibinfo  {journal} {Physical Review Research}\ }\textbf {\bibinfo {volume}
  {4}},\ \bibinfo {pages} {033032} (\bibinfo {year} {2022})}\BibitemShut
  {NoStop}%
\bibitem [{\citenamefont {Jurcevic}\ \emph {et~al.}(2017)\citenamefont
  {Jurcevic}, \citenamefont {Shen}, \citenamefont {Hauke}, \citenamefont
  {Maier}, \citenamefont {Brydges}, \citenamefont {Hempel}, \citenamefont
  {Lanyon}, \citenamefont {Heyl}, \citenamefont {Blatt},\ and\ \citenamefont
  {Roos}}]{Jurcevic2017}%
  \BibitemOpen
  \bibfield  {author} {\bibinfo {author} {\bibfnamefont {P.}~\bibnamefont
  {Jurcevic}}, \bibinfo {author} {\bibfnamefont {H.}~\bibnamefont {Shen}},
  \bibinfo {author} {\bibfnamefont {P.}~\bibnamefont {Hauke}}, \bibinfo
  {author} {\bibfnamefont {C.}~\bibnamefont {Maier}}, \bibinfo {author}
  {\bibfnamefont {T.}~\bibnamefont {Brydges}}, \bibinfo {author} {\bibfnamefont
  {C.}~\bibnamefont {Hempel}}, \bibinfo {author} {\bibfnamefont {B.~P.}\
  \bibnamefont {Lanyon}}, \bibinfo {author} {\bibfnamefont {M.}~\bibnamefont
  {Heyl}}, \bibinfo {author} {\bibfnamefont {R.}~\bibnamefont {Blatt}},\ and\
  \bibinfo {author} {\bibfnamefont {C.~F.}\ \bibnamefont {Roos}},\ }\bibfield
  {title} {\bibinfo {title} {Direct {{Observation}} of {{Dynamical Quantum
  Phase Transitions}} in an {{Interacting Many-Body System}}},\ }\href
  {https://doi.org/10.1103/PhysRevLett.119.080501} {\bibfield  {journal}
  {\bibinfo  {journal} {Physical Review Letters}\ }\textbf {\bibinfo {volume}
  {119}},\ \bibinfo {pages} {080501} (\bibinfo {year} {2017})}\BibitemShut
  {NoStop}%
\bibitem [{\citenamefont {Fl{\"a}schner}\ \emph {et~al.}(2018)\citenamefont
  {Fl{\"a}schner}, \citenamefont {Vogel}, \citenamefont {Tarnowski},
  \citenamefont {Rem}, \citenamefont {L{\"u}hmann}, \citenamefont {Heyl},
  \citenamefont {Budich}, \citenamefont {Mathey}, \citenamefont {Sengstock},\
  and\ \citenamefont {Weitenberg}}]{Flaschner2018}%
  \BibitemOpen
  \bibfield  {author} {\bibinfo {author} {\bibfnamefont {N.}~\bibnamefont
  {Fl{\"a}schner}}, \bibinfo {author} {\bibfnamefont {D.}~\bibnamefont
  {Vogel}}, \bibinfo {author} {\bibfnamefont {M.}~\bibnamefont {Tarnowski}},
  \bibinfo {author} {\bibfnamefont {B.~S.}\ \bibnamefont {Rem}}, \bibinfo
  {author} {\bibfnamefont {D.~S.}\ \bibnamefont {L{\"u}hmann}}, \bibinfo
  {author} {\bibfnamefont {M.}~\bibnamefont {Heyl}}, \bibinfo {author}
  {\bibfnamefont {J.~C.}\ \bibnamefont {Budich}}, \bibinfo {author}
  {\bibfnamefont {L.}~\bibnamefont {Mathey}}, \bibinfo {author} {\bibfnamefont
  {K.}~\bibnamefont {Sengstock}},\ and\ \bibinfo {author} {\bibfnamefont
  {C.}~\bibnamefont {Weitenberg}},\ }\bibfield  {title} {\bibinfo {title}
  {Observation of dynamical vortices after quenches in a system with
  topology},\ }\href {https://doi.org/10.1038/s41567-017-0013-8} {\bibfield
  {journal} {\bibinfo  {journal} {Nature Physics}\ }\textbf {\bibinfo {volume}
  {14}},\ \bibinfo {pages} {265} (\bibinfo {year} {2018})}\BibitemShut
  {NoStop}%
\bibitem [{\citenamefont {Zhang}\ \emph {et~al.}(2017)\citenamefont {Zhang},
  \citenamefont {Pagano}, \citenamefont {Hess}, \citenamefont {Kyprianidis},
  \citenamefont {Becker}, \citenamefont {Kaplan}, \citenamefont {Gorshkov},
  \citenamefont {Gong},\ and\ \citenamefont {Monroe}}]{Zhang2017b}%
  \BibitemOpen
  \bibfield  {author} {\bibinfo {author} {\bibfnamefont {J.}~\bibnamefont
  {Zhang}}, \bibinfo {author} {\bibfnamefont {G.}~\bibnamefont {Pagano}},
  \bibinfo {author} {\bibfnamefont {P.~W.}\ \bibnamefont {Hess}}, \bibinfo
  {author} {\bibfnamefont {A.}~\bibnamefont {Kyprianidis}}, \bibinfo {author}
  {\bibfnamefont {P.}~\bibnamefont {Becker}}, \bibinfo {author} {\bibfnamefont
  {H.}~\bibnamefont {Kaplan}}, \bibinfo {author} {\bibfnamefont {A.~V.}\
  \bibnamefont {Gorshkov}}, \bibinfo {author} {\bibfnamefont {Z.-X.}\
  \bibnamefont {Gong}},\ and\ \bibinfo {author} {\bibfnamefont
  {C.}~\bibnamefont {Monroe}},\ }\bibfield  {title} {\bibinfo {title}
  {Observation of a many-body dynamical phase transition with a 53-qubit
  quantum simulator},\ }\href {https://doi.org/10.1038/nature24654} {\bibfield
  {journal} {\bibinfo  {journal} {Nature}\ }\textbf {\bibinfo {volume} {551}},\
  \bibinfo {pages} {601} (\bibinfo {year} {2017})}\BibitemShut {NoStop}%
\bibitem [{\citenamefont {Guo}\ \emph {et~al.}(2019)\citenamefont {Guo},
  \citenamefont {Yang}, \citenamefont {Zeng}, \citenamefont {Peng},
  \citenamefont {Li}, \citenamefont {Deng}, \citenamefont {Jin}, \citenamefont
  {Chen}, \citenamefont {Zheng},\ and\ \citenamefont {Fan}}]{Guo2019}%
  \BibitemOpen
  \bibfield  {author} {\bibinfo {author} {\bibfnamefont {X.-Y.}\ \bibnamefont
  {Guo}}, \bibinfo {author} {\bibfnamefont {C.}~\bibnamefont {Yang}}, \bibinfo
  {author} {\bibfnamefont {Y.}~\bibnamefont {Zeng}}, \bibinfo {author}
  {\bibfnamefont {Y.}~\bibnamefont {Peng}}, \bibinfo {author} {\bibfnamefont
  {H.-K.}\ \bibnamefont {Li}}, \bibinfo {author} {\bibfnamefont
  {H.}~\bibnamefont {Deng}}, \bibinfo {author} {\bibfnamefont {Y.-R.}\
  \bibnamefont {Jin}}, \bibinfo {author} {\bibfnamefont {S.}~\bibnamefont
  {Chen}}, \bibinfo {author} {\bibfnamefont {D.}~\bibnamefont {Zheng}},\ and\
  \bibinfo {author} {\bibfnamefont {H.}~\bibnamefont {Fan}},\ }\bibfield
  {title} {\bibinfo {title} {Observation of a {{Dynamical Quantum Phase
  Transition}} by a {{Superconducting Qubit Simulation}}},\ }\href
  {https://doi.org/10.1103/PhysRevApplied.11.044080} {\bibfield  {journal}
  {\bibinfo  {journal} {Physical Review Applied}\ }\textbf {\bibinfo {volume}
  {11}},\ \bibinfo {pages} {044080} (\bibinfo {year} {2019})}\BibitemShut
  {NoStop}%
\bibitem [{\citenamefont {Smale}\ \emph {et~al.}(2019)\citenamefont {Smale},
  \citenamefont {He}, \citenamefont {Olsen}, \citenamefont {Jackson},
  \citenamefont {Sharum}, \citenamefont {Trotzky}, \citenamefont {Marino},
  \citenamefont {Rey},\ and\ \citenamefont {Thywissen}}]{Smale2019}%
  \BibitemOpen
  \bibfield  {author} {\bibinfo {author} {\bibfnamefont {S.}~\bibnamefont
  {Smale}}, \bibinfo {author} {\bibfnamefont {P.}~\bibnamefont {He}}, \bibinfo
  {author} {\bibfnamefont {B.~A.}\ \bibnamefont {Olsen}}, \bibinfo {author}
  {\bibfnamefont {K.~G.}\ \bibnamefont {Jackson}}, \bibinfo {author}
  {\bibfnamefont {H.}~\bibnamefont {Sharum}}, \bibinfo {author} {\bibfnamefont
  {S.}~\bibnamefont {Trotzky}}, \bibinfo {author} {\bibfnamefont
  {J.}~\bibnamefont {Marino}}, \bibinfo {author} {\bibfnamefont {A.~M.}\
  \bibnamefont {Rey}},\ and\ \bibinfo {author} {\bibfnamefont {J.~H.}\
  \bibnamefont {Thywissen}},\ }\bibfield  {title} {\bibinfo {title}
  {Observation of a transition between dynamical phases in a quantum degenerate
  {{Fermi}} gas},\ }\href {https://doi.org/10.1126/sciadv.aax1568} {\bibfield
  {journal} {\bibinfo  {journal} {Science Advances}\ }\textbf {\bibinfo
  {volume} {5}},\ \bibinfo {pages} {eaax1568} (\bibinfo {year}
  {2019})}\BibitemShut {NoStop}%
\bibitem [{\citenamefont {Nie}\ \emph {et~al.}(2020)\citenamefont {Nie},
  \citenamefont {Wei}, \citenamefont {Chen}, \citenamefont {Zhang},
  \citenamefont {Zhao}, \citenamefont {Qiu}, \citenamefont {Tian},
  \citenamefont {Ji}, \citenamefont {Xin}, \citenamefont {Lu},\ and\
  \citenamefont {Li}}]{Nie2020}%
  \BibitemOpen
  \bibfield  {author} {\bibinfo {author} {\bibfnamefont {X.}~\bibnamefont
  {Nie}}, \bibinfo {author} {\bibfnamefont {B.-B.}\ \bibnamefont {Wei}},
  \bibinfo {author} {\bibfnamefont {X.}~\bibnamefont {Chen}}, \bibinfo {author}
  {\bibfnamefont {Z.}~\bibnamefont {Zhang}}, \bibinfo {author} {\bibfnamefont
  {X.}~\bibnamefont {Zhao}}, \bibinfo {author} {\bibfnamefont {C.}~\bibnamefont
  {Qiu}}, \bibinfo {author} {\bibfnamefont {Y.}~\bibnamefont {Tian}}, \bibinfo
  {author} {\bibfnamefont {Y.}~\bibnamefont {Ji}}, \bibinfo {author}
  {\bibfnamefont {T.}~\bibnamefont {Xin}}, \bibinfo {author} {\bibfnamefont
  {D.}~\bibnamefont {Lu}},\ and\ \bibinfo {author} {\bibfnamefont
  {J.}~\bibnamefont {Li}},\ }\bibfield  {title} {\bibinfo {title} {Experimental
  {{Observation}} of {{Equilibrium}} and {{Dynamical Quantum Phase
  Transitions}} via {{Out-of-Time-Ordered Correlators}}},\ }\href
  {https://doi.org/10.1103/PhysRevLett.124.250601} {\bibfield  {journal}
  {\bibinfo  {journal} {Physical Review Letters}\ }\textbf {\bibinfo {volume}
  {124}},\ \bibinfo {pages} {250601} (\bibinfo {year} {2020})}\BibitemShut
  {NoStop}%
\bibitem [{\citenamefont {Tian}\ \emph {et~al.}(2020)\citenamefont {Tian},
  \citenamefont {Yang}, \citenamefont {Qiu}, \citenamefont {Liang},
  \citenamefont {Yang}, \citenamefont {Xu},\ and\ \citenamefont
  {Duan}}]{Tian2020}%
  \BibitemOpen
  \bibfield  {author} {\bibinfo {author} {\bibfnamefont {T.}~\bibnamefont
  {Tian}}, \bibinfo {author} {\bibfnamefont {H.-X.}\ \bibnamefont {Yang}},
  \bibinfo {author} {\bibfnamefont {L.-Y.}\ \bibnamefont {Qiu}}, \bibinfo
  {author} {\bibfnamefont {H.-Y.}\ \bibnamefont {Liang}}, \bibinfo {author}
  {\bibfnamefont {Y.-B.}\ \bibnamefont {Yang}}, \bibinfo {author}
  {\bibfnamefont {Y.}~\bibnamefont {Xu}},\ and\ \bibinfo {author}
  {\bibfnamefont {L.-M.}\ \bibnamefont {Duan}},\ }\bibfield  {title} {\bibinfo
  {title} {Observation of {{Dynamical Quantum Phase Transitions}} with
  {{Correspondence}} in an {{Excited State Phase Diagram}}},\ }\href
  {https://doi.org/10.1103/PhysRevLett.124.043001} {\bibfield  {journal}
  {\bibinfo  {journal} {Physical Review Letters}\ }\textbf {\bibinfo {volume}
  {124}},\ \bibinfo {pages} {043001} (\bibinfo {year} {2020})}\BibitemShut
  {NoStop}%
\bibitem [{\citenamefont {Karrasch}\ and\ \citenamefont
  {Schuricht}(2013)}]{Karrasch2013}%
  \BibitemOpen
  \bibfield  {author} {\bibinfo {author} {\bibfnamefont {C.}~\bibnamefont
  {Karrasch}}\ and\ \bibinfo {author} {\bibfnamefont {D.}~\bibnamefont
  {Schuricht}},\ }\bibfield  {title} {\bibinfo {title} {Dynamical phase
  transitions after quenches in nonintegrable models},\ }\href
  {https://doi.org/10.1103/PhysRevB.87.195104} {\bibfield  {journal} {\bibinfo
  {journal} {Physical Review B}\ }\textbf {\bibinfo {volume} {87}},\ \bibinfo
  {pages} {195104} (\bibinfo {year} {2013})}\BibitemShut {NoStop}%
\bibitem [{\citenamefont {Heyl}(2014)}]{Heyl2014}%
  \BibitemOpen
  \bibfield  {author} {\bibinfo {author} {\bibfnamefont {M.}~\bibnamefont
  {Heyl}},\ }\bibfield  {title} {\bibinfo {title} {Dynamical quantum phase
  transitions in systems with broken-symmetry phases},\ }\href
  {https://doi.org/10.1103/PhysRevLett.113.205701} {\bibfield  {journal}
  {\bibinfo  {journal} {Physical Review Letters}\ }\textbf {\bibinfo {volume}
  {113}},\ \bibinfo {pages} {205701} (\bibinfo {year} {2014})}\BibitemShut
  {NoStop}%
\bibitem [{\citenamefont {Vajna}\ and\ \citenamefont
  {D{\'o}ra}(2014)}]{Vajna2014}%
  \BibitemOpen
  \bibfield  {author} {\bibinfo {author} {\bibfnamefont {S.}~\bibnamefont
  {Vajna}}\ and\ \bibinfo {author} {\bibfnamefont {B.}~\bibnamefont
  {D{\'o}ra}},\ }\bibfield  {title} {\bibinfo {title} {Disentangling dynamical
  phase transitions from equilibrium phase transitions},\ }\href
  {https://doi.org/10.1103/PhysRevB.89.161105} {\bibfield  {journal} {\bibinfo
  {journal} {Physical Review B}\ }\textbf {\bibinfo {volume} {89}},\ \bibinfo
  {pages} {161105} (\bibinfo {year} {2014})}\BibitemShut {NoStop}%
\bibitem [{\citenamefont {Andraschko}\ and\ \citenamefont
  {Sirker}(2014)}]{Andraschko2014}%
  \BibitemOpen
  \bibfield  {author} {\bibinfo {author} {\bibfnamefont {F.}~\bibnamefont
  {Andraschko}}\ and\ \bibinfo {author} {\bibfnamefont {J.}~\bibnamefont
  {Sirker}},\ }\bibfield  {title} {\bibinfo {title} {Dynamical quantum phase
  transitions and the {{Loschmidt}} echo: {{A}} transfer matrix approach},\
  }\href {https://doi.org/10.1103/PhysRevB.89.125120} {\bibfield  {journal}
  {\bibinfo  {journal} {Physical Review B}\ }\textbf {\bibinfo {volume} {89}},\
  \bibinfo {pages} {125120} (\bibinfo {year} {2014})}\BibitemShut {NoStop}%
\bibitem [{\citenamefont {Karrasch}\ and\ \citenamefont
  {Schuricht}(2017)}]{Karrasch2017}%
  \BibitemOpen
  \bibfield  {author} {\bibinfo {author} {\bibfnamefont {C.}~\bibnamefont
  {Karrasch}}\ and\ \bibinfo {author} {\bibfnamefont {D.}~\bibnamefont
  {Schuricht}},\ }\bibfield  {title} {\bibinfo {title} {Dynamical quantum phase
  transitions in the quantum {{Potts}} chain},\ }\href
  {https://doi.org/10.1103/PhysRevB.95.075143} {\bibfield  {journal} {\bibinfo
  {journal} {Physical Review B}\ }\textbf {\bibinfo {volume} {95}},\ \bibinfo
  {pages} {075143} (\bibinfo {year} {2017})}\BibitemShut {NoStop}%
\bibitem [{\citenamefont {Jafari}\ and\ \citenamefont
  {Johannesson}(2017{\natexlab{b}})}]{Jafari2017}%
  \BibitemOpen
  \bibfield  {author} {\bibinfo {author} {\bibfnamefont {R.}~\bibnamefont
  {Jafari}}\ and\ \bibinfo {author} {\bibfnamefont {H.}~\bibnamefont
  {Johannesson}},\ }\bibfield  {title} {\bibinfo {title} {Decoherence from spin
  environments: {{Loschmidt}} echo and quasiparticle excitations},\ }\href
  {https://doi.org/10.1103/PhysRevB.96.224302} {\bibfield  {journal} {\bibinfo
  {journal} {Physical Review B}\ }\textbf {\bibinfo {volume} {96}},\ \bibinfo
  {pages} {224302} (\bibinfo {year} {2017}{\natexlab{b}})}\BibitemShut
  {NoStop}%
\bibitem [{\citenamefont {Cheraghi}\ and\ \citenamefont
  {Mahdavifar}(2018)}]{Cheraghi2018}%
  \BibitemOpen
  \bibfield  {author} {\bibinfo {author} {\bibfnamefont {H.}~\bibnamefont
  {Cheraghi}}\ and\ \bibinfo {author} {\bibfnamefont {S.}~\bibnamefont
  {Mahdavifar}},\ }\bibfield  {title} {\bibinfo {title} {Ineffectiveness of the
  {{Dzyaloshinskii}}\textendash{{Moriya}} interaction in the dynamical quantum
  phase transition in the {{ITF}} model},\ }\href
  {https://doi.org/10.1088/1361-648X/aae1c5} {\bibfield  {journal} {\bibinfo
  {journal} {Journal of Physics: Condensed Matter}\ }\textbf {\bibinfo {volume}
  {30}},\ \bibinfo {pages} {42LT01} (\bibinfo {year} {2018})}\BibitemShut
  {NoStop}%
\bibitem [{\citenamefont {Wrze{\'s}niewski}\ \emph {et~al.}(2022)\citenamefont
  {Wrze{\'s}niewski}, \citenamefont {Weymann}, \citenamefont {Sedlmayr},\ and\
  \citenamefont {Doma{\'n}ski}}]{Wrzesniewski2022}%
  \BibitemOpen
  \bibfield  {author} {\bibinfo {author} {\bibfnamefont {K.}~\bibnamefont
  {Wrze{\'s}niewski}}, \bibinfo {author} {\bibfnamefont {I.}~\bibnamefont
  {Weymann}}, \bibinfo {author} {\bibfnamefont {N.}~\bibnamefont {Sedlmayr}},\
  and\ \bibinfo {author} {\bibfnamefont {T.}~\bibnamefont {Doma{\'n}ski}},\
  }\bibfield  {title} {\bibinfo {title} {Dynamical quantum phase transitions in
  a mesoscopic superconducting system},\ }\href
  {https://doi.org/10.1103/PhysRevB.105.094514} {\bibfield  {journal} {\bibinfo
   {journal} {Physical Review B}\ }\textbf {\bibinfo {volume} {105}},\ \bibinfo
  {pages} {094514} (\bibinfo {year} {2022})}\BibitemShut {NoStop}%
\bibitem [{\citenamefont {Puskarov}\ and\ \citenamefont
  {Schuricht}(2016)}]{Puskarov2016}%
  \BibitemOpen
  \bibfield  {author} {\bibinfo {author} {\bibfnamefont {T.}~\bibnamefont
  {Puskarov}}\ and\ \bibinfo {author} {\bibfnamefont {D.}~\bibnamefont
  {Schuricht}},\ }\bibfield  {title} {\bibinfo {title} {Time evolution during
  and after finite-time quantum quenches in the transverse-field {{Ising}}
  chain},\ }\href {https://doi.org/10.21468/SciPostPhys.1.1.003} {\bibfield
  {journal} {\bibinfo  {journal} {SciPost Physics}\ }\textbf {\bibinfo {volume}
  {1}},\ \bibinfo {pages} {003} (\bibinfo {year} {2016})}\BibitemShut {NoStop}%
\bibitem [{\citenamefont {Sharma}\ \emph {et~al.}(2016)\citenamefont {Sharma},
  \citenamefont {Divakaran}, \citenamefont {Polkovnikov},\ and\ \citenamefont
  {Dutta}}]{Sharma2016}%
  \BibitemOpen
  \bibfield  {author} {\bibinfo {author} {\bibfnamefont {S.}~\bibnamefont
  {Sharma}}, \bibinfo {author} {\bibfnamefont {U.}~\bibnamefont {Divakaran}},
  \bibinfo {author} {\bibfnamefont {A.}~\bibnamefont {Polkovnikov}},\ and\
  \bibinfo {author} {\bibfnamefont {A.}~\bibnamefont {Dutta}},\ }\bibfield
  {title} {\bibinfo {title} {Slow quenches in a quantum {{Ising}} chain:
  {{Dynamical}} phase transitions and topology},\ }\href
  {https://doi.org/10.1103/PhysRevB.93.144306} {\bibfield  {journal} {\bibinfo
  {journal} {Physical Review B}\ }\textbf {\bibinfo {volume} {93}},\ \bibinfo
  {pages} {144306} (\bibinfo {year} {2016})}\BibitemShut {NoStop}%
\bibitem [{\citenamefont {Bhattacharya}\ and\ \citenamefont
  {Dutta}(2017)}]{Bhattacharya2017}%
  \BibitemOpen
  \bibfield  {author} {\bibinfo {author} {\bibfnamefont {U.}~\bibnamefont
  {Bhattacharya}}\ and\ \bibinfo {author} {\bibfnamefont {A.}~\bibnamefont
  {Dutta}},\ }\bibfield  {title} {\bibinfo {title} {Interconnections between
  equilibrium topology and dynamical quantum phase transitions in a linearly
  ramped {{Haldane}} model},\ }\href
  {https://doi.org/10.1103/PhysRevB.95.184307} {\bibfield  {journal} {\bibinfo
  {journal} {Physical Review B}\ }\textbf {\bibinfo {volume} {95}},\ \bibinfo
  {pages} {184307} (\bibinfo {year} {2017})}\BibitemShut {NoStop}%
\bibitem [{\citenamefont {Kennes}\ \emph {et~al.}(2018)\citenamefont {Kennes},
  \citenamefont {Schuricht},\ and\ \citenamefont {Karrasch}}]{Kennes2018}%
  \BibitemOpen
  \bibfield  {author} {\bibinfo {author} {\bibfnamefont {D.~M.}\ \bibnamefont
  {Kennes}}, \bibinfo {author} {\bibfnamefont {D.}~\bibnamefont {Schuricht}},\
  and\ \bibinfo {author} {\bibfnamefont {C.}~\bibnamefont {Karrasch}},\
  }\bibfield  {title} {\bibinfo {title} {Controlling dynamical quantum phase
  transitions},\ }\href {https://doi.org/10.1103/PhysRevB.97.184302} {\bibfield
   {journal} {\bibinfo  {journal} {Physical Review B}\ }\textbf {\bibinfo
  {volume} {97}},\ \bibinfo {pages} {184302} (\bibinfo {year}
  {2018})}\BibitemShut {NoStop}%
\bibitem [{\citenamefont {Hou}\ \emph {et~al.}(2022)\citenamefont {Hou},
  \citenamefont {Gao}, \citenamefont {Guo},\ and\ \citenamefont
  {Chien}}]{Hou2022}%
  \BibitemOpen
  \bibfield  {author} {\bibinfo {author} {\bibfnamefont {X.-Y.}\ \bibnamefont
  {Hou}}, \bibinfo {author} {\bibfnamefont {Q.-C.}\ \bibnamefont {Gao}},
  \bibinfo {author} {\bibfnamefont {H.}~\bibnamefont {Guo}},\ and\ \bibinfo
  {author} {\bibfnamefont {C.-C.}\ \bibnamefont {Chien}},\ }\bibfield  {title}
  {\bibinfo {title} {Metamorphic dynamical quantum phase transition in
  double-quench processes at finite temperatures},\ }\href
  {https://doi.org/10.1103/PhysRevB.106.014301} {\bibfield  {journal} {\bibinfo
   {journal} {Physical Review B}\ }\textbf {\bibinfo {volume} {106}},\ \bibinfo
  {pages} {014301} (\bibinfo {year} {2022})}\BibitemShut {NoStop}%
\bibitem [{\citenamefont {Mera}\ \emph {et~al.}(2017)\citenamefont {Mera},
  \citenamefont {Vlachou}, \citenamefont {Paunkovi{\'c}}, \citenamefont
  {Vieira},\ and\ \citenamefont {Viyuela}}]{Mera2017}%
  \BibitemOpen
  \bibfield  {author} {\bibinfo {author} {\bibfnamefont {B.}~\bibnamefont
  {Mera}}, \bibinfo {author} {\bibfnamefont {C.}~\bibnamefont {Vlachou}},
  \bibinfo {author} {\bibfnamefont {N.}~\bibnamefont {Paunkovi{\'c}}}, \bibinfo
  {author} {\bibfnamefont {V.~R.}\ \bibnamefont {Vieira}},\ and\ \bibinfo
  {author} {\bibfnamefont {O.}~\bibnamefont {Viyuela}},\ }\bibfield  {title}
  {\bibinfo {title} {Dynamical phase transitions at finite temperature from
  fidelity and interferometric {{Loschmidt}} echo induced metrics},\ }\href
  {https://doi.org/10.1103/PhysRevB.97.094110} {\bibfield  {journal} {\bibinfo
  {journal} {Physical Review B}\ }\textbf {\bibinfo {volume} {97}},\ \bibinfo
  {pages} {094110} (\bibinfo {year} {2017})}\BibitemShut {NoStop}%
\bibitem [{\citenamefont {Sedlmayr}\ \emph
  {et~al.}(2018{\natexlab{b}})\citenamefont {Sedlmayr}, \citenamefont
  {Fleischhauer},\ and\ \citenamefont {Sirker}}]{Sedlmayr2018b}%
  \BibitemOpen
  \bibfield  {author} {\bibinfo {author} {\bibfnamefont {N.}~\bibnamefont
  {Sedlmayr}}, \bibinfo {author} {\bibfnamefont {M.}~\bibnamefont
  {Fleischhauer}},\ and\ \bibinfo {author} {\bibfnamefont {J.}~\bibnamefont
  {Sirker}},\ }\bibfield  {title} {\bibinfo {title} {The fate of dynamical
  phase transitions at finite temperatures and in open systems},\ }\href
  {https://doi.org/10.1103/PhysRevB.97.045147} {\bibfield  {journal} {\bibinfo
  {journal} {Physical Review B}\ }\textbf {\bibinfo {volume} {97}},\ \bibinfo
  {pages} {045147} (\bibinfo {year} {2018}{\natexlab{b}})}\BibitemShut
  {NoStop}%
\bibitem [{\citenamefont {Abeling}\ and\ \citenamefont
  {Kehrein}(2016)}]{Abeling2016}%
  \BibitemOpen
  \bibfield  {author} {\bibinfo {author} {\bibfnamefont {N.~O.}\ \bibnamefont
  {Abeling}}\ and\ \bibinfo {author} {\bibfnamefont {S.}~\bibnamefont
  {Kehrein}},\ }\bibfield  {title} {\bibinfo {title} {Quantum quench dynamics
  in the transverse field {{Ising}} model at nonzero temperatures},\ }\href
  {https://doi.org/10.1103/PhysRevB.93.104302} {\bibfield  {journal} {\bibinfo
  {journal} {Physical Review B}\ }\textbf {\bibinfo {volume} {93}},\ \bibinfo
  {pages} {104302} (\bibinfo {year} {2016})}\BibitemShut {NoStop}%
\bibitem [{\citenamefont {Lang}\ \emph
  {et~al.}(2018{\natexlab{a}})\citenamefont {Lang}, \citenamefont {Frank},\
  and\ \citenamefont {Halimeh}}]{Lang2018}%
  \BibitemOpen
  \bibfield  {author} {\bibinfo {author} {\bibfnamefont {J.}~\bibnamefont
  {Lang}}, \bibinfo {author} {\bibfnamefont {B.}~\bibnamefont {Frank}},\ and\
  \bibinfo {author} {\bibfnamefont {J.~C.}\ \bibnamefont {Halimeh}},\
  }\bibfield  {title} {\bibinfo {title} {Dynamical {{Quantum Phase
  Transitions}}: {{A Geometric Picture}}},\ }\href
  {https://doi.org/10.1103/PhysRevLett.121.130603} {\bibfield  {journal}
  {\bibinfo  {journal} {Physical Review Letters}\ }\textbf {\bibinfo {volume}
  {121}},\ \bibinfo {pages} {130603} (\bibinfo {year}
  {2018}{\natexlab{a}})}\BibitemShut {NoStop}%
\bibitem [{\citenamefont {Lang}\ \emph
  {et~al.}(2018{\natexlab{b}})\citenamefont {Lang}, \citenamefont {Frank},\
  and\ \citenamefont {Halimeh}}]{Lang2018a}%
  \BibitemOpen
  \bibfield  {author} {\bibinfo {author} {\bibfnamefont {J.}~\bibnamefont
  {Lang}}, \bibinfo {author} {\bibfnamefont {B.}~\bibnamefont {Frank}},\ and\
  \bibinfo {author} {\bibfnamefont {J.~C.}\ \bibnamefont {Halimeh}},\
  }\bibfield  {title} {\bibinfo {title} {Concurrence of dynamical phase
  transitions at finite temperature in the fully connected transverse-field
  {{Ising}} model},\ }\href {https://doi.org/10.1103/PhysRevB.97.174401}
  {\bibfield  {journal} {\bibinfo  {journal} {Physical Review B}\ }\textbf
  {\bibinfo {volume} {97}},\ \bibinfo {pages} {174401} (\bibinfo {year}
  {2018}{\natexlab{b}})}\BibitemShut {NoStop}%
\bibitem [{\citenamefont {Kyaw}\ \emph {et~al.}(2020)\citenamefont {Kyaw},
  \citenamefont {Bastidas}, \citenamefont {Tangpanitanon}, \citenamefont
  {Romero},\ and\ \citenamefont {Kwek}}]{Kyaw2020}%
  \BibitemOpen
  \bibfield  {author} {\bibinfo {author} {\bibfnamefont {T.~H.}\ \bibnamefont
  {Kyaw}}, \bibinfo {author} {\bibfnamefont {V.~M.}\ \bibnamefont {Bastidas}},
  \bibinfo {author} {\bibfnamefont {J.}~\bibnamefont {Tangpanitanon}}, \bibinfo
  {author} {\bibfnamefont {G.}~\bibnamefont {Romero}},\ and\ \bibinfo {author}
  {\bibfnamefont {L.-C.}\ \bibnamefont {Kwek}},\ }\bibfield  {title} {\bibinfo
  {title} {Dynamical quantum phase transitions and non-{{Markovian}}
  dynamics},\ }\href {https://doi.org/10.1103/PhysRevA.101.012111} {\bibfield
  {journal} {\bibinfo  {journal} {Physical Review A}\ }\textbf {\bibinfo
  {volume} {101}},\ \bibinfo {pages} {012111} (\bibinfo {year}
  {2020})}\BibitemShut {NoStop}%
\bibitem [{\citenamefont {Starchl}\ and\ \citenamefont
  {Sieberer}(2022)}]{Starchl2022}%
  \BibitemOpen
  \bibfield  {author} {\bibinfo {author} {\bibfnamefont {E.}~\bibnamefont
  {Starchl}}\ and\ \bibinfo {author} {\bibfnamefont {L.~M.}\ \bibnamefont
  {Sieberer}},\ }\bibfield  {title} {\bibinfo {title} {Relaxation to a
  {{Parity-Time Symmetric Generalized Gibbs Ensemble}} after a {{Quantum
  Quench}} in a {{Driven-Dissipative Kitaev Chain}}},\ }\href
  {https://doi.org/10.1103/PhysRevLett.129.220602} {\bibfield  {journal}
  {\bibinfo  {journal} {Physical Review Letters}\ }\textbf {\bibinfo {volume}
  {129}},\ \bibinfo {pages} {220602} (\bibinfo {year} {2022})}\BibitemShut
  {NoStop}%
\bibitem [{\citenamefont {Naji}\ \emph {et~al.}(2022)\citenamefont {Naji},
  \citenamefont {Jafari}, \citenamefont {Jafari},\ and\ \citenamefont
  {Akbari}}]{Naji2022}%
  \BibitemOpen
  \bibfield  {author} {\bibinfo {author} {\bibfnamefont {J.}~\bibnamefont
  {Naji}}, \bibinfo {author} {\bibfnamefont {M.}~\bibnamefont {Jafari}},
  \bibinfo {author} {\bibfnamefont {R.}~\bibnamefont {Jafari}},\ and\ \bibinfo
  {author} {\bibfnamefont {A.}~\bibnamefont {Akbari}},\ }\bibfield  {title}
  {\bibinfo {title} {Dissipative {{Floquet Dynamical Quantum Phase
  Transition}}},\ }\href {https://doi.org/10.1103/PhysRevA.105.022220}
  {\bibfield  {journal} {\bibinfo  {journal} {Physical Review A}\ }\textbf
  {\bibinfo {volume} {105}},\ \bibinfo {pages} {022220} (\bibinfo {year}
  {2022})}\BibitemShut {NoStop}%
\bibitem [{\citenamefont {Kawabata}\ \emph {et~al.}(2022)\citenamefont
  {Kawabata}, \citenamefont {Kulkarni}, \citenamefont {Li}, \citenamefont
  {Numasawa},\ and\ \citenamefont {Ryu}}]{Kawabata2022}%
  \BibitemOpen
  \bibfield  {author} {\bibinfo {author} {\bibfnamefont {K.}~\bibnamefont
  {Kawabata}}, \bibinfo {author} {\bibfnamefont {A.}~\bibnamefont {Kulkarni}},
  \bibinfo {author} {\bibfnamefont {J.}~\bibnamefont {Li}}, \bibinfo {author}
  {\bibfnamefont {T.}~\bibnamefont {Numasawa}},\ and\ \bibinfo {author}
  {\bibfnamefont {S.}~\bibnamefont {Ryu}},\ }\href@noop {} {\bibinfo {title}
  {Dynamical quantum phase transitions in {{SYK Lindbladians}}}} (\bibinfo
  {year} {2022}),\ \Eprint {https://arxiv.org/abs/2210.04093} {arXiv:2210.04093
  [cond-mat, physics:hep-th, physics:quant-ph]} \BibitemShut {NoStop}%
\bibitem [{\citenamefont {Halimeh}\ \emph {et~al.}(2017)\citenamefont
  {Halimeh}, \citenamefont {{Zauner-Stauber}}, \citenamefont {McCulloch},
  \citenamefont {{de Vega}}, \citenamefont {Schollw{\"o}ck},\ and\
  \citenamefont {Kastner}}]{Halimeh2017a}%
  \BibitemOpen
  \bibfield  {author} {\bibinfo {author} {\bibfnamefont {J.~C.}\ \bibnamefont
  {Halimeh}}, \bibinfo {author} {\bibfnamefont {V.}~\bibnamefont
  {{Zauner-Stauber}}}, \bibinfo {author} {\bibfnamefont {I.~P.}\ \bibnamefont
  {McCulloch}}, \bibinfo {author} {\bibfnamefont {I.}~\bibnamefont {{de
  Vega}}}, \bibinfo {author} {\bibfnamefont {U.}~\bibnamefont
  {Schollw{\"o}ck}},\ and\ \bibinfo {author} {\bibfnamefont {M.}~\bibnamefont
  {Kastner}},\ }\bibfield  {title} {\bibinfo {title} {Prethermalization and
  persistent order in the absence of a thermal phase transition},\ }\href
  {https://doi.org/10.1103/PhysRevB.95.024302} {\bibfield  {journal} {\bibinfo
  {journal} {Physical Review B}\ }\textbf {\bibinfo {volume} {95}},\ \bibinfo
  {pages} {024302} (\bibinfo {year} {2017})}\BibitemShut {NoStop}%
\bibitem [{Note1()}]{Note1}%
  \BibitemOpen
  \bibinfo {note} {We note here however that the discontinuities in the rate
  function seen in Ref.~\cite {Hou2022} are hard to understand from a unitary
  time evolution perspective, and we see no evidence for such behavior in our
  results.}\BibitemShut {Stop}%
\bibitem [{\citenamefont {Lieb}\ \emph {et~al.}(1961)\citenamefont {Lieb},
  \citenamefont {Schultz},\ and\ \citenamefont {Mattis}}]{Lieb1961}%
  \BibitemOpen
  \bibfield  {author} {\bibinfo {author} {\bibfnamefont {E.}~\bibnamefont
  {Lieb}}, \bibinfo {author} {\bibfnamefont {T.}~\bibnamefont {Schultz}},\ and\
  \bibinfo {author} {\bibfnamefont {D.}~\bibnamefont {Mattis}},\ }\bibfield
  {title} {\bibinfo {title} {Two soluble models of an antiferromagnetic
  chain},\ }\href {https://doi.org/10.1016/0003-4916(61)90115-4} {\bibfield
  {journal} {\bibinfo  {journal} {Annals of Physics}\ }\textbf {\bibinfo
  {volume} {16}},\ \bibinfo {pages} {407} (\bibinfo {year} {1961})}\BibitemShut
  {NoStop}%
\bibitem [{\citenamefont {Kitaev}(2001)}]{Kitaev2001}%
  \BibitemOpen
  \bibfield  {author} {\bibinfo {author} {\bibfnamefont {A.~Y.}\ \bibnamefont
  {Kitaev}},\ }\bibfield  {title} {\bibinfo {title} {Unpaired {{Majorana}}
  fermions in quantum wires},\ }\href@noop {} {\bibfield  {journal} {\bibinfo
  {journal} {Physics-Uspekhi}\ }\textbf {\bibinfo {volume} {44}},\ \bibinfo
  {pages} {131} (\bibinfo {year} {2001})}\BibitemShut {NoStop}%
\bibitem [{\citenamefont {Sachdev}(2011)}]{Sachdev2011}%
  \BibitemOpen
  \bibfield  {author} {\bibinfo {author} {\bibfnamefont {S.}~\bibnamefont
  {Sachdev}},\ }\href@noop {} {\emph {\bibinfo {title} {Quantum {{Phase
  Transitions}}}}}\ (\bibinfo  {publisher} {{Cambridge University Press}},\
  \bibinfo {address} {{Cambridge}},\ \bibinfo {year} {2011})\BibitemShut
  {NoStop}%
\bibitem [{\citenamefont {Carr}(2011)}]{Carr2011}%
  \BibitemOpen
  \bibfield  {author} {\bibinfo {author} {\bibfnamefont {L.~D.}\ \bibnamefont
  {Carr}},\ }\href@noop {} {\emph {\bibinfo {title} {Understanding Quantum
  Phase Transitions}}}\ (\bibinfo  {publisher} {{CRC Press}},\ \bibinfo
  {address} {{Boca Raton, FL}},\ \bibinfo {year} {2011})\BibitemShut {NoStop}%
\bibitem [{\citenamefont {Fisher}(1965)}]{Fisher1965}%
  \BibitemOpen
  \bibfield  {author} {\bibinfo {author} {\bibfnamefont {M.}~\bibnamefont
  {Fisher}},\ }\href@noop {} {\emph {\bibinfo {title} {Boulder {{Lectures}} in
  {{Theoretical Physics}}}}},\ Vol.~\bibinfo {volume} {7}\ (\bibinfo
  {publisher} {{University of Colorado}},\ \bibinfo {address} {{Boulder}},\
  \bibinfo {year} {1965})\BibitemShut {NoStop}%
\bibitem [{\citenamefont {Yang}\ and\ \citenamefont {Lee}(1952)}]{Yang1952}%
  \BibitemOpen
  \bibfield  {author} {\bibinfo {author} {\bibfnamefont {C.~N.}\ \bibnamefont
  {Yang}}\ and\ \bibinfo {author} {\bibfnamefont {T.~D.}\ \bibnamefont {Lee}},\
  }\bibfield  {title} {\bibinfo {title} {Statistical theory of equations of
  state and phase transitions. {{I}}. {{Theory}} of condensation},\ }\href
  {https://doi.org/10.1103/PhysRev.87.404} {\bibfield  {journal} {\bibinfo
  {journal} {Physical Review}\ }\textbf {\bibinfo {volume} {87}},\ \bibinfo
  {pages} {404} (\bibinfo {year} {1952})}\BibitemShut {NoStop}%
\bibitem [{\citenamefont {Zanardi}\ and\ \citenamefont
  {Paunkovi{\'c}}(2006)}]{Zanardi2006}%
  \BibitemOpen
  \bibfield  {author} {\bibinfo {author} {\bibfnamefont {P.}~\bibnamefont
  {Zanardi}}\ and\ \bibinfo {author} {\bibfnamefont {N.}~\bibnamefont
  {Paunkovi{\'c}}},\ }\bibfield  {title} {\bibinfo {title} {Ground state
  overlap and quantum phase transitions},\ }\href
  {https://doi.org/10.1103/PhysRevE.74.031123} {\bibfield  {journal} {\bibinfo
  {journal} {Physical Review E}\ }\textbf {\bibinfo {volume} {74}},\ \bibinfo
  {pages} {031123} (\bibinfo {year} {2006})}\BibitemShut {NoStop}%
\bibitem [{\citenamefont {Zanardi}\ \emph {et~al.}(2007)\citenamefont
  {Zanardi}, \citenamefont {Quan}, \citenamefont {Wang},\ and\ \citenamefont
  {Sun}}]{Zanardi2007b}%
  \BibitemOpen
  \bibfield  {author} {\bibinfo {author} {\bibfnamefont {P.}~\bibnamefont
  {Zanardi}}, \bibinfo {author} {\bibfnamefont {H.~T.}\ \bibnamefont {Quan}},
  \bibinfo {author} {\bibfnamefont {X.}~\bibnamefont {Wang}},\ and\ \bibinfo
  {author} {\bibfnamefont {C.~P.}\ \bibnamefont {Sun}},\ }\bibfield  {title}
  {\bibinfo {title} {Mixed-state fidelity and quantum criticality at finite
  temperature},\ }\href {https://doi.org/10.1103/PhysRevA.75.032109} {\bibfield
   {journal} {\bibinfo  {journal} {Physical Review A}\ }\textbf {\bibinfo
  {volume} {75}},\ \bibinfo {pages} {032109} (\bibinfo {year}
  {2007})}\BibitemShut {NoStop}%
\bibitem [{\citenamefont {Heyl}\ \emph {et~al.}(2018)\citenamefont {Heyl},
  \citenamefont {Pollmann},\ and\ \citenamefont {D{\'o}ra}}]{Heyl2018}%
  \BibitemOpen
  \bibfield  {author} {\bibinfo {author} {\bibfnamefont {M.}~\bibnamefont
  {Heyl}}, \bibinfo {author} {\bibfnamefont {F.}~\bibnamefont {Pollmann}},\
  and\ \bibinfo {author} {\bibfnamefont {B.}~\bibnamefont {D{\'o}ra}},\
  }\bibfield  {title} {\bibinfo {title} {Detecting equilibrium and dynamical
  quantum phase transitions in {{Ising}} chains via out-of-time-ordered
  correlators},\ }\href {https://doi.org/10.1103/RevModPhys.86.153} {\bibfield
  {journal} {\bibinfo  {journal} {Physical Review Letters}\ }\textbf {\bibinfo
  {volume} {121}},\ \bibinfo {pages} {16801} (\bibinfo {year}
  {2018})}\BibitemShut {NoStop}%
\end{thebibliography}

%Control: production of eprint (1) enabled
%

\end{document}